\documentclass[12pt, reqno, a4paper]{amsart} 
\usepackage{latexsym}
\usepackage[english]{babel}
\usepackage{fancyhdr}
\usepackage[mathscr]{eucal}
\usepackage{amsfonts}
\usepackage{amssymb}

\usepackage{amscd}
\usepackage{bbm}
\usepackage{latexsym}
\usepackage{color}
\usepackage[dvipsnames]{xcolor}

\theoremstyle{plain}
\newtheorem{theorem}{Theorem}[section]
\newtheorem{remark}[theorem]{Remark}
\newtheorem{lemma}[theorem]{Lemma}
\newtheorem{corollary}[theorem]{Corollary}
\newtheorem{proposition}[theorem]{Proposition}

\theoremstyle{definition}
\newtheorem{definition}[theorem]{Definition}

\newcommand {\absleq} {{\leq_{|\, \cdot\, |}\, }}
\def\tD {{\tilde{\mathcal D}}}
\def\tE {{\tilde{E}}}
\def\tA {{\tilde{A}}}

\def\ta {{\tilde{a}}}

\def\Cg{{\mathcal C}}
\def\Dg{{\mathcal D}}
\def\Fg{{\mathcal F}}
\def\Gg{{\mathcal G}}
\def\Hg {{\mathcal H}}

\def\Lg {{\mathcal L}}
\def\Mg{{\mathcal M}}
\def\Ng{{\mathcal N}}

\def\Og{{\mathcal O}}
\def\Qg {{\mathcal Q}}

\def\Rg {{\mathcal R}}

\def\Xg {{\mathcal X}}
\def\Yg {{\mathcal Y}}
\def\Zg {{\mathcal Z}}

\def\tA {{\tilde{A}}}
\def\tB {{\tilde{B}}}

\def\tG {{\tilde{G}}}
\def\tL {{\tilde{L}}}

\def\tT {{\tilde{T}}}
\def\ta {{\tilde{a}}}
\def\tb {{\tilde{b}}}
\def\tc {{\tilde{c}}}
\numberwithin{equation}{section}
\def\HL {{L^2}}
\def\Lam{{\Lambda}}

\def\tf{{\tilde{f}}}
\def\tg{{\tilde{g}}}
\def\tu{{\tilde{u}}}

\def\ben{\begin{enumerate}}
\def\een{\end{enumerate}}
\def\bgdf{\begin{definition}}
\def\eddf{\end{definition}}
\def\bglm{\begin{lemma}}
\def\edlm{\end{lemma}}
\def\bgpf{\begin{proof}}
\def\edpf{\end{proof}}
\def\bgth{\begin{theorem}}
\def\edth{\end{theorem}}
\def\bgcor{\begin{corollary}}
\def\edcor{\end{corollary}}
\def\bgprop{\begin{proposition}}
\def\edprop{\end{proposition}}
\def\bgrm{\begin{remark}}
\def\edrm{\end{remark}}

\def\lbeq(#1){\label{eqn:#1}}
\def\refeq(#1){{\rm (\ref{eqn:#1})}}
\def\refeqs(#1,#2){{\rm (\ref{eqn:#1}) and (\ref{eqn:#2})}}
\def\refeqss(#1,#2,#3){{\rm (\ref{eqn:#1}),\ (\ref{eqn:#2}) and (\ref{eqn:#3})}}
\def\refeqsss(#1,#2,#3,#4){{\rm (\ref{eqn:#1}),\ (\ref{eqn:#2}),\ 
(\ref{eqn:#3}) and (\ref{eqn:#4})}}
\def\lbth(#1){\label{th:#1}}
\def\refth(#1){{\rm Theorem \ref{th:#1}}}
\def\refths(#1,#2){{\rm Theorems \ref{th:#1} and \ref{th:#2}}}
\def\refthb(#1){{\bf Theorem \ref{th:#1}}}
\def\lblm(#1){\label{lm:#1}}
\def\reflm(#1){{\rm Lemma \ref{lm:#1}}}
\def\reflms(#1,#2){{\rm Lemmas \ref{lm:#1} and \ref{lm:#2}}}
\def\reflmss(#1,#2,#3){{\rm Lemmas \ref{lm:#1}, \ref{lm:#2} and \ref{lm:#3}}}
\def\reflmsss(#1,#2,#3,#4){{\rm Lemmas \ref{lm:#1},\, \ref{lm:#2},\, \ref{lm:#3} and \ref{lm:#4}}}
\def\reflmb(#1){{\bf Lemma \ref{lm:#1}}}
\def\lbprop(#1){\label{prp:#1}}
\def\refprop(#1){{\rm Proposition \ref{prp:#1}}}
\def\refprops(#1,#2,#3,#4){{\rm Propositions \ref{prp:#1},\, \ref{prp:#2},
\, \ref{prp:#3} \, and \ref{prp:#4}}}
\def\refpropb(#1){{\bf Proposition \ref{prp:#1}.}}
\def\lbcor(#1){\label{cor:#1}}
\def\refcor(#1){{\rm Corollary \ref{cor:#1}}}
\def\refcors(#1,#2){{\rm Corollaries \ref{cor:#1} and \ref{cor:#2}}}
\def\lbrm(#1){\label{rm:#1}}
\def\refrm(#1){{\rm Remark \ref{rm:#1}}}
\def\lbass(#1){\label{ass:#1}}
\def\refass(#1){{\rm Assumption \ref{ass:#1}}}
\def\lbdf(#1){\label{df:#1}}
\def\refdf(#1){{\rm Definition \ref{df:#1}}}
\def\refdfs(#1,#2){{\rm Definitions \ref{def:#1} and \ref{def:#2}}}
\def\lbsec(#1){\label{s:#1}}
\def\refsec(#1){{\rm \S\ref{s:#1}}}
\def\lbsubsec(#1){\label{ss:#1}}
\def\refsubsec(#1){{\rm \S\ref{ss:#1}}}

\newcommand{\lam}{\lambda}
\def\ab{{\bf a}}

\def\Bb{{\bf B}}

\def\ph{{\varphi}}

\def\bqn{\begin{equation}}
\def\eqn{\end{equation}}
\def\C{{\mathbb C}}
\def\N{{\mathbb N}}

 \def\Cb{{\overline{\mathbb C}}}

\def\R{{\mathbb R}}
\def\a{\alpha}
\def\b{\beta}
\def\c{\gamma}
\def\Ga{\Gamma}
\def\d{\delta}

\def\Sg{{\mathcal S}}
\def\Vg{{\mathcal V}}
\def\D{\Delta}

\def\ep{\varepsilon}
\def\z{\zeta}
\def\th{\theta}

\def\m{\mu}
\def\n{\nu}
\def\r{\rho}
\def\s{\sigma}
\def\t{\tau}
\def\w{\omega}
\def\W{\Omega}
\def\la{\langle}
\def\ra{\rangle}

\def\lap{\Delta}
\def\ax{{\la x \ra}}
\def\ay{{\la y \ra}}
\def\az{{\la z \ra}}
\def\aw{{\la w \ra}}
\def\pa{{\partial}} 

\def\tv{\tilde{v}}

\def\br{\begin{array}}
\def\er{\end{array}}

\def\Ker{\rm Ker\,}
\def\rank{{\rm rank\,}}

\def\tW{\tilde{W}}

\begin{document}

\title[$L^p$-boundedness of wave operators for $\lap^2 + V$ on $\R^4$]
{The $L^p$-boundedness of wave operators for fourth order  
Schr\"odinger operators on $\R^4$}

\author[A.~Galtbayar]{Artbazar Galtbayar}
\address{Center of Mathematics for Applications
and Department of Applied Mathematics \\
National University of Mongolia \\
University Street 3, Ulaanbaatar (Mongolia). \\ 
\footnote{Supported by project P2020-3984 of
the National University of Mongolia}}
\email{galtbayar@num.edu.mn}

\author[K.~Yajima]{Kenji Yajima}
\address{Department of Mathematics \\ Gakushuin University 
\\ 1-5-1 Mejiro \\ Toshima-ku \\ Tokyo 171-8588 (Japan). \\ 
\footnote{Supported by JSPS grant in aid for scientific research No. 19K03589}}
\email{kenji.yajima@gakushuin.ac.jp}

\begin{abstract} 
We prove that the wave operators of scattering theory 
for the fourth order Schr\"odinger operator $\lap^2 + V(x)$ in $\R^4$ 
are bounded in $L^p(\R^4)$ for the set of $p$'s of $(1,\infty)$ 
depending on the kind of spectral singularities of $H$ at zero 
which can be described by the space of bounded solutions of 
$(\lap^2 + V(x))u(x)=0$. 
\end{abstract}

\maketitle

\section{Introduction}
Let $H = \lap^2 + V$, $\lap=\pa^2/\pa{x}_1^2+ \cdots+ \pa^2/\pa{x}_4^2$, 
be the fourth order Schr\"odinger operator 
on $\R^4$ with real potentials $V(x)$ which satisfy the short-range 
condition that  
\bqn 
\lbeq(short)
\sup_{y \in \R^4} (1+|y|)^{\d} \|V(x)\|_{L^q(|x-y|<1)}<\infty \ 
\mbox{for a $q>1$ and $\d>1$}.
\eqn 
The operator $H$ is defined via 
the closed and bounded from below quadratic form 
$q(u) = \int_{\R^4}(|\lap u(x)|^2 + V(x)|u(x)|^2)dx$ 
with domain $D(q)=H^2(\R^4)$ and is selfadjoint in $L^2(\R^4)$ 
(cf. \cite{Ka}). The spectrum of $H$ consists of the absolutely 
continuous (AC for short) part $[0,\infty)$ and the bounded set of 
eigenvalues which are discrete in $\R\setminus \{0\}$ 
and accumulate  possibly  at zero; 
the wave operators $W_{\pm}$ defined by the strong limits in $L^2(\R^4)$: 
\[
W_{\pm}= \lim_{t \to \pm\infty} e^{itH} e^{-itH_0}, \quad H_0 = \lap^2 
\]
exist and ${\textrm{Range}}\, W_{\pm} = L^2_{ac}(H)$, 
the AC subspace of $L^2(\R^4)$ for $H$ (\cite{Ku}). 

The wave operators  satisfy the intertwining property:  
\bqn \lbeq(inter) 
f(H)P_{ac}(H) = W_\pm f(H_0) W_{\pm}^\ast  
\eqn 
for Borel functions $f$ on $\R$, where $P_{ac}(H)$ is 
the projection to $L^2_{ac}(H)$. 
It follows that, if $W_\pm$ are bounded in 
$L^p(\R^4)$ for $p$ in a subset $I $ of 
$\{p\colon 1\leq p \leq \infty\}$ and $I^\ast=\{p/(p-1)\colon p \in I\}$, 
then 
\bqn \lbeq(inter-estimate)
\|f(H)P_{ac}(H)\|_{\Bb(L^{q},L^p)}\leq C \|f(H_0)\|_{\Bb(L^{q},L^p)}
\eqn  
for $p\in I$ and $q\in I^\ast$ with the constant $C$  
which is independent of $f$ and, $L^p$-mapping properties of 
$f(H)P_{ac}(H)$, the AC part of $f(H)$, may be deduced from 
those of $f(H_0)$ which is the Fourier multiplier by $f(|\xi|^4)$. 
Here for Banach spaces $\Xg$ and $\Yg$, $\Bb(\Xg, \Yg)$ 
is the Banach space 
of bounded operators from $\Xg$ to $\Yg$ and $\Bb(\Xg)=\Bb(\Xg, \Xg)$. 

In this paper, we study whether or not $W_\pm$ originally defined on 
$(L^2\cap L^p)(\R^4)$, $1\leq p\leq \infty$ can be extended 
to bounded operators 
in $L^p(\R^4)$ for $p$ in a certain range of $p\in [1,\infty]$. 
For $1\leq p \leq \infty$ 
and $D \subset \R^4$, $\|u\|_{L^p(D)}$ is the norm of $L^p(D)$, 
$\|u\|_p= \|u\|_{L^p(\R^4)}$, $\|u\|=\|u\|_2$ and $(u,v)$ is 
the inner product of $L^2(\R^4)$; the notation $(u,v)$ will be 
used whenever the 
integral $\int_{\R^4} u(x)\overline{v(x)}dx$ makes sense, e.g. for 
$u \in \Sg(\R^4)$ and $v \in \Sg'(\R^4)$;  
\[
L^{p}_{loc,u}(\R^4)=\{u \colon  
\|u\|_{L^p_{loc, u}} \colon = \sup\{\|u(x)\|_{L^p(|x-y|\leq 2)} \colon 
y\in \R^4\}<\infty\}.  
\]
We define the Fourier transform $\Fg u(\xi)$ or $\hat{u}(\xi)$ of $u$ by 
\[
\hat{u}(\xi)= \Fg u(\xi)
= \frac1{(2\pi)^2}\int_{\R^4} e^{-ix\xi}u(x) dx;  
\]
the multiplication operator with $f(\xi)$ is denoted by $M_f$;  
\[
f(D)\colon = \Fg^\ast M_{f} \Fg    
\]
is the Fourier multiplier defined by $f$.  

Let $\chi_{\leq}(\lam)$ and $\chi_{\geq}(\lam)$ be smooth functions 
on $[0,\infty)$ such that 
\[
\chi_{\leq}(\lam)= \left\{\br{ll} 1, \ & \lam\leq 1, 
\\ 0, \ & \lam \geq 2, \er\right. \quad 
\chi_{\leq}(\lam)+\chi_{\geq}(\lam)=1 
\]
and, for $a>0$, $\chi_{\leq{a}}(\lam)= \chi_{\leq}(\lam/a)$ and 
$\chi_{\geq{a}}(\lam)= \chi_{\geq}(\lam/a)$. 
We define the ``high'' and the ``low'' energy parts of $W_{\pm}$ by 
$W_{\pm}\chi_{\geq a}(|D|)$ and $W_{\pm}\chi_{\leq a}(|D|)$ 
respectively. 

For the high energy part we have 
following theorems. Let $\ax=(1+|x|^2)^{1/2}$ for $x \in \R^d$, $d\in \N$. 

\bgth \lbth(small) Suppose 
$V \in L^{q}_{loc,u}(\R^4)$ for a $q>1$ and 
$\la \log |x|\ra^{2} V \in L^1(\R^4)$. Let $a>0$ and $1<p<\infty$.  
Then, there exists constant $c_0$ such that 
$W_{\pm} \chi_{\geq{a}}(|D|)$ are bounded in $L^p(\R^4)$ 
if $V$ satisfies 
$\|V\|_{L^{q}_{loc,u}}+ \|\la \log |x|\ra^{2} V\|_{L^1}\leq c_0$. 
\edth 

\bgrm 
In \refth(small) $V$ does not actually satisfy 
\refeq(short), however, for any $a>0$, $|V|^\frac12$ is Kato-smooth 
on $[a,\infty)$ for small $c_0$ 
(\cite{Kato-s}, \cite{RS3}, see \reflm(first-est)) and 
$W_{\pm}\chi_{\leq {a}}(|D|)= \chi_{\leq {a}}(\sqrt{H}) W_\pm $
exist. 
\edrm 

The same result holds for larger $V$ if $V$ decays faster at infinity; 
$V$ in \refth(high) satisfies the assumption \refeq(short).  

\bgth \lbth(high) Suppose that 
$\ax^{3} V \in  L^1 (\R^3)$, $V\in L^q(\R^4)$ for a $q>1$ 
and that $H$ has no positive eigenvalues. Then, for any $a>0$, 
$W_{\pm}\chi_{\geq a}(|D|)$ are bounded in $L^p(\R^4)$ 
for $1<p<\infty$.
\edth 

We remark that $H$ can have positive eigenvalues for ``very nice'' 
potentials $V$ (\cite{FSW,MWY}) in contrast to the case of 
ordinary Schr\"odinger operators $-\lap + V$ which 
have no positive eigenvalues for the large class of short-range 
potentials (\cite{IJ,KT}). We refer to \cite{FSW}, \cite{MWY} 
and reference therein 
for more informations on positive eigenvalues for $(-\lap)^m + V$, 
$m=2, 3, \dots$. We shall assume in this paper  
positive eigenvalues are absent from $H$. 

The range of $p$ for which the low energy parts 
$W_\pm\chi_{\leq a}(|D|)$ are bounded in $L^p(\R^4)$ 
depends on the space of zero energy resonances:    
\bqn \lbeq(Ng-def)
\Ng_\infty (H) \colon 
= \{u \colon u \in L^\infty(\R^4) \colon (\lap^2 + V(x))u=0\}. 
\eqn  
In \refsec(resonance) we shall prove   
the following lemma which is the version of the result in \cite{GT-1}.  

\bglm \lblm(reso-intro)  
Suppose $\la \log |x| \ra^2 \ax^3 V \in (L^1 \cap L^q)(\R^4)$ for a $q>1$.
Then, $\Ng_\infty (H)$ is finite-dimensional real vector space and, 
for any real-valued $\ph\in \Ng_\infty(H)$, there exist 
$c_0 \in \R$, $\ab \in \R^4$ and real symmetric matrix $A$ such that 
\bqn 
\ph(x) = - c_0 + \frac{\ab \cdot {x}}{|x|^2} 
+ \frac{Ax \cdot x}{|x|^4} + O(|x|^{-3})\quad (|x|\to \infty)\,.
\lbeq(reso-intro)
\eqn
\edlm 
We call $\ph\in \Ng_\infty(H)\setminus\{0\}$ $s$-wave, $p$-wave or 
$d$-wave resonance respectively if $c_0\not=0$, $c_0=0$ and $\ab\not={\bf 0}$ 
or $c_0=0$, $\ab={\bf 0}$ and $A\not=0$; if  
$c_0=0$, $\ab={\bf 0}$, $A=0$, then $\ph$ is zero energy eigenfunction 
of $H$.

\bgth \lbth(main) Assume that $H$ has no positive eigenvalues. Let $q>1$.
\ben
\item[{\rm (1)}] 
Suppose that 
$\la \log|x| \ra^2 \ax^{8}V \in (L^1\cap L^q)(\R^4)$. Let 
$\Ng_{\infty}(H)=\{0\}$ or $\Ng_{\infty}(H)$ consist only of 
$s$-wave resonances. Then, 
$W_\pm$ are bounded in $L^p(\R^4)$ for all $1<p<\infty$. 
\item[{\rm (2)}] 
Suppose that $\la \log |x| \ra^2 \ax^{12}V \in (L^1\cap L^q)(\R^4)$. 
Let $\Ng_{\infty}(H)$ consist only of $s$- and $p$-wave resonances.   
Then, $W_{\pm}$ are bounded in $L^p(\R^4)$ for $1<p<4$ and are unbounded 
if $4\leq p \leq \infty$. 
\item[{\rm (3)}] Suppose that $\la \log |x| \ra^2 \ax^{16}V \in 
(L^1 \cap L^q)(\R^4)$. Let $\Ng_\infty(H)$ contain $d$-wave resonances. 
Then, $W_{\pm}$ are bounded in $L^p(\R^4)$ for $1<p\leq 2$.
\item[{\rm (4)}] Suppose that $\la \log |x| \ra^2 \ax^{16} V \in 
(L^1 \cap L^q)(\R^4)$. Let $\Ng_\infty(H)$ consist only of 
$s$-, $p$-wave resonances and zero energy eigenfunctions. Then, 
$W_{\pm}$ are bounded in $L^p(\R^4)$ for $1<p<4$.  
\een
\edth 

\bgrm We believe that $W_{\pm}$ are unbounded in $L^p(\R^4)$ for $p>2$ in 
{\rm (3)} and for $p\geq 4$ in {\rm (4)}, however, we were not able to 
prove this.
\edrm  
We rephrase \refth(main) in terms of the singularities of the resolvent 
$R(\lam) = (H-\lam^4)^{-1}$ at $\lam=0$, which is more directly connected 
to the proof given below. For stating this version 
of the theorem, we need some more notation. Let for 
$z \in \C^{++}=\{z \in \C: \Re z>0, \Im z>0\}$   
\[
R_0(z^4)= (H_0-z^4)^{-1},  \ G_0(z)=(-\lap-z^2)^{-1} 
\]
be resolvents of $H_0$ and $-\lap = \Fg^\ast M_{|\xi|^2} \Fg$; 
$R_0^{+}(\lam^4)=\lim_{\ep \downarrow 0}R_0(\lam^4 +i\ep)$, 
$G_0(\lam), G_0(i\lam)$, $\lam>0$ are their boundary 
values. For $z\in \Cb^{++}=\{z \in \C: \Re z\geq 0, \Im z\geq 0\}$, 
$z\not=0$, we have \bqn 
R_0(z^4)u(x)
= \frac{1}{2z^2}(G_0(z) - G_0(iz))u(x), 
\quad u \in \Dg_{\ast}.
\lbeq(1-1)
\eqn
It is well known (e.g. \cite{AS}) that $G_0(z)$ 
is the convolution operator with 
\bqn 
\Gg_{z}(x) = \frac1{(2\pi)^{4}}\int_{\R^4}\frac{e^{ix\xi}d\xi}
{|\xi|^2 - z^2}= 
\frac{i}{4}\left(\frac{z}{2\pi |x|}\right) H^{(1)}_1 (z|x|)\,, 
\lbeq(Green) 
\eqn 
where $H^{(1)}_{1}(z)$ is the Hankel function and $\Gg_{z}(x)$ 
has the expansion:  
\bqn 
\frac1{4\pi^2|x|^2}+ 
\frac{z^2}{4\pi}\sum_{n=0}^\infty 
\left(g(z)+\frac{c_n}{2\pi}-\frac{\log |x|}{2\pi}\right)
\frac{(-z^2|x|^2/4)^n}{n!(n+1)!}\,. \lbeq(Green-a)  
\eqn 
Here $c_n=1/(2(n+1))+ \sum_{j=1}^n j^{-1}$ and, with the principal branch, 
\bqn 
g(z)= -\frac1{2\pi}\log\left(\frac{z}{2}\right)-\frac{\c}{2\pi} 
+ \frac{i}{4}\,,  \lbeq(g)
\eqn 
$\c$ being Euler's constant. Thus, $R_0^{+}(\lam^4)$, $\lam>0$, 
is the convolution with 
\[ 
\Rg_{\lam}(x)= \frac{1}{2\lam^2}(\Gg_{\lam}(x)- \Gg_{i\lam}(x))
\] 
and \refeq(Green-a) implies $\Rg_{\lam}(x)= \Rg(\lam|x|)$ with   
$\Rg(\lam)$ being equal to   
\bqn  \lbeq(Rg0-rev) 
\frac{1}{4\pi}\sum_{n, even} 
\left(g(\lam)+\frac{c_n}{2\pi}-\frac{i}{8}\right)
\frac{(\lam^2/4)^n}{n!(n+1)!}
- \frac{i}{32\pi}\sum_{n, odd} \frac{(\lam^2/4)^n}{n!(n+1)!}\,.
\eqn%
Reordering \refeq(Rg0-rev) in the descendent order as 
$\lam \to 0$, we obtain  
\begin{gather}
\Rg_{\lam}(x) =\tg_0(\lam)-\frac{\log |x|}{8\pi^2} 
-i \frac{\lam^2|x|^2}{4^4\pi} + \frac{\lam^4\tg_2(\lam) |x|^4 }{3\cdot 4^3}
-\frac{\lam^4|x|^4\log |x| }{6\cdot 4^4\pi^2} + \cdots \notag \\
=\tg_0(\lam)+ N_0(x) + \lam^2 G_2(x)  
+ \lam^4 \tg_2(\lam)G_4(x)+ \lam^4 G_{4,l}(x) + \cdots\,, \lbeq(descent) \\
\tilde{g}_n(\lam) 
= \frac1{4\pi}\left(g(\lam)+\frac{c_n}{2\pi}-\frac{i}{8}\right), \ 
n=0, 1, \dots,  \lbeq(tgn)
\end{gather}  
where \refeq(descent) is the definition and where, if $n$ is odd, 
$G_{2n,l}(x)=0$ and no factor $\tg_{2n}(\lam)$ in front 
of $G_{2n}(x)$ appears.  

Let $N_0, G_{2n}, G_{2n,l}$ be the convolutions  with 
$N_0(x), G_{2n}(x), G_{2n,l}(x)$  for $n=1,2, \dots$ and 
with $v(x)\colon = |V(x)|^{\frac12}$
\[ 
N_0^{(v)} = M_v N_0 M_v, \ G_{2n}^{(v)}= M_v G_{2n} M_v, \ 
G_{2n,l}^{(v)}= M_v G_{2n,l} M_v.
\] 
Let ${\textrm{sign}}\, a = \pm 1$ if $\pm a>0$  and 
${\textrm{sign}}\, a =0$ if $a=0$,  
\[ 
U(x) = {\textrm{sign}}\, V (x), \ \mbox{and} \ w(x)= U(x) v(x) \ 
\] 
so that $V(x) = v(x) w(x)$. Define 
$g_0(\lam) = \|V\|_1 \tg_0(\lam)$, $\tv= \|v\|_2^{-1} v$;  
\[
P= \tv\otimes \tv, \ Q=1-P, \ \mbox{and} \ T_0 = M_U + N_0^{(v)}.
\]
Define the function $\Mg(\lam^4)$ of $\lam>0$ with values in 
$\Bb(\HL)$ by 
\bqn \lbeq(Mdef)
\Mg^{+}(\lam^4)= M_U + M_vR_0^{+}(\lam^4)M_v. 
\eqn 
From \refeq(descent) we have 
\bqn \lbeq(Uplus)
\Mg^{+}(\lam^4)= T_0+ g_0(\lam)P 
+\lam^2 G_2^{(v)} + \lam^4 \tg_2(\lam)G_4^{(v)} + \lam^4 G_{4,l}^{(v)} + \cdots.
\eqn 
It follows from the absence of positive eigenvalues gfrom $H$, 
$\Mg^{+}(\lam^4)^{-1}$ exists for $\lam>0$ in $\Bb(\HL)$ 
and is locally H\"older continuous (\cite{Ku}). 
Here and hereafter we write $\HL$ for $\HL(\R^4)$. 
The operator $M_v \Mg^{+}(\lam^4)^{-1}M_v $ will 
play the central role in the paper and we introduce the 
short hand notation 
\bqn \lbeq(Qgv)
\Qg_v (\lam)=M_v \Mg^{+}(\lam^4)^{-1}M_v .
\eqn 
As is well known   
\bqn \lbeq(off)
R^{+}(\lam^4)=R_0^{+}(\lam^4)- R_0^{+}(\lam^4)\Qg_v(\lam) R_0^{+}(\lam^4).
\eqn 

The following definition is due to \cite{GT-1}. 
It is shown that ${\rm Ker}_{Q\HL}QT_0Q$ is finite-dimensional 
(cf. \reflm(finite-dim)) and operators in \refdf(singularities) 
are bounded (\reflm(members));  
$\Hg_2$ is the Hilbert space of Hilbert-Schmidt operators on $\HL$.  
As can be seen from \refeq(Uplus), \refeq(Qgv) and \refeq(off), 
\refdf(singularities) is closely connected to the singularities 
of $\Mg(\lam^4)^{-1}$ and $R^{+}(\lam^4)$ at $\lam =0$.

\bgdf \lbdf(singularities) 
{\rm (1)} We say that $H$ is regular at zero if $QT_0 Q$ 
is invertible in $Q\HL$ and is singular at zero otherwise. 
If $H$ is singular at zero, let $S_1$ be the projection in $Q\HL$ to 
${\rm Ker}_{Q\HL}QT_0Q$. \\[3pt]
{\rm (2)} Suppose that $H$ is singular at zero. We define that:
\ben 
\item[{\rm (2-1)}]  $H$ has singularity of the first kind 
if $T_1=S_1T_0 P T_0 S_1$ is invertible in $S_1\HL$. 

\item[{\rm (2-2)}] If $T_1\vert_{S_1\HL}$ is not invertible, 
let $S_2$ be the projection in $S_1\HL$ to 
${\rm Ker}_{S_1\HL} T_1\vert_{S_1\HL}$. 
$H$ has singularity of the second kind if  
$T_2=S_2 G_2^{(v)}S_2$ is invertible in $S_2\HL$. Here  
$T_1\vert_{S_1\HL}$ is the part of $T_1$ in $S_1\HL$.

\item[{\rm (2-3)}] If $T_2\vert_{S_2\HL}$ is not invertible, 
let $S_3$ be the projection in $S_2\HL$ to 
${\rm Ker}_{S_2\HL}T_2\vert_{S_2\HL}$. 
$H$ has singularity of the third kind if 
$T_3=S_3G_4^{(v)}S_3$ is invertible in $S_3\HL$.

\item[{\rm (2-4)}] If $T_3\vert_{S_3\HL}$ is not invertible, 
we say $H$ has singularity of the fourth kind. 
Let $S_4$ be the projection in $S_3\HL$ to ${\rm Ker}_{S_3\HL} T_3$ and  
$T_4=S_4G^{(v)}_{4,l}M_vS_4$. 
\een
\eddf
It is known (\cite{GT-1}) that $T_4$ is invertible in $S_4 \HL$. 
We clearly have $Q= \colon S_0 \supset S_1 \supset \cdots  \supset S_4$.
We denote the extension of $S_j$ to $\HL$ defined by setting 
as the zero operator 
on $\HL \ominus S_j \HL$ by the same letter $S_j$. 
The nature of the singularities of $H$ at zero is closely connected to the 
structure of $\Ng_\infty(H)$. The following lemma is a slight improvement 
of a result of \cite{GT-1} and will be proved in \refsec(resonance). 

\bglm \lblm(resonance) 
{\rm (1)} Let $\la \log |x| \ra^2 V \in (L^1 \cap L^q)(\R^4)$ 
for a $q>1$. Then, the operator $H$ is singular at zero 
if and only if $\Ng_\infty(H) \not=\{0\}$. In this case  the map $\Phi$ 
defined by 
\bqn \lbeq(inverse)
\Phi(\z)= N_0 M_v\z-\|v\|^{-2} (PT_0 \z, v), \quad \z \in S_1\HL(\R^4) 
\eqn 
is isomorphic from $S_1\HL$ to $\Ng_\infty(H)$ and 
$\Phi^{-1}(\ph)=- w \ph$.\\[3pt]
{\rm (2)} Let $V$ be as in {\rm (1)}. 
Suppose  $H$ has singularity of the first kind, then 
$\rank S_1= 1$ and $H$ has only $s$-wave resonances. \\[3pt]
{\rm (3)} Let $\la \log |x| \ra^2 \ax^3 V \in (L^1 \cap L^q)(\R^4)$. 
Then, $\Phi$ maps 
$\z\in S_1L^2\ominus S_2L^2$, $S_2L^2\ominus S_3L^2$, 
$S_3L^2\ominus S_4L^2$ and $S_4L^2$ to 
$s$-wave, $p$-wave, $d$-wave resonance and 
zero energy eigenfunction, respectively. 
\edlm 

By virtue of \reflm(resonance) \refth(main) can be rephrased as follows. 

\bgth \lbth(main-rephrase)
Assume that $H$ has no positive eigenvalues. Let $q>1$.
\ben 
\item[{\rm (1)}] 
Suppose $\la \log|x| \ra^2 \ax^{8}V \in (L^1\cap L^q)(\R^4)$. If  
$H$ is regular or has singularity of the first kind at zero, 
then $W_\pm$ are bounded in $L^p(\R^4)$ for $1<p<\infty$. 
\item[{\rm (2)}] Suppose  
$\ax^{12}\la \log |x| \ra^2 V \in (L^1\cap L^q)(\R^4)$. If  
$H$ has singularity of the second kind at zero, then 
$W_{\pm}$ are bounded in $L^p(\R^4)$ for $1<p<4$ and are unbounded 
for $4\leq p \leq \infty$. 
\item[{\rm (3)}] Suppose $\ax^{16}\la \log |x|\ra^2 V \in 
(L^1 \cap L^q)(\R^4)$. If $H$ has singularity of the 
third kind at zero, then $W_{\pm}$ are bounded in $L^p(\R^4)$ 
for $1<p\leq 2$.
\item[{\rm (4)}] Suppose  $\ax^{16}\la \log |x|\ra^2 V \in 
(L^1 \cap L^q)(\R^4)$. If $H$ has singularity of the 
fourth kind at zero. Then, $W_{\pm}$ are bounded in $L^p(\R^4)$ for 
for $1<p\leq 2$ if $T_3\not=0$ and for $1<p<4$ if otherwise.  
\een
\edth 

Because of the intertwining property \refeq(inter), 
the problem of $L^p$ boundedness of wave operators has attracted interest 
of many authors and, for the ordinary Schr\"odinger operators $H=-\lap +V$, 
various results depending on the dimensions and on the singularities of $H$ 
at zero have been obtained.  For some more information, we refer to the 
introduction of 
\cite{Ya-2dim-new,Ya-4dim-new} and the references therein, 
\cite{Y-odd-sing,FY,GG,GG-4,KY,EGG,Y-3d-sing} among others. 
  
For wave operators for $H=\lap^2 + V(x)$ the investigation 
started only recently and the following results have been obtained 
under suitable conditions on the decay at infinity and the smoothness 
of $V(x)$ in addition to the absence of positive eigenvalues of $H$. 
When $d=1$, $W_\pm$ are bounded in $L^p(\R^1)$ 
for $1<p<\infty$ but not for $p=1$ and $p=\infty$; they are bounded 
from the Hardy space $H^1$ to $L^1$ and from $L^1$ to $L^1_{w}$
(\cite{MWY}); if $d=3$ and 
$\Ng_\infty=\{u \in L^\infty(\R^3): (\lap^2 + V)u=0\}=0$ 
then $W_\pm$ are bounded in $L^p(\R^3)$ for $1<p<\infty$ (\cite{GG-16}); 
if $d \geq 5$ and 
$\Ng_\infty
=\cap_{\ep>0}
\{u \in \ax^{-\frac{d}{2}+2+\ep}L^2(\R^d): (\lap^2 + V)u=0\}=0$, 
then they are bounded in $L^p(\R^1)$ for all $1\leq p \leq \infty$ 
(\cite{EG,EG-1}). However, no results on $L^p$-boundness of $W_\pm$ are 
known when $d=2, 4$. We should mention, however, 
detailed study on dispersive estimates 
has been carried out by Li-Soffer-Yao \cite{LSY} for $d=2$ and 
Green-Toprak \cite{GT,GT-1} for $d=4$.

The rest of the paper is devoted to the proof of the theorems.
We explain here the basic idea of the proof,  
introducing some more notation and displaying the plan of the paper. 
Various constants whose specific values are not important 
will be denoted by 
the same letter $C$ and it may differ from line to line. 
We prove the theorems only for $W_{-}$ because the complex conjugation 
changes $W_{-}$ to $W_{+}$. 
We often identify integral operators with their 
kernels and say integral operator $K(x,y)$ for 
the operator defined by $K(x,y)$;  
we say $\m(\lam)$, $\lam>0$ is {\it good multiplier} (GMU for short) if 
$\m(|D|)$ is a bounded in $L^p(\R^4)$ for all $1<p<\infty$; 
If $|\mu^{(j)}(\lam)|\leq C \lam^{-j}$ for $0\leq j \leq 3$, 
then $\m(\lam)$ is a GMU (\cite{Stein-old}, p.96). 
We use the function space 
$\Dg_\ast=\{u \in \Sg(\R^4) \colon \hat{u}
\in C_0^\infty(\R^4\setminus\{0\})\}$ which is dense in  
$L^p(\R^4)$ for all $1\leq p <\infty$.

In \S2  we first prove that operators 
in \refdf(singularities) are bounded (\reflm(members)) 
and  give some estimates on the remainders of the series 
\refeq(descent). We then introduce the spectral projection 
$\Pi(\lam)$ for $H_0$ at $\lam^4$ by  
\bqn  \lbeq(spect-proj-def)
\Pi(\lam)u(x)= \frac{2}{\pi{i}}\lim_{\ep \downarrow 0}
(R_0(\lam^4-i\ep)-R_0(\lam^4+i\ep))u(x) \,.
\eqn 
Let $\t_a$ be the translation by $a\in \R^4$: $\t_a{u}(x)=u(x-a)$. Then 
\bqn 
\Pi(\lam)u(x) = \frac1{(2\pi)^2}
\int_{{\mathbb S}^3}
e^{{i\lam}{x\w}}\hat{u}(\lam \w)d\w 
= (\Pi(\lam)\tau_{-x}u)(0), \lbeq(spect-proj)
\eqn  
and the $x$-dependence of $\Pi(\lam) u(x)$ 
is attributed to that of $\tau_{-x}u$, which simplifies some estimates 
in later sections; 
$\Pi(\lam)$ transforms the multiplications 
to the Fourier multipliers   
\bqn \lbeq(mult)
f(\lam) \Pi(\lam)u(x) = \Pi(\lam) f(|D|)u (x), 
\eqn 
which is particularly useful when $f(\lam)$ is GMU. 
Note that, for $u \in \Dg_\ast$, 
$\Pi(\lam)u(x)=0$ for $\lam$ outside a compact interval of $(0,\infty)$. 

We then introduce 
the {\it stationary representation formula} of $W_{-}$:
\bqn \lbeq(sta-00)
W_{-}u = u - 
\int_0^\infty R_0^{+}(\lam^4)\Qg_v(\lam)\Pi(\lam)u \lam^3 d\lam 
\eqn 
(cf. \cite{RS3}) 
which is valid under the assumption of the theorems 
(except \refth(small) where the restriction to the high energy part is 
necessary)   
and is the starting point of the proof of theorems. 
As we shall 
exclusively deal with $W_{-}$, {\it we shall often 
omit the superscript $+$ from $R_0^{+}(\lam^4)$ and $\Mg^{+}(\lam^4)$. }
Finally in \S2 we prove that the Fourier multiplier 
defined via $\Rg(\lam)$ satisfies 
\bqn \lbeq(Rg-Lp)
\|\Rg(|y||D|)\chi_{\geq {a}}(|D|)\|_{\Bb(L^p)}
\leq C (1+ |\log |y||), \quad 
1<p<\infty.
\eqn

We say an operator is {\it good operator} (GOP for short) 
if it is bounded in $L^p(\R^4)$ for all $1<p<\infty$; 
operator $T$ or operator-valued function 
$T(\lam)$ of $\lam>0$ is {\it good producer} (GPR for short) if 
the operators defined by \refeq(sta-00) 
with $T$ or $T(\lam)$ in place of $\Qg_v(\lam)$: 
\begin{gather} 
\W(T)u \colon = \int_0^\infty R_0(\lam^4)\,T\, \Pi(\lam)u \lam^3 d\lam, 
\lbeq(OmegaT)
\\
\tilde{\W}(T(\lam))u 
\colon 
= \int_0^\infty R_0(\lam^4)\,T(\lam)\, \Pi(\lam)u \lam^3 d\lam 
\lbeq(OmegaTlam)
\end{gather} 
respectively are GOPs. 

We define the operator $K$ for $u \in \Dg^\ast$ by   
\bqn \lbeq(K-first)
Ku(x) = \int_0^\infty \Rg_\lam(x) (\Pi(\lam)u)(0) \lam^3 d\lam\,. 
\eqn 
We shall prove in \refsec(integral) that $K$ is GOP; 
$K$ is the building block for many operators, e.g. for $T=T(x,y)$, 
$\W(T)$ is the superposition of translations of $K$b by $T(y,z)$:
\bqn \lbeq(T-K)
\W(T)u(x)= \iint_{\R^8} T(y,z)(\tau_{y} K \tau_{-z}u)dz dy 
\eqn 
and, by Minkowski's inequality, 
\bqn 
\|\W(T)u\|_p \leq C \|T\|_{\Lg^1}\|u\|_p, 
\quad \Lg^1= L^1(\R^4 \times \R^4) 
\lbeq(WT-estimate) 
\eqn 
(cf. \reflm(T-K)), from which we shall derive \refprop(R-theo) that 
\bqn 
\|\tilde{\W}(T(\lam))u\|_p 
\leq C \sum_{j=0}^3 \int_0^\infty \la \lam \ra^{j-1}
\|T^{(j)}(\lam)\|_{\Lg^1} \|u\|_p d\lam\,. \lbeq(WTlam-estimate)
\eqn 
Here and hereafter $f^{(j)}(\lam) = (d/d\lam)^j f(\lam)$ for $j=0,1, \dots$. 

\bgdf We say $T(\lam,x,y)$ is 
{\it variable separable} ($\Vg\Sg$ for short) if it has the form  
$T(\lam,x,y)= \sum_{j=1}^N \m_j(\lam)T_j(x,y)$; 
it is {\it good variable separable} ($\Gg\Vg\Sg$ for short) if $\m_j$ are 
GMU and $T_j (x,y) \in \Lg^1$ for $j=1, \dots, N$.
A $\Gg\Vg\Sg$ is GPR. 
\eddf 

In \refsec(high) we shall prove \refth(small) and \refth(high). 
Let $1<p<\infty$. For high energy part we have from \refeq(sta-00) that 
\bqn \lbeq(sta-high)
W_{-}\chi_{\geq{a}}(|D|)u = 
\chi_{\geq{a}}(|D|) u - 
\int_0^\infty R_0^{+}(\lam^4)\Qg_v(\lam)\Pi(\lam)u \lam^3 
\chi_{\geq{a}}(\lam)d\lam 
\eqn 
Expanding as $\Qg_v(\lam)=V - VR_0(\lam^4)V +\cdots $ in \refeq(sta-high)  
produces the well known Born series for 
$W_{-}\chi_{\geq{a}}(|D|)$:  
\begin{align}  
& W_{-}\chi_{\geq{a}}(|D|)u = \chi_{\geq{a}}(|D|)u - 
W_1 \chi_{\geq{a}}(|D|)u + \cdots, \lbeq(Born) \\
& W_n \chi_{\geq{a}}(|D|)u = \int_0^\infty R_0(\lam^4) (M_VR_0(\lam^4))^{n-1}M_V 
\Pi(\lam)u \lam^3 \chi_{\geq{a}}(\lam)d\lam\,. \lbeq(Wn-0)
\end{align}  
Then, $W_1 \chi_{\geq{a}}(|D|)u = \W(M_V)\chi_{\geq{a}}(|D|)$ 
and, since $M_V$ maybe considered as the integral operator with 
$V(x)\d(x-y)\in \Lg^1$, $W_1 \chi_{\geq{a}}(|D|)$ is GOP by 
\refeq(WT-estimate). Let $V_y^{(2)}(x) = V(x) V(x-y)$. 
We shall show 
\bqn \lbeq(W2-express)
W_2 \chi_{\geq a}(|D|)u = \int_{\R^4} 
\W(M_{V^{(2)}_y})\Rg(|y||D|)\chi_{\geq a}(|D|)\tau_y u dy.  
\eqn 
Then, since 
$\|\W(M_{V^{(2)}_y})\|_{\Bb(L^p)}\leq C \int_{\R^4} |V(x)V(x-y)|dx$ 
by \refeq(WT-estimate)  
and  
$\|\Rg(|y||D|)\chi_{\geq{a}}(|D|)\|_{\Bb(L^p)}\leq C(1+|\log |y||)$ 
by \refeq(Rg-Lp),  
\begin{align}\lbeq(W2-intro)
\|W_2 \chi_{\geq a}(|D|)u\|_p 
& \leq C \int_{\R^8} |V(x)V(x-y)|(1+|\log |y||)dxdy  \\
& \leq C(\|V\|_{L^{q}_{loc,u}}+ \|\la \log |x|\ra^{2} V\|_{L^1})^2 \|u\|_p 
\notag.
\end{align}
Iterating the estimate \refeq(W2-intro), we show that for $n=3,4, \dots$ 
\bqn \lbeq(small-effect-0)
\|W_n \chi_{\geq{a}}(|D|) u\|_p \leq C^n 
(\|V\|_{L^{q}_{loc,u}}
+ \|\la \log |x|\ra^{2} V\|_{L^1})^n \|u\|_p 
\eqn 
with $C>0$ independent of $V$ and $n$. This yields \refth(small).  

For proving \refth(high), we expand $\Qg_v(\lam)$ with remainder:
\begin{align} 
\lbeq(with) 
\Qg_v(\lam) & = \sum_{n=0}^{N-1}(-1)^{n}M_V (R_0(\lam^4)M_V)^n 
+ (-1)^N D_N (\lam), \\
D_N(\lam) & = M_v (M_w R_0(\lam^4)M_v)^N(1+M_w R_0(\lam^4)M_v)^{-1} M_w. 
\lbeq(with-2) 
\end{align}
The first term on the right of \refeq(with) 
produces $\sum_{n=0}^{N-1}(-1)^n W_n \chi_{\geq a}(|D|)$ which is GOP by 
\refeq(small-effect-0).  
The decay of $\Rg_\lam(x)$ as $\lam \to \infty$ yields   
\bqn  \lbeq(R0-deri)
\|\pa_{\lam}^j D_N(\lam)\|_{\Lg_1} \leq 
C {\lam^{-\frac{2N}{q'}}}(\|\ax^{(2j-3)_{+}} V\|_{L^1}+ \|V\|_{L^q})^N
\eqn  
for $j=0,1, \dots$, where $1< q <4/3$ and $4<q'=q/(q-1)<\infty$. 
If we take $N$ large enough such that $2N/q'>3$, 
then $\chi_{\geq a}(\lam)D_N(\lam)$ becomes GPR for $a>0$ 
by \refeq(WTlam-estimate) and \refth(high) follows. 

In \refsec(low1) we begin studying the low energy part  
and prove \refth(main-rephrase) for the case that 
$H$ is regular at zero. From \refeq(sta-00) we have 
\bqn \lbeq(sta-low)
W_{-}\chi_{\leq {a}}(|D|)u =  
\chi_{\leq {a}}(|D|)u - \int_0^\infty R_0(\lam^4) \Qg_v(\lam)
\Pi(\lam) u \lam^3 \chi_{\leq {a}}(\lam) d\lam .
\eqn 
We use the following notation. 

\bgdf 
For a Banach space $\Xg$, an integer $k \geq 0$ and 
a function $f(\lam)>0$ for small $\lam>0$, 
$\Og_{\Xg}^{(k)}(f)$ is the space of $\Xg$-valued $C^k$ functions 
$T(\lam)$ defined for small $\lam>0$ such that 
$\|(d/d\lam)^j T(\lam)\|_{\Xg}\leq C_j \lam^{-j} |f(\lam)|$ for 
$j=0, \dots, k$. We shall abuse notation and write $\Og_{\Xg}^{(k)}(f)$ 
for an element of $\Og_{\Xg}^{(k)}(f)$. 
\eddf

We write $\Rg_{\lam, 2n}(x)$ for the remainder of 
\refeq(descent):
\bqn \lbeq(remain-R0)
\Rg_{\lam, 2n}(x)= \lam^{2n}\tg_{2n}(\lam)G_{2n}(x)+ \lam^{2n}G_{2n,l}(x)+ 
\cdots, \ , n=1,2, \dots 
\eqn 
and $\Rg_{\lam, 0}(x)  = \Rg_\lam(x)$. Let $R_{2n}(\lam^4)$ and 
$R_{2n}^{(v)}(\lam^4)$ be the corresponding operators defined by 
\begin{gather} \lbeq(M2ndef)
R_{2n}(\lam^4) = 
\lam^{2n}\tg_n(\lam)G_{2n} +\lam^{2n} G_{2n,l} + \cdots, \\
R_{2n}^{(v)}(\lam^4) = 
M_v(\lam^{2n}\tg_n(\lam)G_{2n} +\lam^{2n} G_{2n,l} + \cdots )M_v,
\lbeq(M2n-vdef)
\end{gather}
where we understand $G_0$ is the identity operator and $G_{0,l}(x)=N_0(x)$. 
Note that $R_0^{(v)}(\lam^4)=\Mg(\lam^4)$. 
By virtue of \reflm(first-est) and \refeq(Uplus)  
\bqn \lbeq(m4-mlam4)
\Mg(\lam^4)= T_0 + g_0(\lam) P + \lam^2 G^{(v)}_2+ R_4^{(v)}(\lam), 
\quad 
R_4^{(v)}(\lam) \in \Og_{\Lg^1}^{(4)}(\lam^4 \log\lam). 
\eqn 
If $H$ is regular at zero, then we obtain (\reflm(abs-bound)) 
via Feshbach formula that, with $D_0 = Q(QT_0Q)^{-1}Q \in \Lg^1$ 
and $L_0$ of rank two, 
\[
(T_0+ g_0(\lam)P)^{-1}=h(\lam)L_0 + D_{0}, \quad h(\lam)= (g_0(\lam) + c_1)^{-1}
\]
It follows by the perturbation expansion that 
$\Qg_v(\lam)$ is the sum of $\Gg\Vg\Sg$ and 
$\Og_{\Lg^1}^{(4)}(\lam^4 \log\lam)$ and, hence, 
$W_{-}\chi_{\leq {a}}(|D|)$ 
is GOP for sufficiently small $a>0$ if $H$ is regular at zero. 

In \refsec(resonance) we shall begin studying the case that 
$H$ has singularities at zero. In \refsec(resonance) we 
show that $\Ng_\infty(H)\not=\{0\}$ if and only if $H$ 
is singular at zero and, then prove \reflm(reso-intro) and 
\reflm(resonance).  In \S7 and \S8 we shall prove \refth(main-rephrase) 
for the case that singularities are of the first and second kinds 
respectively. In \S9 we shall prove it  
when the singularities are of the third and fourth kinds. In these cases, 
$\Mg(\lam^4)^{-1}$ is singular at $\lam=0$ and the singularities 
become stronger as the order of the types increases 
from the first to the fourth. We shall study them by 
repeatedly and inductively applying \reflm(JN) due to Jensen and Nenciu,  
which again becomes more complicated as the order of the types increases.  

If $H$ has singularity of the first kind at zero, then 
$\Mg(\lam^4)^{-1}$ has only logarithmic singularity 
at zero: If $\z\in S_1\HL$ is the basis vector of the one dimensional  
$S_1\HL$ (cf. \reflm(resonance)), then 
with constants $a,b$, 
\bqn 
\Qg_v(\lam) \equiv \a(\lam)(v\z\otimes v\z), \quad 
\a(\lam)=(a \log \lam + b) 
\eqn 
modulo GPR (\reflm(s-mod)) and modulo GOP 
$W_{-}\chi_{\leq a}(|D|)u(x) $ becomes 
\bqn \lbeq(33)
- \int_0^\infty R_0(\lam^4)(v \z)(x) 
(v\z, \Pi(\lam)u) \a(\lam)\lam^3 \chi_{\leq a}(\lam)d\lam\,.
\eqn 
The point here is that the  
singularity of $\a(\lam)$ can be cancelled by the property of 
the basis vector $\z \in S_1\HL$ that 
\bqn \lbeq(mom-prop)
\int_{\R^4} v(x)\z(x)dx=0
\eqn 
which implies $(v\z, \Pi(\lam)u)=(v\z, \Pi(\lam)u(x)-\Pi(\lam)u(0))$, and 
\begin{align}
& \Pi(\lam)u(x)-\Pi(\lam)u(0) \notag \\
& = \int_{{\mathbb S}^3} (e^{i\lam{x\w}}-1)\hat{u}(\lam\w)\frac{d\w}{(2\pi)^2} 
= i\lam \sum_{l=1}^4 x_l \int_0^1 \Pi(\lam)R_l u(\th{x})d\th
  \lbeq(reason) 
\end{align} 
producing the factor $\lam$ which makes $\m(\lam)= \lam \a(\lam)$ GMU,  
where $R_j$, $1\leq j \leq 4$ are Riesz transforms. Then, \refeq(33) becomes 
$-i \sum_{l=1}^4  \int_0^1 d\th$ of 
\begin{align}\lbeq(intro-35)
& \int_0^\infty R_0(\lam^4)(M\z \otimes (x_l v\z))  
\Pi(\lam)(\tau_{-\th{x}}R_l \m(|D|)u)(0)\lam^3 d\lam \\
& = \int_{\R^8} (v\z)(y) z_l (v\z)(z)
\t_{y} (K \tau_{-\th{z}}R_l \m(|D|)u)(x) dy dz\,  \notag 
\end{align} 
and \refeq(WT-estimate) implies that 
$W_{-}\chi_{\leq a}(|D|)$ is GOP. 

If $H$ has singularity of the second kind, then 
$\Qg_v(\lam)$ has two types of singularities at $\lam=0$ 
and modulo $\Gg\Sg\Vg$
\bqn \lbeq(second-sing)
\Qg_v(\lam) \equiv 
\sum_{j,k=1}^n \log\lam \a_{jk} (\z_j \otimes \z_k)+ 
\sum_{j,k=1}^m \lam^{-2} \b_{jk}(\z_j \otimes \z_k)\,,
\eqn 
where $\{\z_1, \dots, \z_n\}$ is the basis of $S_1 \HL$ 
such that $\{\z_1, \dots, \z_m\}$, $m \leq n$ spans $S_2 \HL$ 
(\reflm(second-Mlam)) and each of $\z_1, \dots, \z_n$ satisfies 
\refeq(mom-prop).  
We substitute \refeq(second-sing) in \refeq(sta-low). Then,   
the argument above shows that the first sum with the 
log-singularity produces GOP. 
To deal with the second term which has $\lam^{-2}$-singularity, 
we expand $e^{i\lam{x\w}}$ to the second order in 
\refeq(reason). Then, $\Pi(\lam)u(x)-\Pi(\lam)u(0)$ becomes   
\bqn \lbeq(Pi-sec)
\sum_{l=1}^4 i \lam x_l (\Pi(\lam)R_l u)(0) 
+ \sum_{m,l=1}^4 {x_m x_l \lam^2}\int_0^1 (1-\th)
(\Pi(\lam)\t_{-\th{x}}R_m R_l u)(0)d\th, 
\eqn 
which we substitute for $\Pi(\lam)u$ in \refeq(sta-low). 
Because of the factor $\lam^2$, 
the second term of \refeq(Pi-sec) produces GOP;  
the first produces the sum over $1\leq j,k \leq m$ 
and $1\leq l \leq 4$ of 
\bqn \lbeq(B-op)
W_{B,jkl} u(x) = i \la x_l v, \z_k\ra 
\int_0^\infty (R_0 (\lam^4)M_v\z_j)(x) 
(\Pi(\lam) R_l u)(0)\lam^2 \chi_{\leq{a}}(\lam)d\lam\,.
\eqn 
Ignoring $i\la x_l v, \z_k\ra $ and $R_l$ and omitting the indices, 
we consider the typical operator 
\bqn
W_B u= \int_0^\infty R_0 (\lam^4)\ph(x) 
(\Pi(\lam) u)(0)\lam^2 \chi_{\leq{a}}(\lam)d\lam  \lbeq(B-decomp-0)
\eqn 
where $\ph(x)= v(x)\z(x)$ and $\z\in S_2 \HL$. It is evident that 
\bqn 
W_B u =\chi_{\geq 4a}(|D|)W_Bu + \chi_{\leq 4a}(|D|)W_Bu \lbeq(B-decomp)
\eqn 
and we transfer $\chi_{\geq 4a}(|D|)$ and $\chi_{\leq 4a}(|D|)$ to the 
inside of the integral of \refeq(B-decomp-0). 
Let $\m(\xi)= \chi_{\geq 4a}(|\xi|)|\xi|^{-4}$. Then, 
\[
\chi_{\geq 4a}(|D|)R_0(\lam^4)\ph(x)= 
\m(D)\ph(x)+ \lam^4 \m(D)R_0(\lam^4)\ph(x) .
\] 
Thanks to the factor $\lam^4$ the second member on the right produces 
GOP for $\chi_{\geq 4a}(|D|)W_B$; the first does the rank one operator  
\[
\m(|D|)\ph(x)(u, f), \quad 
f(x)=\Fg(\chi_{\leq{a}}(\xi)|\xi|^{-1})(x)\,.
\]
Here $\m(D)\ph(x)\in L^p(\R^4)$ for $1\leq p\leq \infty$ (cf. \reflm(mu)) 
and $f \in L^q(\R^4)$ if and only if $4/3<q \leq \infty$. 
Thus, $\chi_{\geq 4a}(|D|)W_B$ is bounded in $L^p(\R^4)$ 
for $1\leq p<4$ and is  unbounded for $p\geq 4$. 
This proves that $W_{-}$ is unbounded in $L^p(\R^4)$ if $p\geq 4$. 
b
Since $\hat{\ph}(0)=0$, 
$\chi_{\leq 4a}(|D|)R_0(\lam^4)\ph(x)$ is the limit $\ep \downarrow 0$ of 
\[
 \sum_{m=1}^4 \frac{-i}{(2\pi)^4}
\int_0^1 \int_{\R^4} z_m {\ph}(z) \tau_{{\th}z} R_m 
\left(\int_{\R^4}e^{ix\xi} \frac{|\xi| \chi_{\leq 4a}(|\xi|)}
{(|\xi|^4-\lam^4-i\ep)}d\xi\right) dz d\th
\]
(cf. \refeq(dx-2ch)) and  
\begin{gather*}
\chi_{\leq 4a}(|D|)W_B u(x)= \sum_{m=1}^4 \frac{-i}{(2\pi)^4}
\int_0^1 \int_{\R^4} z_m {\ph}(z) \tau_{{\th}z} R_m Y u (x) dz d\th, \\ 
Yu(x) = \int_0^\infty 
\left(\int_{\R^4}e^{ix\xi} \frac{|\xi| \chi_{\leq 4a}(|\xi|)}
{(|\xi|^4-\lam^4-i0)}d\xi\right)
(\Pi(\lam) u)(0)\lam^2 \chi_{\leq{a}}(\lam)d\lam.
\end{gather*} 
Substituting 
\bqn 
\frac{|\xi| }{(|\xi|^4-\lam^4-i0)}   
=\frac{\lam}{(|\xi|^4-\lam^4-i0)}
+ \frac{1}{(|\xi|+\lam) (|\xi|^2+\lam^2)}
\lbeq(mult-10)
\eqn 
in the integral by $d\xi$ yields that 
\bqn 
Yu(x) = \chi_{\leq 4a}(|D|)K \chi_{\leq {a}}(|D|)u(x) + 
\int_{\R^4} L(x,y)u(y)dy \,, 
\eqn 
where $L$ is the integrval operator with the kernel 
\[
L(x,y) = \iint_{\R^8}
\frac{e^{ix\xi-iy\eta} \chi_{\leq 4a}(|\xi|)\chi_{\leq{a}}(|\eta|)}
{(|\xi|^2+|\eta|^2)(|\xi|+|\eta|)|\eta|}d\xi{d\eta}\,.
\]
Evidently $\chi_{\leq 4a}(|D|)K \chi_{\leq {a}}(|D|)$ 
is GOP and we shall prove in Appendix that 
$L$ is bounded in $L^p(\R^4)$ for $1<p<4$. Hence, $Y$ is bounded 
in $L^p(\R^4)$ for $1<p<4$ and so is $\chi_{\leq 4a}(|D|)W_B$. 
Thus $W_{-}$ is bounded in 
$L^p(\R^4)$ for $1<p<4$ but unbounded for $p\geq 4$ in this case. 

If $H$ has singularities of the third or the fourth kind at zero, 
then leading singularities of $\Qg_v(\lam)$ as $\lam \to 0$ 
are of order of $\lam^{-4}(\log \lam)^{-1}$ and $\lam^{-4}$ respectively. 
However, they act in subspaces $S_3\HL$ and $S_4 \HL$ 
and $\z \in S_3\HL$ and  $S_4 \HL$  satisfy the additional moment 
property $(x^\a v, \z)=0$ for $|\a|\leq 1$ and $|\a|\leq 2$ 
respectively, which partly cancels the singularities as previously. 
Thus, we can proceed by following the line of ideas of previous 
sections, however, the argument will become more complicated. 
We shall avoid outlining it here and proceed to the text 
as we do not want to make the introduction too long .

\section{Preliminaries }

\subsection{Free resolvents} 
In this subsection we present some estimates related to 
$R_0^{+}(\lam^4)$ or the expansion \refeq(descent). We begin with the 
following lemma. 

\bglm \lblm(members) Let $q>1$. For members on the right of 
\refeq(Uplus) we have  
\begin{align} 
& \|N_0^{(v)}\|_{\Hg_2}
\leq C(\|V\|_{L^{q}_{loc,u}} + \|\la \log |x|\ra^{2}V\|_1).  \lbeq(N0v) \\
& \|G_{2j}^{(v)}\|_{\Hg_2} \leq C \|\ax^{4j}V\|_1.  \lbeq(G2jv) \\
& \|G_{2j,l}^{(v)}\|_{\Hg_2} \leq 
C(\|\ax^{4j}V\|_{L^{q}_{loc,u}} + \|\la \log |x|\ra^{2}\ax^{4j}V\|_1). 
\lbeq(G2jlv) 
\end{align} 
\edlm 
\bgpf Let $q'=q/(q-1)$. Then, H\"older's inequality implies that   
\begin{align}
& \int_{|x-y|\leq 2}|V(x)(\log |x-y|)^2 V(y)| dx dy \lbeq(q-prime)
\\
& \leq \|V\|_1 \|V\|_{L^q_{loc,u}} \|\log |x|\|_{L^{2q'}(|x|\leq 2)}^{2} 
\leq C(\|V\|_1^2+ \|V\|_{L^q_{loc,u}}^2) \notag 
\end{align}
If $|x-y|\geq 2$, then $\log |x-y|\leq \log \ax + \log \ay $ and 
\[
\int_{|x-y|\geq 2}|V(x)(\log |x-y|)^2 V(y)| dx dy 
\leq C\|(\log \ax)^2 V\|_1 \|V\|_1 .
\]
This proves \refeq(N0v). We omit the proof for \refeq(G2jv) which is obvious 
and the one for \refeq(G2jlv) 
which is similar to that of \refeq(N0v). 
\edpf 

For $\Gg_{z}(x)$, $z \in \Cb^{+}$, we have the 
integral representation (\cite{DLMF} 10.9.21):
\begin{equation}\lbeq(re-i0) 
\Gg_{z}(x)= \frac{e^{iz|x|}}{2(2\pi)^{\frac{3}2}
\Ga \left(\frac{3}{2}\right) |x|^{2}} 
\int_0^\infty e^{-t} t^{\frac{1}2}
\left(\frac{t}2 -iz|x| \right)^{\frac{1}2}dt\,,
\end{equation} 
where $z^\frac12$ is the branch such that $z^\frac12>0$ for $z>0$. 
Thus, if we let   
\bqn \lbeq(R-H)
\Hg(\lam) = \frac{e^{i\lam}}{4(2\pi)^{\frac{3}2}
\Ga \left(\frac{3}{2}\right)\lam^2}
\int_0^\infty e^{-t}t^{\frac12}\left(\frac{t}2-i\lam \right)^{\frac12}dt, 
\quad \lam \in \Cb^{+}, 
\eqn 
then, $\Rg(\lam) = \Hg(\lam) - \Hg(i\lam)$ for $\lam>0$ and 
\bqn \lbeq(viaHankel)
\Rg_\lam(x)= \Hg(\lam|x|) - \Hg(i\lam|x|)\,.
\eqn 

Recall the definitions \refeq(remain-R0) and \refeq(M2n-vdef) 
for $\Rg_{\lam,2n}(x)$ and $R_{2n}^{(v)}(\lam^4)$, $n=0,1, \dots$ 
respectively. 

\bglm \lblm(first-est) 
\ben 
\item[{\rm (1)}] For $j=0,1, \dots$, 
\bqn 
|\pa_\lam^j \Rg_{\lam}(x) |\leq C_j \left\{
\br{l} 
\la \log \lam |x|\ra  \lam^{-j}\,,\  0<\lam|x| \leq 1, \\
|x|^j \la \lam |x| \ra^{-\frac32}\,, \ \ 1 \leq \lam|x| \er \right. 
\lbeq(R-0)
\eqn 
\item[{\rm (2)}] If $V \in (L^q_{loc,u}\cap L^r)(\R^4)$ for some $q>1$ and 
$1\leq r \leq 8/5$. Then $M_v$ is 
$H_0$-smooth on $[a,\infty)$ for any $a>0$. 
\item[{\rm (3)}] Let $j=0, \dots, 2n$ and $a>0$. Then, for $0<\lam<a$, 
\bqn  
|\pa_\lam^j \Rg_{\lam,2n}(x) |
\leq C_j \la \log \lam |x|\ra  \lam^{2n-j}|x|^{2n}\,, \lbeq(R-2n) 
\eqn 
where, if $n$ is odd, $\la \log \lam |x|\ra$  
should be removed from the right. 
\item[{\rm (4)}] For the operator $R_{2n}^{(v)}(\lam^4)$ we have 
\bqn 
\|(d/d\lam)^j R_{2n}^{(v)}(\lam^4)\|_{\Hg_2} 
\leq C \lam^{2n-j}\la \log \lam \ra
\|\ax^{4n} \la \log |x|\ra ^2 V\|_1  \lbeq(remainder-est)
\eqn 
where, if $n$ is odd, $\la \log \lam \ra$  
should be removed from the right.   
\een
\edlm 
\bgpf
(1) From \refeq(Rg0-rev) we have \refeq(R-0) for $0<\lam |x|\leq 1$;   
for $\lam |x|\geq 1$ it follows from the representation 
\refeq(re-i0) which implies that for $\lam |x| \geq 1$    
\begin{align} 
& \left|\pa_\lam^j (\lam^{-2}\Gg_{\lam}(x))\right|
\leq C_j |x|^j \la \lam |x| \ra^{-\frac32} , \quad j=0 , 1, \dots,  
\lbeq(Glamn) \\  
& \left|\pa_\lam^j(\lam^{-2} \Gg_{i\lam}(x))\right|\leq C 
e^{-\lam|x|}|x|^{j} \la \lam |x| \ra^{-\frac32}, 
\quad j=0,1, \dots. \lbeq(Gilamn)
\end{align}
(2) Let $\lam>a$. We estimate 
\[
\|\Mg_v(\lam^4)\|_{\Hg_2}^2= 
\left(\int_{\lam|x-y|\leq 1}+ \int_{\lam|x-y|\geq 1}\right)
|V(x)\Rg_\lam(x-y)^2 V(y)|dx dy 
\]
by using \refeq(R-0).  Then, the integral over
$\lam|x-y|\geq 1$ is evidently bounded by $C\|V\|_1^2$. 
Since $\la \log \lam |x|\ra\leq C_\ep (\lam|x|)^{-\ep}$ for 
an arbitrary small $\ep>0$ for $\lam |x|\leq 1$, the integral 
over $\lam|x-y|\leq 1$ is bounded by using H\"older's inequality by
\[
{C_\ep}{\lam^\ep}\int_{\R^4} |V(x)|
\left(\sup_{x\in \R^4}\int_{|x-y|\leq 1/a} 
\frac{|V(y)|dy}{|x-y|^\ep}\right)dx \leq C\|V\|_{L^q_{loc,u}} \|V\|_1.
\]
Thus, $\Mg_v^{\pm}(\lam^4)$ is uniformly bounded in $\Bb(\HL)$ for $\lam>a$ 
(note that $\Mg_v^{-}(\lam^4)=\Mg_v^{+}(\lam^4)^\ast$) and, then,  
it is well known that $M_v$ is $H_0$-smooth on $[a,\infty)$ (\cite{RS3}). 
\\[3pt]
(3) 
If $\lam |x|\leq 1$, then \refeq(R-2n) is obvious. 
Let $\lam |x|\geq 1$ and $\lam<a$, then $|x|\geq 1/a$ 
and the right side of \refeq(R-0) is bounded by that of 
\refeq(R-2n) if $j\leq 2n$. Let $\tG_{2m, \lam}(x)$ denote 
$\lam^{2m}|x|^{2m}$ for odd $m$ and 
$\lam^{2m}\tg_m(\lam|x|)|x|^{2m}$ for even $m$. Then 
for $0\leq m\leq n-1$
\[
|\pa_\lam^j \tG_{2m, \lam}(x) | \leq C 
\lam^{2m-j}\la \log \lam |x| \ra |x|^{2m} \leq C \lam^{2n-j} |x|^{2n}. 
\] 
and  $\Rg_{\lam, 2n}(x)= \Rg_{\lam,0}(x) - \tg_0(\lam|x|) - \lam^2 G_2(x)- 
\cdots- 
\tG_{2(n-1), \lam}(x)$ also satisfies \refeq(R-2n) for $\lam |x|\geq1$. 
\\[3pt]
(4) By virtue of \refeq(R-2n) 
and the estimate 
$|\log (\lam |x|)|\leq \la \log \lam \ra \la \log |x| \ra$, 
the obvious modification of the proof of \refeq(N0v) implies 
\refeq(remainder-est). 
\edpf 

For shortening formula for $\Mg(\lam^4)$, we define  for $1\leq m < n$,  
\begin{gather*}  
R_{2m\to 2n}(\lam)=
\lam^{2m}\tg_m(\lam)G_{2m}+ \cdots+ \lam^{2n} G_{2n,l}, \\
R_{2m\to 2n}^{(v)}(\lam)=
M_v(\lam^{2m}\tg_m(\lam)G_{2m}+ \cdots+ \lam^{2n} G_{2n,l})M_v 
\end{gather*}
where we understand that $\tg_k(\lam)G_{2k,l}=0$ if $k$ is odd.

\subsection{Stationary representation formula}

\bglm \lblm(spect-proj) 
Let $\Pi(\lam)$ be the spectral projection defined by 
\refeq(spect-proj-def). Then, $\Pi(\lam)$ satisfies 
\refeq(spect-proj) and \refeq(mult).
\edlm 
\bgpf Let $u \in \Dg_\ast$. Express the right of 
\refeq(spect-proj-def) via Fourier transform , 
use polar coordinates and change the variables. Then,  
\bqn 
\Pi(\lam)u(x) = 
\lim_{\ep \to 0} \frac{\ep}{4\pi^3} 
\int_{0}^\infty 
\frac{d\s}{(\s-\lam^4)^2+ \ep^2}
\left(\int_{{\mathbb S}^3}e^{i{\s}^\frac14 {x\w}}\hat{u}
({\s}^\frac14 \w)d\w \right) .
\eqn 
The first  of \refeq(spect-proj) follows by Poisson's formula. 
The second  of \refeq(spect-proj) 
and \refeq(mult) are obvious from  the first. 
\edpf 

Under the condition of \refth(high) or \refth(main), 
$M_v$ is $H_0^\frac12$-compact and $M_v R_0^{+}(\lam^4) M_v$ 
is $\Hg_2$-valued function of $\lam>0$ of class $C^1$ by 
virtue of \refeq(R-0). Moreover, the absence of positive eigenvalues 
from $H$ implies $\Mg^{+}(\lam^4)$, $\lam>0$ 
is invertible in $\Bb(\HL)$ (\cite{Ku}). Hence,  
$\Mg^{+}(\lam^4)^{-1}$ is $C^1$ with values in $\Bb(\HL)$ and 
the following theorem is well known. 

\bgth Suppose $V$ satisfies the short range condition \refeq(short) 
and $\la \log |x| \ra^2 V \in L^1(\R^4)$. 
Then, for $u \in \Dg_\ast$, we have the stationary representation formula 
\refeq(sta-00) for $W_{-}u$. 
\edth  

\bgrm \reflm(first-est) {\rm (2)} implies that $M_v$ is $H_0$-smooth on 
$[a,\infty)$ for any $a>0$ under the condition of \refth(small); 
if $V$ is small it is also $H$-smooth on $[a,\infty)$ 
and we have the representation formula \refeq(sta-high) for high 
energy part $W_{-}\chi_{\geq{a}}(|D|)u$ (see \cite{RS3}).
\edrm

\subsection{Fourier multiplier defined by the resolvent kernel}

We need the following lemma in \refsec(high).   
In what follows $a \absleq b$ means $|a|\leq |b|$.

\bglm \lblm(H02y) Let $a>0$ and $1<p<\infty$. Then, 
there exists a 
constant $C_{a,p}$ independent of $y \in \R^4$ such that 
\bqn 
\|\Rg (|y||D|)\chi_{\geq{a}}(|D|)\|_{\Bb(L^p)} 
\leq C_{a,p} (1+ |\log|y||)\,. \lbeq(H02y-a)
\eqn 
\edlm

For the proof we use Peral's theorem (\cite{Peral}):

\bglm \lblm(peral) Let $\psi(\xi) \in C^\infty(\R^n)$ be such that 
$\psi(\xi)=0$  in a neighbourhood of $\xi= 0$ and 
$\psi(\xi)=1$ for $|\xi|>a$ for an $a>0$. Then, 
the translation invariant Fourier 
integral operator 
\[
\frac{1}{(2\pi)^{n/2}}\int_{\R^n} 
e^{ix\xi+i|\xi|}\frac{\psi(\xi)}{|\xi|^b} \hat{f}(\xi)d\xi,
\]
is bounded in $L^p(\R^n)$ if and only if
\[
\left|\frac1{p}-\frac12\right| < \frac{b}{n-1}\,.
\]
\edlm 

\paragraph{\bf Proof of \reflmb(H02y)}  Recall \refeq(viaHankel) that 
$\Rg(\lam|x|)=\Rg_{\lam}(x)= \Hg(\lam|x|)- \Hg({i\lam}|x|)$ 
and $\Hg(\lam)$ has the integral representation \refeq(R-H). Let  
for $a>0$ 
\[
\Hg_{<a}(\lam)= \chi_{<{a}}(\lam)\Hg(\lam), \quad 
\Hg_{\geq {a}}(\lam){=} \chi_{\geq{a}}(\lam)\Hg(\lam).  
\] 
and 
\[
\Rg_{<{a}}(\lam)=\Hg_{<{a}}(\lam)- \Hg_{<{a}}({i\lam}), \  
\Rg_{\geq{a}}(\lam)=\Hg_{\geq{a}}(\lam)- \Hg_{\geq{a}}({i\lam}).
\]
(1) If we write \refeq(R-H) for $z=\lam>0$ in the form 
\[
\Hg(\lam) = \frac{e^{i\lam}}{4(2\pi)^{\frac{3}{2}}\Ga \left(\frac{3}{2}\right) 
 \lam^{\frac32}}F(\lam), 
\quad F(\lam)= 
\int_0^\infty e^{-t}t^{\frac12}\left(\frac{t}{2\lam}-i\right)^{\frac12}dt, 
\]
then $|\pa_\lam^j (F(\lam)\chi_{\geq a}(\lam))| \leq C \lam^{-j}$ for 
$0\leq j \leq 3$ 
and $F(|D|)\chi_{\geq a}(|D|)$ is a GOP.  Peral's theorem implies that 
$e^{i|D|} |D|^{-\frac{3}{2}}\chi_{\geq{a}}(|D|)$ is also GOP, hence, 
so is $\Hg_{\geq{a}}(|D|)$ and, the norm 
$\|\Hg_{{\geq{a}}}(|y||D|)\|_{\Bb(L^p)}$ is independent of $|y|$ 
by the scaling. From  
\[
\Hg(i\lam) = \frac{-e^{-\lam}}
{4(2\pi)^{\frac{3}{2}}\Ga \left(\frac{3}{2}\right)\lam^{\frac32}}
F(i\lam), \quad 
F(i\lam)= \int_0^\infty e^{-t}t^{\frac12}
\left(\frac{t}{2\lam}+1\right)^{\frac12}dt
\]
it is obvious that the Fourier transform of 
$\Hg_{\geq {a}}(i|\xi|)$ is in $\Sg(\R^4)$ and 
$\Hg_{\geq{a}}(i|D|)\in \Bb(L^p(\R^4))$ for all 
$1\leq p \leq \infty$ with  
$\|\Hg_{\geq{a}}(i|y||D|)\|_{\Bb(L^p)}$ being independent 
of $|y|$. Thus, $\Rg_{\geq {a}}(|y||D|)$ satisfies 
\bqn 
\|\Rg_{\geq{a}}(|y||D|)\|_{\Bb(L^p)} \leq C_p. \lbeq(H02y-add)
\eqn  
(2) \refeq(Rg0-rev) implies 
\[
\pa_\lam^j \left\{
\chi_{\leq a}(\lam) \left(\Rg(\lam)+\frac{1}{8\pi^2}\log \lam \right) \right\}
\absleq C_j, \quad 0\leq j \leq 3.  
\]
It follows by Mikhlin's theorem that for any $1<p<\infty$ 
\[
\left\|\Rg_{\leq{a}}(|y||D|)+ \frac1{8\pi^2}\log (|y||D|)\chi_{\leq{a}}(|y||D|)
\right\|_{\Bb(L^p)} \leq C_p 
\]
with $y$-independent $C_p$. Thus, we need estimate the $\Bb(L^p)$-norm of 
\begin{multline*}
\log(|y||D|)\chi_{\leq a}(|y||D|)\chi_{>2a}(|D|) \\
\leq \log |y| \chi_{\leq a}(|y||D|)\chi_{>2a}(|D|)
+ \log |D|\chi_{\leq a}(|y||D|)\chi_{>2a}(|D|). 
\end{multline*}
The first term on the right is evidently bounded in $\Bb(L^p)$ 
by  $C|\log|y||$. To estimate the second let   
$f(\lam,y)= (\log \lam) \chi_{\leq a} (|y|\lam)\chi_{>2a}(\lam)$. 
We have  
\bqn \lbeq(Fj)
|f^{(j)}(\lam,y)|\leq C\lam^{-j}(1 + |\log |y||). \quad  0 \leq j \leq 3; 
\eqn 
Indeed $f(\lam,y)\not=0$ only if $|y|<2$ and $a <\lam <2a/|y|$ and,  
\[
|f(\lam,y) |\leq \max(|\log a|, |\log 2a/|y||) 
\leq (|\log |y|| + C_a),  
\]
which implies \refeq(Fj) for $j=0$. The proof for $j=1,2,3$ is 
similar. Thus, $\|f(|D|,y)\|_{\Bb(L^p)}\leq C\la \log |y| \ra$ and  
\bqn 
\|\Rg_{<a}(|y||D|)\chi_{>2a}(|D|)\|_{\Bb(L^p)} 
\leq C_{a,p} (1+ |\log|y||).
\lbeq(H02y) 
\eqn 
Estimates \refeq(H02y-add) and \refeq(H02y) imply \refeq(H02y-a). 
\qed

\section{Integral operators} \lbsec(integral) 

\subsection{Operator $K$}
Define the operator $K$ by \refeq(K-first) and  
\begin{align}  
K_{1} u(x) & = \int_0^\infty  \Gg_\lam (x)(\Pi(\lam)u)(0) \lam d\lam\,, 
\lbeq(K-1) \\
K_{2} u(x) & = \int_0^\infty  \Gg_{i\lam} (x)(\Pi(\lam)u)(0) \lam d\lam\,
\lbeq(K-2) 
\end{align} 
for $u \in \Dg_\ast$ so that, by virtue of  \refeq(1-1),
\bqn 
K u = \frac1{2}(K_{1}-K_{2})u.
\eqn 
Since $(\Pi(\lam)u)(0)\in C_0^\infty(0,\infty)$, \refeq(R-2n) 
implies that integrals \refeq(K-1) and \refeq(K-2) converge for $x\not=0$ 
and they are smooth functions of $x\in \R^4\setminus \{0\}$.

\bglm \lblm(3-1) Let $1<p<2$ and let for $\ep>0$ and $u \in \Dg_\ast$  
\[
K_{1,\ep}u(x)= 
\frac{-1}{(4\pi^2)^2 (|x|^2+i\ep)} 
\int_{\R^4} \frac{u(y)}{|x|^2-|y|^2 +i\ep}dy. 
\]
Then, with a constant $C_p>0$ independent of $\ep>0$ 
\begin{gather} 
K_{1}u(x) = \lim_{\ep \to 0} K_{1,\ep}u(x), \  
\mbox{pointwise for }\ x \not=0,  \lbeq(T11) \\
\|K_{1,\ep}u\|_p \leq C_p \|u\|_p, \lbeq(K1-bound) \\
\lim_{\ep \to 0} \|K_{1,\ep} u - K_1 u\|_p =  0, \lbeq(K1-Lp)
\end{gather} 
in particular, $K_1$ is bounded in $L^p(\R^4)$.  
\edlm

\bgpf Let $u, \ph \in \Dg_\ast$. Then, by \refeq(spect-proj) 
and Fubini's theorem 
\[  
(K_{1}u, \ph) = \frac{1}{(2\pi)^2} \int_0^\infty \left(
\int_{\R^4}\Gg_{\lam}(x)\overline{\ph(x)}dx 
\right) 
\left(
\int_{{\mathbb S}^3} \hat{u}(\lam\w) d\w \right) \lam d\lam\,. 
\] 
Since the limit converges uniformly for $\lam$ in compacts 
of $(0,\infty)$ and   
\[
\int_{\R^4}\Gg_{\lam}(x)\overline{\ph(x)}dx
=\lim_{\ep \downarrow 0}\frac{1}{(2\pi)^2}
\int_{\R^4}\frac{
\overline{\hat{\ph}(\xi)}}{\xi^2-\lam^2-i\ep} d\xi\,,   
\]
we obtain by using polar coordinates 
$\eta= \lam{\w}$, $\lam>0$, $w\in {\mathbb S}^3$ that 
\bqn 
(K_{1}u, \ph) = \lim_{\ep \downarrow 0}
\frac{1}{(2\pi)^4}
\iint_{\R^8}\frac{\hat{u}(\eta)\overline{\hat{\ph}(\xi)}}
{(|\xi|^2 -|\eta|^2 -i\ep)|\eta|^2}d{\xi}d{\eta} \,.  \lbeq(2-2)
\eqn 
On substituting 
\bqn \lbeq(exp-t)
\frac{1}{|\xi|^2 -|\eta|^2 -i\ep}= i \int_0^\infty 
e^{-it(|\xi|^2-|\eta|^2-i\ep)}dt,  \quad \ep>0,
\eqn 
and by using the Fubini theorem, we see that \refeq(2-2) is equal to 
\bqn \lbeq(2-3)
\lim_{\ep \downarrow 0}
\frac{i}{(2\pi)^4}
\int_0^\infty e^{-\ep{t}}
\left(
\int_{\R^4} e^{it\eta^2}
\hat{u}(\eta)
\frac{d\eta}{|\eta|^2}
\right) 
\overline{\left(
\int_{\R^4} e^{it|\xi|^2}
\hat{\ph}(\xi)
d{\xi}
\right)} dt.
\eqn 
By the Parseval identity we have for the $d\xi$ integral second   
\bqn 
\frac1{(2\pi)^2}
\overline{\int_{\R^4} e^{it|\xi|^2}\hat{\ph}(\xi)d{\xi}}
= \frac{-1}{(4\pi{t})^2} \int_{\R^4} e^{\frac{i|x|^{2}}{4t}}\overline{\ph(x)} dx 
\absleq \frac{C}{\la t \ra^{2}}.
\lbeq(1st)
\eqn 
For the $d\eta$-integral, substitute 
$e^{it|\eta|^2}= 1+ i |\eta|^2 \int_0^ t e^{is|\eta|^2} ds$. 
Applying the Parseval identity, we have 
\begin{align*} 
& \frac1{(2\pi)^2}\int_{\R^4} e^{it\eta^2}\hat{u}(\eta)\frac{d\eta}{|\eta|^2} 
\\
& = \frac1{(2\pi)^2} \int_{\R^4} \frac{u(y) dy}{|y|^2} 
- i \lim_{\ep \downarrow 0}
\int_{0}^t 
\left(\int_{\R^4} 
\frac{e^{-\frac{i|y|^2}{4s}}}{(4\pi{s})^2}u(y) dy \right) e^{-\frac{\ep}{s}}ds.   \end{align*} 
where we have inserted the harmless factor $e^{-\frac{\ep}{s}}$ in the 
second term for later purpose. Then,  explicitly computing the $s$-integral 
implies 
\bqn \lbeq(2nd-e)
\frac1{(2\pi)^2}\int_{\R^4} e^{it\eta^2}\hat{u}(\eta)\frac{d\eta}{|\eta|^2} 
= \frac1{(2\pi)^2} \int_{\R^4} (1-e^{-\frac{i|y|^2}{4t}})
\frac{u(y)}{|y|^2} dy.
\eqn 
Since \refeq(2nd-e) is bounded by 
$C \||y|^{-2}{u}\|_1$, the integral with respect to $t$ 
of \refeq(2-3) is absolutely convergent 
without the factor $e^{-\ep{t}}$ and the limit is unchanged 
if $e^{-\ep{t}}$ is replaced by $e^{-\ep/{t}}$. 
Eqns \refeqs(1st,2nd-e) then imply that 
$(K_{1}u, \ph)$ is equal to the $\lim_{\ep \downarrow 0}$ of  
\begin{align*} 
& \frac{-i}{4(4\pi^2)^2}\int_0^\infty 
\left(
\int_{\R^4}(1-e^{-
\frac{i|y|^2}{4t}
})
\frac{u(y)}{|y|^2} dy 
\right)
\left(
\int_{\R^4} e^{\frac{i|x|^{2}}{4t}}\overline{\ph(x)} dx
\right) 
e^{-\frac{\ep}{4t}}
t^{-2}{dt} \notag \\ 
& =
\frac{-i}{4(4\pi^2)^2}
\int_{\R^8}        
\left(\int_0^\infty 
\big(
e^{\frac{i|x|^{2}}{4t}}-  e^{\frac{i(|x|^{2}-y^2)}{4t}}
\big)
e^{-\frac{\ep}{4t}} t^{-2}dt
\right) 
\frac{\overline{\ph(x)} u(y)}{|y|^2}dy dx  \notag
\end{align*}
If we compute the inner integral explicitly, this becomes   
\bqn  
\int_{\R^8} \frac{-1}{(4\pi^2)^2}
\frac{\overline{\ph(x)} u(y)dy dx}{(|x|^{2}+i\ep)(|x|^{2}-|y|^2+i\ep)} 
=(K_{1,\ep} u, \ph)\,.
\lbeq(2-4)
\eqn  
Thus we have shown that for any $u, \ph \in \Dg_{\ast}$
\bqn \lbeq(weak-K1)
(K_1u,\ph) = \lim_{\ep \to 0}\ (K_{1,\ep}u, \ph).
\eqn

It is obvious that $K_{1,\ep}u(x)$ is spherically symmetric and, 
if we write $K_{1,\ep}u(x)=K_{1,\ep}u(\r)$ if $|x|=\r$ and  
\bqn \lbeq(spher)
Mu(r) = \frac1{\c_3}\int_{{\mathbb S}^3} u(r\w) d\w, 
\quad \c_3= |{\mathbb S}^3|,   
\eqn 
then, 
\[
K_{1,\ep}u(\r) =
\frac{-\c_3}{(4\pi^2)^2 (\r^2+i\ep)} 
\int_{\R^4} \frac{Mu(r)r^3}{\r^2-r^2 +i\ep}dr. 
\]
and change of variable implies 
\bqn \lbeq(12-1)
K_{1,\ep}u(\sqrt{\r})=
\frac{-\c_3}{2(4\pi^2)^2(\r+i\ep) } 
\int_{0}^\infty \frac{Mu(\sqrt{r})r}{\r-r +i\ep}dr. 
\eqn 
For $u \in \Dg_\ast$, $Mu(r)$ is $C^\infty$ in $(0,\infty)$. It is then 
well-known that the right side of \refeq(12-1) converges uniformly along with 
derivarives on compacts of $(0,\infty)$. Since $K_1u(x)$ is 
also smooth for $x\not=0$, then \refeq(weak-K1) implies 
$K_1u(x) = \lim_{\ep \to 0}K_{1,\ep}u(x)$ for all $x\not=0$.  
 
Moreover, the maximal Hilbert transform 
(cf. Theorem 1.4 and Lemma 1.5 of Chapter 6 of \cite{SW} (pp. 218-219))  
implies that, if we set $f(r)= Mu(\sqrt{r})r$ then 
\bqn 
F(\sqrt{\r})\colon =\sup_{\ep>0} |K_{1,\ep}u(\sqrt{\r})|\leq 
\frac{C}{\r}\big(\Mg_f(\r)+\Mg_{\tilde{f}}(\r)\big),
\eqn 
where $\Mg_f(\r)$ is the Hardy-Littlewood maximal function of $f$ and 
$\tilde{f}$ is the Hilbert transform of $f$.  Define  
$F(x)=F(|x|)$  for $x \in \R^4$. Since 
$\r^{1-p}$ is $1$-dimensional $(A)_p$ weight for $1<p<2$ 
(\cite{Stein}, page 218), we obtain 
by the weighted inequality for the maximal functions that for $1<p<2$ 
\begin{align}
& \int_{\R^4}|F(x)|^p dx 
=(\c_3/2) \int_{0}^\infty |F(\sqrt{\r})|^p \r d\r \notag \\
& \leq C \int_{0}^\infty (|\Mg_f(\r)|^p +|\Mg_{\tilde{f}}(\r)|^p)\r^{1-p} d\r 
\notag \\
& \leq C \int_{0}^\infty (|f(r)|^p +|\tilde{f}(r)|^p)r^{1-p} d\r \lbeq(2-6).
\end{align}
If we apply the weighted inequality for the Hilbert transform, then  
\begin{multline*}
\refeq(2-6)\leq 
C_1 \int_{0}^\infty |f(r)|^p r^{1-p} dr 
= C_1 \int_{0}^\infty |Mu(\sqrt{r})r|^p r^{1-p} dr \\
= 2C_1 \int_{0}^\infty |Mu({r})|^p r^{3} dr 
\leq C \|u\|_p^p\,.
\end{multline*} 
Thus, $\|\sup_{\ep>0} |K_{1,\ep}u(x)|\|_p \leq C \|u\|_p$ 
and the dominated convergence theorem implies 
$\|K_{1,\ep}u(x)-K_{1}u(x)\|_p \to 0$ as $\ep \to 0$ for $1<p<2$.  
\edpf

\bgrm Operator $K_{1}$ is unbounded in $L^p(\R^4)$ for $2<p<\infty$. 
To see this we note 
\begin{align} \lbeq(2-4-)
& \frac1{(|x|^2+i\ep)(|x|^2-|y|^2 +i\ep)}= 
\frac1{(|y|^2+i\ep)(|x|^2-|y|^2 +i\ep)}\\ 
& \hspace{1.5cm} 
- \frac{|x|^2-|y|^2}{(|x|^2+i\ep)(|y|^2+i\ep)(|x|^2-|y|^2 +i\ep)} 
\notag 
\end{align}
and recall that the integral operator produced by 
the first term on the right of \refeq(2-4-) 
is uniformly bounded in $L^p(\R^4)$ for $\ep>0$ if 
$2<p<\infty$ (cf. {\rm{Lemma 3.4}} of \cite{Ya-4dim-new}). 
Hence, if $K_1$ were bounded in $L^p(\R^4)$ for a $p \in (2,\infty)$, 
then it must be that for $u,w\in C_0^\infty(\R^4)$ 
\[
\lim_{\ep\downarrow 
0}\left|\iint_{\R^4 \times \R^4}
\frac{(|x|^2-|y|^2)u(y)w(x) dx dy}
{(|x|^2+i\ep)(|y|^2+i\ep)(|x|^2-|y|^2 +i\ep)} \right|
\leq C \|u\|_p \|w\|_q 
\]
for a constant $C>0$, $q=p/(p-1)$. However, this is impossible because 
the left hand side is equal to  
\[
\left| \iint_{\R^4 \times \R^4}
\frac{u(y)w(x) dx dy}{|x|^2|y|^2} \right|
\]
which cannot be bounded by $ C \|u\|_p \|w\|_q $.
\edrm 

\bglm \lblm(3-2)  For $u \in \Dg_\ast$, 
$K_2 u(x) = \lim_{\ep \to 0} K_{2,\ep}u(x)$, 
\bqn \lbeq(T12)
K_{2,\ep}u(x)
=\frac{1}{(4\pi^2)^2 (|x|^2+ i\ep)} 
\int_{\R^4} \frac{u(y)}{(|x|^2+|y|^2+ i\ep)}dy; 
\eqn 
$K_2$ is bounded in $L^p(\R^4)$ for $1<p<2$ and unbounded for 
$2<p<\infty$.
\edlm 
\bgpf Since 
\[
G_0(i\lam)\overline{\ph}(x)= 
\lim_{\ep \downarrow {0}} \frac1{(2\pi)^2}\int_{\R^4} 
\frac{e^{ix\xi}\overline{\hat{\ph}(\xi)}d\xi}{|\xi|^2+ \lam^2-i\ep}
\]
converges uniformly with respect $\lam$ in compact subsets of $\R$, we 
obtain 
\begin{align}
(K_{2}u, \ph)& = \frac1{(2\pi)^2} \int_{\R^4} 
\left(  
\int_0^\infty \Gg_{i\lam}(x)
\left(\int_{{\mathbb S}^3} \hat{u}(\lam\w) d\w \right)
\lam d\lam\right) \overline{\ph(x)}dx  
\notag \\ 
& = \lim_{\ep \downarrow {0}}
\frac{1}{(2\pi)^4}
\int_{\R^8}\frac{\hat{u}(\eta)\overline{\hat{\ph}(\xi)}}
{(|\xi|^2 +|\eta|^2-i\ep)|\eta|^2}d{\xi}d{\eta}\,.   \lbeq(3-3-3)
\end{align}
We compute \refeq(3-3-3) by repeating the argument 
in the proof of \reflm(3-1) by replacing $-|\eta|^2$ by $|\eta|^2$, 
which implies $K_2 u(x) = \lim_{\ep \to 0} K_{2,\ep}u(x)$. 
It is obvious that $K_2 u(x)$ 
is rotationally symmetric. We write $K_2 u(x) =K_2 u(\r)$, $\r=|x|$. 
Denote $Mu(\r^\frac14)=f(\r)$. Then 
\[
|(K_{2}u)(\r^\frac14)|\leq 
\frac{1}{4(4\pi^2)^2 \r^{\frac12}} \int_{0}^\infty \frac{|f(r)|}
{\r^{\frac12}+r^{\frac12}}dr
= \frac{1}{4(4\pi^2)^2} \int_{0}^\infty \frac{|f(r\r)|}{1+r^{\frac12}}dr. 
\]
Minkowski's inequality implies 
\begin{align*}
&  \|(K_{2}u)(\r^\frac14)\|_{L^p((0,\infty), d\r)} \leq 
\int_{0}^\infty \frac{\|f(r\r)\|_{L^p([0,\infty),d\r)}} 
{4(4\pi^2)^2(1+r^{\frac12})}dr \\
& \leq C 
\int_{0}^\infty \frac{\|f\|_p}{r^{1/p}(1+r^{\frac12})}dr
\leq C \|f\|_{L^p((0,\infty)}
\end{align*} 
and, by H\"older's inequality,
\begin{align*}
\|(K_{2}u)(x)\|_p 
& 
= 
\left(\frac{\c_3}{4} \int_0^\infty |(K_{2}u)(\r^\frac14)|^p d\r \right)^{1/p}
\\
& \leq C \left(\int_0^\infty |(Mu)(r^\frac14)|^p dr\right)^{1/p}  
\leq C \|u\|_p^p . 
\end{align*}
This proves the lemma. 
\edpf 

\bglm \lblm(allp) The operator $K$ is bounded in $L^p(\R^4)$ for all $1<p<\infty$. \edlm 
\bgpf \reflms(3-1,3-2) prove that 
$K$ is bounded in $L^p(\R^4)$ for $1<p<2$. 
We shall prove the same for $2<p<\infty$. Then, the lemma will follow 
by the interpolation. 
Define $K_{\ep}u= 2(K_{1,\ep}u- K_{2,\ep}u)$ for $\ep>0$. 
By virtue of \reflms(3-1,3-2), 
$Ku (x) = \lim_{\ep \downarrow {0}}K_{\ep}u(x)$ for $x\not=0$ and 
a simple computation implies 
\bqn 
K_{\ep}u(x) = \frac{1}{2(4\pi^2)^2}(F_{-,\ep}u(x)- F_{+,\ep}u(x)),
\eqn 
where $F_{\pm ,\ep}u(x)$ are rotationally invariant functions given by 
\[
F_{\pm,\ep}u(x)=
\int_{\R^4}\frac{u(y)dy}{(|x|^2\pm |y|^2 +i\ep)|y|^2}.
\] 
Notice that the dangerous term $(1/2\pi^4)|x|^{-2}|y|^{-2}$ has been  
cancelled.  We denote $F_{\pm,\ep}u(x)= f_{\pm,\ep}(\r)$, $\r=|x|$.
Then,
\[
\tilde{f}_{+}(\r)= \sup_{\ep>0}|f_{+,\ep}(\r)| 
\leq \frac{\c_3}{8\pi^4}\int_0^\infty \frac{|Mu(r)|r^3 dr}{r^2 (\r^2+r^2)} 
=\frac{\c_3}{8\pi^4}\int_0^\infty \frac{|Mu(r\r)|r dr}{(1+r^2)} 
\] 
and Minkowski's inequality implies for any $2<p<\infty$ that 
\begin{align*}
& \|\tilde{f}_{+} (|x|)\|_{L^p(\R^4)} 
\leq C\int_0^\infty \frac{\|Mu(r\r)\|_{L^p((0,\infty),\r^3d\r)}rdr}{(1+r^2)} \\
& = C \|Mu(\r)\|_{L^p((0,\infty),\r^3d\r)} \int_0^\infty 
\frac{r^{1-4/p} dr}{(1+r^2)}
\leq C \|u\|_p\,. 
\end{align*}
It follows that $F_{+\ep}(x)$ converges as $\ep \to 0$ to 
\[
F_{+}u(x)=\int_{\R^4} \frac{u(y)dy}{(|x|^2+|y|^2)|y|^2}.
\]
in $L^p(\R^4)$ for all $2<p<\infty$. 

It is shown in  Lemma 3.4 
in \cite{Ya-4dim-new} via the same 
argument as in the proof of 
\reflm(3-1) that,  for $2<p<\infty$, 
$F_{-,\ep}$ is uniformly bounded in $\Bb(L^p)$ 
for $\ep>0$ and $F_{-,\ep}u(x)$ converges as $\ep \to 0$ for $x\not=0$ 
and simultaneously in $L^p(\R^4)$. 
Hence $K$ is bounded in $L^p(\R^4)$ for $2<p<\infty$ as well 
and the lemma follows. \edpf 

\subsection{Good operators}

Here we prove \refeq(WT-estimate) and \refeq(WTlam-estimate). 

\bglm \lblm(T-K)  Let $\W(T)$ be defined by \refeq(OmegaT). 
Suppose that $T$ is integral operator with kernel 
$T(x,y)\in \Lg^1$. Then, we have \refeq(T-K) 
for almost all $x \in \R^4$ and $\W(T)$ satisfies 
\refeq(WT-estimate) for all $1<p<\infty$: 
$ \|\W(T)u\|_p \leq C \|T\|_1 \|u\|_p$.
\edlm 
\bgpf By the definition 
\begin{align*}
\W(T)u(x)& = 
\int_0^\infty \left(\iint_{\R^8}
\Rg_\lam(x-y)T(y,z) (\Pi(\lam)u)(z) dy dz\right)\lam^3 d\lam \\
& = 
\int_0^\infty \left(\iint_{\R^8}T(y,z)\tau_y 
\Rg_\lam(x)(\Pi(\lam)\tau_{-z} u)(0) dy dz\right)\lam^3 d\lam \,.
\end{align*}
If we may change the order of integrations, $\W(T)u (x)$ becomes 
\bqn \lbeq(T-K-2)
\iint_{\R^8}T(y,z)\t_y \left(\int_0^\infty 
\Rg_\lam(x)(\Pi(\lam)\tau_{-z} u)(0) \lam^3 d\lam \right) dy dz\, 
\eqn 
which implies \refeq(T-K). By virtue 
of \reflm(allp) we have  $\|\t_y K \t_{-z}u\|_p =\|K \t_{-z}u\|_p 
\leq C \|u\|_p$ and Minkowski's lemma implies \refeq(T-K).  

To show that the change of order of integrations is possible for 
almost all $x\in \R^4$, it suffices by virtue of Fubini's theorem  
to show that 
$\Rg_\lam(x-y)T(y,z) (\Pi(\lam)u)(z)\lam^3$ is (absolutely) integrable  
with respect to 
$(x,y,z,\lam)\in B_R(0) \times \R^4 \times \R^4 \times (0,\infty)$ 
for any $R>0$, where $B_R(0)=\{x \colon |x|<R\}$. 
However, this is obvious since $\Pi(\lam)u(z)=0$ for $\lam$ outside 
a compact interval $[\a,\b]\Subset (0,\infty)$,    
$|\Pi(\lam)u(z)|\leq C \az^{-3/2}$ uniformly 
for $\lam\in [\a,\b]$ and  
$\int_{B(0,R)}|\Rg_\lam(x-y)|dx$ is uniformly bounded for $y \in \R^4$ 
and $\lam \in [\a,\b]$. This completes the proof.  
\edpf    

\bgcor \lbcor(vari-sep) 
Any $T(\lam,x,y) \in \Gg\Vg\Sg$ is GPR. 
\edcor 

The following is the variant of Proposition 3.9 of 
\cite{Ya-4dim-new}.  
We take advantage of this chance to point out that 
Proposition 3.9 of \cite{Ya-4dim-new} has an error 
and it has to be replaced by the following proposition 
and that some obvious modifications are necessary 
in the part of \cite{Ya-4dim-new} which uses that proposition. 
Let $a_{+}= \max(a,0)$.

\bgprop \lbprop(R-theo) Let $\Lam>0$ and $F(\lam,x,y)$ be 
an $\Lg^1$-valued function of 
$\lam\in(0,\infty)$ of class $C^2$ such that 
$F^{(2)}(\lam)$ is absolutely continuous on compact intervals 
of $(0,\infty)$. Let 
$F(\lam)$ be the integral operator with the kernel $F(\lam,x,y)$.
\ben 
\item[{\rm (1)}] Suppose that 
$\lim_{\lam\to {\infty}}\lam^{j}\|F^{(j)}(\lam)\|_{\Lg^1}=0$ for $j=0,1,2$ 
and that 
$\int_\Lam^\infty \lam^2 \|F^{(3)}(\lam)\|_{\Lg^1}d\lam <\infty$. 
Then, for any $a>\Lam$ 
\bqn \lbeq(T-K-extra)
\tilde{\W}{\geq{a}}(F(\lam))u= \int_{0}^\infty
R_0(\lam^4)F(\lam) \Pi(\lam)u \lam^3 \chi_{\geq{a}}(\lam)d\lam 
\eqn 
is GOP.   

\item[{\rm (2)}] Suppose 
$\int_0^\Lam \|F^{(3)}(\lam)\|_{\Lg^1}\lam^2 d\lam <\infty$.
Then, for any $0<a<\Lam$
\bqn \lbeq(WgT)
\tilde{\W}_{\leq a}(F(\lam)) u= \int_0^\infty R^{+}_0 (\lam^4)F(\lam)\Pi(\lam)u \lam^3 
\chi_{\leq{a}}(\lam)d\lam 
\eqn 
is GOP. 
\een
\edprop 
\bgpf (1) Let $u \in \Dg_\ast$ and  
$\Pi(\lam)u(z)= (\Pi(\lam)\tau_z u)(0)=0$ outside $[\a,\b] \Subset (0,\infty)$.  
By Taylor's formula we have 
\bqn \lbeq(F-3)
F(\lam)= \frac{-1}{2} \int_{0}^\infty ((\r-\lam)_{+})^2 F^{(3)}(\r)d\r.
\eqn 
Let $B(\lam)=((1-\lam)_{+})^2= ((1-\lam^2)_{+})^2(1+\lam)^{-2}$. 
It is known (cf. \cite{Stein}, p.389) that the Fourier transform 
of $B(|\xi|)$ is integrable on $\R^4$, $B(|D|)$ is bounded in $L^p(\R^4)$ 
for all $1\leq p \leq \infty$ and $\|B(|D|/\r)\|_{\Bb(L^p)}$ is independent 
of $\r$. Substitute \refeq(F-3) and change the order 
of the integrals. Then, \refeq(T-K-extra) becomes 
\bqn 
\frac{-1}{2} \int_{0}^\infty \left( 
\int_0^\infty ((\r-\lam)_{+})^2 
R_0(\lam^4)F^{(3)}(\r)\Pi(\lam)\lam^3 \chi_{\geq{a}}(\lam)u d\lam \right) d\r.
\lbeq(sr-0)
\eqn 
and, by virtue of \refeq(mult) and \refeq(OmegaT), 
the inner integral does  
\begin{align} 
& \r^2 
\int_0^\infty 
R_0(\lam^4)F^{(3)}(\r)\Pi(\lam)B(|D|/\r)\chi_{\geq{a}}(|D|) u \lam^3 d\lam 
\notag \\
& = {\r^2}\W(F^{(3)}(\r))B(|D|/\r)\chi_{\geq{a}}(|D|)u. 
\lbeq(sr)
\end{align}  
Thus, Minkowski's inequality and \reflm(T-K) imply 
\bqn 
\|\refeq(sr-0)\|_p \leq C \int_a^\infty 
\r^2 \|F^{(3)}(\r)\|_{\Lg^1} \|u\|_p d\r \leq C \|u\|_p.
\eqn 
This proves the first statement. 

(2) The proof of the second statement is a slight modification of that of (1). 
By Taylor's formula we have 
\begin{align}
F(\lam)& = F(0) + \lam F'(0)+ \frac12 \lam^2 F''(0) + R_F(\lam), \lbeq(F-first) \\
R_F (\lam)& = \frac12\int_0^\infty ((\lam-\r)_{+})^2 F^{(3)}(\r) d\r 
\lbeq(F-remainder)
\end{align}
which we substite in \refeq(WgT). 
Since $F(0), F'(0)$ and $F''(0)$ are integral operators with kernels in 
$\Lg^1$ and $\lam^j \chi_{\leq{a}}(\lam)$ are GMUs,  
\reflm(T-K) implies that the operator produced by 
$F(0) + \lam F'(0)+ \frac12 \lam^2 F''(0)$ is GOP. 

We need consider the operator produced by $R_F (\lam)$. 
Changing the order of integration and inserting $\chi_{\leq{2a}}(\r)$ 
which satisfies  $\chi_{\leq{2a}}(\r)=1$ 
if  $((\lam-\r)_{+})^2 \chi_{\leq a}(\lam)\not=0$, the operator 
produced by $R_F(\lam)$ is equal to 
\[
\frac12 \int_0^\infty 
\left(
\int_0^\infty ((\lam-\rho)_{+})^2
R_0(\lam^4)F^{(3)}(\r)\Pi(\lam)\chi_{\leq {a}}(|D|)u \lam^3 d\lam 
\right)\chi_{\leq{2a}}(\r) d\r.
\]
We substitute $((\lam-\r)_{+})^2 = (\lam-\r)^2-((\r-\lam)_{+})^2$. 
Then, $(\lam-\rho)^2$ produces  
\begin{align*}
Su & 
= 
\frac12\int_0^\infty (S_0(\r)u - 2\r S_1(\r)u + \r^2 S_2(\r)u) 
\chi_{\leq{2a}}(\r) d\r 
\\
S_j(\r)u & = 
\int_0^\infty 
R_0(\lam^4)F^{(3)}(\r)\Pi(\lam)|D|^{2-j}\chi_{\leq {a}}(|D|)u  \lam^3 d\lam, 
\quad j=0,1,2.
\end{align*}
Since $\lam^{2-j} \chi_{\leq {a}}(\lam)$, $j=0,1,2$ are GMUs, 
\reflm(T-K) implies  
\[
\|S_j(\r) u \|_p \leq C_j \|F^{(3)}(\r)\|_{\Lg^1} \|u\|_p , 
\quad j=0,1,2. 
\]
It follows by Minkowski's inequality that 
\[
\|Su\|_p \leq C \sum_{j=0}^2 
\int_0^\infty \r^{j} \|F^{(3)}(\r)\|_{\Lg^1}\|u\|_p \chi_{\leq{2a}}(\r) d\r 
\leq C \|u\|_p.
\]
Let $B(\lam)=((1-\lam)_{+})^2$. Then the function 
$((\rho-\lam)_{+})^2$ produces  
\bqn \lbeq(R-3)
\frac12 \int_0^\infty 
\left(
\int_0^\infty 
R_0(\lam^4)F^{(3)}(\r)\Pi(\lam)B(|D|/\r)\chi_{\leq {a}}(|D|)u \lam^3 d\lam 
\right)\r^2 \chi_{\leq{2a}}(\r) d\r 
\eqn 
and the inner integral is equal to 
$\W(F^{(3)}(\r))B(|D|/\r)\chi_{\leq {a}}(|D|)u$. Since 
$B(|D|/\r)$ is uniformly bounded in 
$L^p(\R^4)$ for all $1\leq p \leq \infty$ as previously, 
\reflm(T-K) and Minkowski's inequality imply 
\[
\|\refeq(R-3)\|_{p} \leq 
\int_0^\infty \|F^{(3)}(\r)\|_{L^1}\|u\|_p \r^2 
\chi_{\leq {2a}}(\r) d\r \leq C \|u\|_p\,.
\]
Combining these estimates, we obtain the second statement. 
\edpf

\section{High energy estimate} \lbsec(high) 

We prove here \refths(small,high). 

\subsection{Proof of \refthb(small). Small potentials} 
From what is explained in the introduction, 
we have only to prove \refeq(small-effect-0) for $n=1,2, \dots$ 
for $W_{n}\chi_{\geq{a}}(|D|)u$ defined by \refeq(Wn-0). 

\noindent 
(1) We have already shown that 
$\|W_1 \chi_{\geq{a}}(|D|)u\|_p \leq C \|V\|_1 \|u\|_p$.  

\noindent 
(2) Let $n=2$ and $V_y^{(2)}(x)  = V(x)V(x-y)$. If we have proven 
\refeq(W2-express), then 
Minkowski's inequality, \refeq(T-K) and \refeq(H02y-a) imply  
\begin{align*}
\|W_{2}\chi_{\geq{a}}(|D|)u\|_p
& \leq 
\int_{\R^4} \|\W(M_{V_y^{(2)}})\|_{\Bb(L^p)}
\|\Rg(|y||D|)\chi_{\geq{a}}(|D|)\tau_y u \|_p dy  \notag \\
& \leq  C \|u\|_p \int_{\R^8}|V(x)V(x-y)| \la \log |y| \ra dx dy. 
\end{align*}
Then we may estimate the right side as in the proof of \reflm(members) (1),  
which yields the desired estimate for $n=2$:  
\[
\|W_{2}\chi_{\geq{a}}(|D|)u\|_p\leq C 
(\|V\|_{L^q_{loc,u}} + \|\la \log |x|\ra V\|_{L^1})^2 \|u\|_p .
\]
We now prove \refeq(W2-express). We have by changing variables that   
\begin{align}
& M_VR_0({\lam}^4)M_V u(x) 
= \int_{\R^4} V(x)\Rg(\lam|y|)V(x-y) u(x-y)dy \notag \\  
&= \int_{\R^4} V^{(2)}_y(x)\Rg(\lam|y|)(\tau_y u) (x) dy \lbeq(wG0w)
\end{align} 
which we substitute in \refeq(Wn-0). 
Since $\tau_y$ commutes with $\Pi(\lam)$ and 
\bqn \lbeq(mult-appl)
\Rg(\lam|y|)\Pi(\lam)= \Pi(\lam)\Rg(|y||D|)
\eqn 
by virtue of \refeq(mult), this implies that 
$W_{2}\chi_{\geq{a}}(|D|)u(x)$ is equal to  
\[
\int_0^\infty \left(\int_{\R^8} 
\Rg({\lam}|x-y|)V^{(2)}_z(y)(\Pi(\lam)\Rg (|z||D|)\t_z u)(y) 
dzdy \right)\lam^{3} \chi_{\geq{a}}(\lam) d\lam.
\]
Change the order of integrations as in the proof of \reflm(T-K) 
and use the definition \refeq(OmegaT). Then, 
$W_{2}\chi_{\geq{a}}(|D|)u(x)$ becomes 
\begin{align*}
& 
\int_{\R^4}  \left(\int_0^\infty 
\big(R_0({\lam}^4)M_{V^{(2)}_z} 
\Pi(\lam)\Rg(|z||D|)\chi_{\geq{a}}(|D|)\t_z u\big)(x) 
\lam^{3} d\lam \right) dz  \\
& = \int_{\R^4}  \big(
\W(M_{V^{(2)}_z})\Rg(|z||D|)\chi_{\geq{a}}(|D|)\t_z u\big)(x) dz 
\end{align*} 
as desired. 

\noindent 
(3) Let $n \geq 3$ and 
$V^{(n)}_{y_1,\dots,y_{n-1}}(x)= 
V(x) V(x-y_1) \cdots V(x-y_1-\cdots-y_{n-1})$. 
Repeating the argument used for \refeq(wG0w) implies  
\begin{multline*}
M_w(M_vR_0(\lam^4)M_w)^{n-1} M_v u(x)= (M_V R_0(\lam^4))^{n-1} M_V u(x) \\ 
= \int_{\R^{4(n-1)}}
V^{(n)}_{y_1,\dots, y_{n-1}}(x)
\left(\prod_{j=1}^{n-1} \Rg(\lam|y_j|)\right)\tau_{y_1+\cdots+y_{n-1}}u(x) 
dy_1 \dots dy_{n-1} .
\end{multline*}
It follows that $W_{n}\chi_{\geq{a}}(|D|)u(x)$ is equal to 
\begin{multline}
\int_0^\infty \int_{\R^4} \int_{\R^{4(n-1)}}\Rg(\lam |x-y|) 
V^{(n)}_{y_1,\dots, y_{n-1}}(y) \\
\times \left(\prod_{j=1}^{n-1} \Rg(\lam|y_j|)\right)
\Pi(\lam)\tau_{y_1+\cdots+y_{n-1}}u(y) \lam^3 \chi_{\geq{a}}(\lam)
dy_1 \cdots dy_{n-1} dy d\lam\,.  \lbeq(Wn)
\end{multline} 
As in the proof of \reflm(T-K) the integral on the right-hand side is 
integrable for a.e. $x\in \R^4$ and we may integrate \refeq(Wn) by $d\lam$ first. 
We do this after applying \refeq(mult) to 
$(\prod_{j=1}^{n-1}\Rg(\lam|y_j|))\Pi(\lam)$. 
Then \refeq(OmegaT) implies that the right of \refeq(Wn)
can be written as   
\[
\int_{\R^{4(n-1)}} \W(M_{V^{(n)}_{y_1,\dots,y_{n-1}}})
 \chi_{\geq{a}}(|D|)\prod_{j=1}^{n-1}
\Rg (|y_j||D|) \t_{y_1+\cdots+ y_{n-1}} {u} dy_1 \dots dy_{n-1}.   
\]
Here 
$\chi_{\geq{a}}(|D|)= \chi_{\geq{a}}(|D|)\chi_{\geq{a}/2}(|D|)^{n-2}$. 
Then, 
Minkowski's inequality, \reflms(T-K,H02y) imply  
that $\|W_n \chi_{\geq{a}}(|D|)u\|_p$ is bounded by 
\begin{align*}
& C_{a,p}^n 
\int_{\R^{4(n-1)}}  
\|V^{(n)}_{y_1,\dots,y_{n-1}}\|_{L^1(\R^4)} 
\prod_{j=1}^{n-1} \la \log |y_j|\ra \|u\|_p dy_1 \dots dy_{n-1} 
\notag \\
&= C_{a,p}^n 
\int_{\R^{4n}}  
|V(x_0)| 
\prod_{j=1}^{n-1} 
|V(x_j)| \la |\log |x_{j-1}-x_j|\ra  
\|u\|_p dx_0 \dots dx_{n-1} 
\end{align*}
where the change of variables $y_j=x_{j-1}-x_j$, $j=1, \dots, n-1$ was made. 
We estimate the integral inductively by using Schwarz' and  H\"older's 
inequalities $n$-times by 
\begin{align*}
& \|V\|_1^\frac12 
\left(\int_{\R^4} V(x_0)\la \log |x_0-x_1|\ra^2 V(x_1)dx_0 dx_1 \right)^\frac12 
\\ 
& \times \cdots \times 
\left(\int_{\R^4} V(x_{n-2})\la \log |x_{n-2}-x_{n-1}|\ra^2 
V(x_{n-1})dx_{n-2}dx_{n-1}\right)^\frac12 \|V\|_1^\frac12 \\
& \leq C^n \|V\|_1 (\|V\|_{L^q_{loc,u}}+ \|\la \log|x|\ra^{2} V\|_{L^1})^{n-1}
\\
& \leq C^n (\|V\|_{L^q_{loc,u}}+ \|\la \log|x|\ra^{2} V\|_{L^1})^{n}.
\end{align*}
This proves \refeq(small-effect-0) for $n\geq 3$ and completes 
the proof of  \refth(small). 
\qed

\subsection{\bf Proof of \refthb(high)}

For the proof we need a lemma.  

\bglm \lblm(HS) Let $1<q<4/3$, $q'=q/(q-1)$ and $j=0,1, \dots$. 
Suppose $\ax^{(2j-3)_{+}}V \in L^1(\R^4)$ and $V \in L^q(\R^4)$. Then, 
for any $a>0$, $M_vR_0(\lam^4)M_w$ is $\Hg_2$-valued 
$C^j$ function of $\lam>a$ and, for $n=1,2, \dots$
\bqn \lbeq(HS-derivative)
\|\pa_{\lam}^j (M_vR_0(\lam^4)M_w)^n\|_{\Hg_2} \leq 
\frac{C_{n}}{\lam^{\frac{n}{2q'}}}
((\|\ax^{(2j-3)_{+}} V\|_{1}+ \|V\|_q)^n .
\eqn  
\edlm 
\bgpf Let $N_{1,j}$ and $N_{2,j}$, $j=0,1, \dots$ be convolution operators with 
\bqn   
N_{1,j}(x) = \lam^{-j} \la \log \lam|x|\ra \chi_{\leq{1}}(\lam|x|), \quad 
N_{2,j}(x) = \frac{\chi_{\geq 1}(\lam|x|)|x|^j}{\lam^\frac32 |x|^\frac32}\,.
\eqn  
It follows from \refeq(R-0) that   
$\pa_\lam^j \Rg(\lam^4)(x) \absleq C(N_{1,j}(x) + N_{2,j}(x))$, $j=0,1,\dots$.  
By repeating estimate \refeq(q-prime), we have for any $1\leq q\leq \infty$ 
that  
\begin{align*}
& \|M_v N_{1,j} M_w \|_{\Hg_2}^2 \leq C \lam^{-2j} 
\int_{\lam|x-y|\leq 1} |V(x)|\la \log \lam|x-y|\ra^2 |V(y)| dx dy 
\\
& \leq C \lam^{-2j} \|V\|_{1} \|V\|_{L^q_{loc,u}} 
\|\la \log \lam|x|\ra \|_{L^{2q'}(\lam|x|<1)}^2 
\end{align*}
and  $\|M_v N_{1,j} M_w \|_{\Hg_2} 
\leq C\lam^{-j-\frac{2}{q'}}\|v\|_{2} \|w\|_{L^{2q}_{loc,u}}$.

By H\"older's inequality, for $1\leq q<4$,
\begin{align*}
& \|vN_{2,0} w \|_{\Hg_2}^2 \leq 
C 
\int_{\lam|x-y|\geq 1 }\frac{|V(x)||V(y)|}{(\lam |x-y|)^{3}}dx dy \\
& \leq C \|V\|_1 \|V\|_q 
\left(\int_{\lam|x|\geq 1 }\frac{dx}{(\lam |x|)^{3q'}}\right)^{\frac1{q'}} 
\leq C \lam^{-\frac{4}{q'}}\|V\|_1 \|V\|_q 
\end{align*}
Likewise for $1\leq q <4/3$, 
\[
\|vN_{2,1} w \|_{\Hg_2}^2 \leq 
C \frac1{\lam^2}
\int_{\lam|x-y|\geq 1 }\frac{|V(x)||V(y)|}{\lam |x-y|}dx dy 
\leq C \|V\|_1 \|V\|_q \lam^{-2-\frac{4}{q'}}
\]
For $j\geq 2$, we evidently have 
$\|vN_{2,j} w \|_{\Hg_2}\leq C \lam^{-\frac32}\|\ax^{2j-3}V\|_1$. 
Combining these estimates together, we obtain for $1\leq q <\frac43$ that 
\[
\|\pa_{\lam}^j M_vR_0(\lam^4)M_w\|_{\Hg_2} 
\leq C \lam^{-\min\left(j+\frac{2}{q'}, \frac32\right)} 
(\|\ax^{(2j-3)_{+}} V\|_{1}+ \|V\|_q)
\]
This implies \refeq(HS-derivative) for $n=1$. For $n\geq 2$, 
we compute 
$\pa_\lam^j (M_vR_0(\lam^4)M_w)^n$ via Leibniz' formula 
and estimate each factor via \refeq(HS-derivative) for $n=1$. 
The lemma follows.  
\edpf

\paragraph{\bf Proof of \refthb(high).} 
We may assume $1<q<4/3$. Let $N$ be such that $2N/q'>3$. 
We substitute  \refeq(with) with \refeq(with-2) 
in the stationary formula \refeq(sta-high) for the high energy part. 
Then, $W_{-} \chi_{\geq{a}}(|D|)$ becomes 
\[
\sum_{n=0}^{N-1} (-1)^n W_n \chi_{\geq{a}}(|D|)u 
+ (-1)^N \int_0^\infty R_0^{+}(\lam^4)D_N(\lam)\Pi(\lam)u \lam^3 
\chi_{\geq{a}}(\lam)d\lam, 
\]
where we set $W_0 u= u$. By virtue of \refeq(small-effect-0)  
$\sum_{n=0}^{N-1} (-1)^n W_n \chi_{\geq{a}}(|D|)u$ is GOP;    
\reflm(HS) implies that $D_N(\lam,x,y)$ is $\Lg^1$-valued function 
of $\lam\in (a,\infty)$ of class $C^3$ and  
\[
\|\pa_\lam^j D_N(\lam)\|_{\Lg^1}\leq C \lam^{-\frac{2N}{q'}}
(\|\ax^{3} V\|_{L^1}+ \|V\|_{L^2})^N, 
\quad 0\leq j \leq 3.
\] 
Hence the operator $D_N(\lam)$ is GPR for \refeq(sta-high) 
by \refprop(R-theo) and \refth(high) follows. 
\qed

\section{Low energy estimates 1. The case $H$ is regular at zero} \lbsec(low1)
In what follows we shall study the low energy 
part of the wave operator $W_{-}\chi_{\leq {a}}(|D|)$ or the operator 
defined by the integral    
\bqn \lbeq(sta0-low)
\W_{low,a}u= \int_0^\infty R_0 (\lam^4)
\Qg_v(\lam)\Pi(\lam) u \lam^3 
\chi_{\leq{a}}(\lam)d\lam  
\eqn  
for a sufficiently small $a>0$. As previously we define 
\begin{gather} \lbeq(sta0-low-T)
\W_{low,a}(T)u= \int_0^\infty R_0 (\lam^4)
T \Pi(\lam) u \lam^3 \chi_{\leq{a}}(\lam)d\lam, \\
\W_{low,a}(T(\lam))u= \int_0^\infty R_0 (\lam^4)
T(\lam) \Pi(\lam) u \lam^3 \chi_{\leq{a}}(\lam)d\lam \,.
\lbeq(sta0-low-Tlam)
\end{gather}  
We shall abuse notation and say that {\it  $T$ or $T(\lam)$ is GPR 
for small $\lam>0$ if 
$\W_{low,a}(T)$ or $\W_{low,a}(T(\lam))$ is GOP for sufficiently small $a$.}  
{\it We irrespectively write $\Rg_{em}(\lam)$ for GPR which is 
the operator valued function which satisfies the condition of \refprop(R-theo)}. 

In this section, we shall prove \refth(main-rephrase) when 
$H$ is regular at zero. We assume 
$\la \log |x|\ra^2 \ax^{8}V \in (L^1\cap L^q)(\R^4)$ for a $q>1$.

\subsection{Feshbach formula}
For studying $\Mg(\lam^4)^{-1}$ as $\lam \to 0$, we shall often use 
the following lemma. 
Let $A$ be the operator matrix
\[
A = 
\begin{pmatrix} a_{11}  & a_{12} \\ a_{21}  & a_{22} \end{pmatrix} 
\] 
on the direct sum of Banach spaces $\Yg= \Yg_1 \oplus \Yg_2$. 

\bglm[Feshbach formula] \lblm(FS) Suppose $a_{11}$, $a_{22}$ are closed 
and $a_{12}$, $a_{21}$ are bounded operators. 
Suppose that the bounded inverse $a_{22}^{-1}$ exists. 
Then $A^{-1}$ exists if and only if 
$d= (a_{11}- a_{12}a_{22}^{-1}a_{21})^{-1}$ exists. In this case we have 
\bqn \lbeq(FS-formula)
A^{-1} = \begin{pmatrix}  d & -d a_{12} a_{22}^{-1} \\
-a_{22}^{-1}a_{21} d & a_{22}^{-1}a_{21}d a_{12} a_{22}^{-1} + a_{22}^{-1}
\end{pmatrix}.
\eqn 
\edlm 

\subsection{Regular case}  
Recall \refeq(m4-mlam4). The inverse 
$(QT_0Q)^{-1}$ exists in $Q\HL$ since $H$ is regular at zero. 
Let  
\bqn \lbeq(gPT0-inv) 
D_{0}= Q(QT_0Q)^{-1}Q \ \mbox{and}\ L_0  = \begin{pmatrix} P & -PT_0Q D_{0} \\
- D_{0} QT_0P  &  D_{0} QT_0PT_0Q D_{0} 
\end{pmatrix}
\eqn 
in the decomposition $\HL= P \HL \oplus Q\HL$. 

\bglm \lblm(t0-inv)
For small $\lam>0$, $T_0+ g_0(\lam) P$ is invertible and 
\bqn \lbeq(t0-inv)
(T_0+g_0(\lam)P)^{-1}=D_0 + h(\lam)L_0 , \quad h(\lam)=(g_0(\lam)+ c_1)^{-1}
\eqn  
where  $c_1$ is a real constant. The operator $L_0$ is of rank two.
\edlm 
\bgpf In the decomposition $\HL= P \HL \oplus Q\HL$  
\bqn \lbeq(Decom-FS)
T_0+ g_0(\lam)P = 
\begin{pmatrix} g_0(\lam)  + PT_0P & PT_0Q \\ QT_0P & Q T_0 Q \end{pmatrix} 
=\colon \begin{pmatrix} a_{11} & a_{12} \\ a_{21} & a_{22} \end{pmatrix}.
\eqn
Here $a_{22}{=} QT_0Q$ is invertible in $Q\HL$ and 
\[
a_{11}- a_{12}a_{22}^{-1}a_{21}= 
g_0(\lam)P + PT_0 P - PT_0 D_0 T_0P = (g_0(\lam)+ c_1)P,
\]
where $c_1= (({v}, T_0{v})- (QT_0 {v}, D_0 QT_0 {v} ))\|V\|_1^{-1}$. 
This is invertible in $PL^2(\R^2)$ for small $\lam>0$ and 
$d=(g_0(\lam)+ c_1)^{-1}P$. 
Then, \reflm(FS) implies that $(T_0+g_0(\lam)P)^{-1}$ exists 
and is given by \refeq(t0-inv).
\edpf 

Let $\tD_{0}(\lam)= (g_0(\lam)P+T_0)^{-1}=h(\lam) L_0 + D_{0}$. 
\bglm \lblm(reg-M) 
For small $\lam>0$,  $\Mg(\lam^4)$ is invertible in $\HL$ and 
\bqn \lbeq(M-reg-case)
\Mg(\lam^4)^{-1}= \tD_{0}(\lam)- \lam^2 \tD_{0}(\lam)G_2^{(v)} \tD_{0}(\lam) + 
\Og_{\Hg_2}^{(4)}(\lam^4 \log\lam).
\eqn  
\edlm 
\bgpf We have $M^{(v)}_2(\lam)\tD_{0}(\lam)\in \Og^{(4)}_{\Hg_2}(\lam^{2})$ by 
\reflm(first-est) and  
$\Mg(\lam^4)= (1+ M_2^{(v)}(\lam)\tD_{0}(\lam))(g_0(\lam)P+T_0)$ 
by \reflm(t0-inv). It follows that $\Mg(\lam^4)$ is invertible 
for small $\lam>0$ and
\begin{align}
\Mg(\lam^4)^{-1} & = 
\tD_{0}(\lam)- \tD_{0}(\lam) M^{(v)}_2(\lam)\tD_{0}(\lam) \notag  \\
& + \tD_{0}(\lam)(M^{(v)}_2(\lam)\tD_{0}(\lam))^2
(1+ M^{(v)}_2(\lam)\tD_{0}(\lam))^{-1} \lbeq(oturi)
\end{align} 
where the last term on the right belongs to $\Og_{\Hg_2}^{(4)}(\lam^4)$. 
On substituting  $M^{(v)}_2(\lam)= \lam^2 G^{(v)}_2+ M^{(v)}_4(\lam)$, 
we obtain  \refeq(M-reg-case).    
\edpf 

Following Schlag \cite{Schlag}, we say operator $K$ 
is absolutely bounded (ABB for short) 
if $|K(x,y)|$ defines a bounded operator in $\HL$.

\bglm \lblm(abs-bound) 
{\rm (1)} The operator $D_{0}$ is ABB. 

\noindent 
{\rm (2)} If $K$ is ABB and $v,w\in L^2(\R^4)$, then  
$v(x)K(x,y)w(y)\in \Lg^1$.
\edlm  
\bgpf (1) is proved by Green-Toprak (\cite{GT-1}). 
(2) is evident by the Schwarz inequality. 
\edpf 

\paragraph{\bf Proof of \refthb(main-rephrase) when $H$ is regular at zero} 
We show $\Qg_v(\lam)$ is GPR for small $\lam>0$. 
Multiply \refeq(M-reg-case) by $M_v$ from both sides. 
Then, $M_v \tD_{0}(\lam)M_v$ is $\Gg\Vg\Sg$ since 
$L_0$ is of rank 2,   
$M_v D_0 M_v\in \Lg^1$ by \reflm(abs-bound) and $h(\lam)$ is GMU; 
$\lam^2 M_v \tD_{0}(\lam)G_2^{(v)}\tD_{0}(\lam)M_v$ 
is also $\Gg\Vg\Sg$ since $G_2^{(v)}\in \Hg_2$; we evidently have 
$M_v \Og_{\Hg_2}^{(4)}(\lam^4\log\lam)M_v 
= \Rg_{em}\in \Og_{\Lg^1 }^{(4)}(\lam^4\log\lam)$. 
Thus, $\Qg_v(\lam)$ is GPR.  
\qed

\section{Low energy estimate 2. Resonances.} \lbsec(resonance)

In this section we prove \reflm(reso-intro) and \reflm(resonance). 
We assume in this section that 
$\la \log |x| \ra^2 V \in (L^1 \cap L^q)(\R^4)$ for a $q>1$ unless 
otherwise stated. We begin with the following lemma. 
Recall that  $S_1$ is the projection in $Q\HL$ to 
$\Ker_{Q\HL} QT_0Q$.  We shall often write 
$\Ng_\infty$ for $\Ng_\infty(H)$. 

\bglm \lblm(finite-dim)
The projection $S_1$ is of finite rank. The operator 
$QT_0 Q +S_1$ is invertible in $Q\HL$ and we denote 
$D_{0}=(QT_0Q +S_1)^{-1}$. 
\edlm 
\bgpf The operator $QT_0 Q= QUQ + QN_0^{(v)}Q$ is selfadjoint in the 
Hilbert space $Q\HL$. Since ${\bf 1}= U^2$, we have by comparing 
\[
{\bf 1}= \begin{pmatrix} Q & 0 \\ 0 & P \end{pmatrix}, \quad 
U^2 = 
\begin{pmatrix} QUQ & QUP \\ PUQ & PUP \end{pmatrix}^2 
\]
that $(QUQ)^2 = Q-QUPUQ$. Since ${\rm rank}\, QUPUQ =1$, 
Weyl's theorem implies $\s_{\rm ess}((QUQ)^2)= \s_{\rm ess}(Q)=\{1\}$ 
on $Q\HL$ and $\s_{\rm ess}(QUQ) \subset \{1,-1\}$. 
The operator $N_0^{(v)}$ is compact in $\HL$ by \reflm(members) 
and hence so 
is $QN_0^{(v)}Q$ in $Q\HL$. Thus, 
$\s_{\rm ess}(QT_0 Q)\vert_{QL^2} \subset \{1,-1\}$ 
by Weyl's theorem once more and 
$0$ is an isolated eigenvalue of $QT_0 Q\vert_{QL^2}$ of finite multiplicity. 
The rest of the lemma follows by the Riesz-Schauder theorem 
\cite{Yosida}. 
\edpf 

\paragraph{\bf Proof of \reflmb(resonance) (1)} 
Let $\z \in S_1 \HL \setminus\{0\}$. 
Then, $Q\z=\z$, $QT_0\z= 0$ and  
$T_0 \z= c_0 v$ for a constant $c_0$  or $(U+ N_0^{(v)})\z= c_0 v$.
The constant $c_0$ is given by 
\bqn \lbeq(c-0)
c_0 =\|v\|_2^{-2}(T_0\z, v)\,.
\eqn 
Thus, if we define 
$\ph=\Phi(\z)$ by \refeq(inverse) or 
\bqn \lbeq(z-ph)
\ph =-c_0 + N_0 M_v \z=
- c_0 - \frac{1}{8\pi^2} \int_{\R^4}\log |x-y| v(y) \z(y) dy, 
\eqn 
then, $U\z+ v\ph=0$, $\z= - w\ph$ and  
$\lap^2 \ph = v\z = -V\ph$ or $(\lap^2 + V)\ph=0$. 

We next  show that $\ph \in L^\infty(\R^4)$, which show that 
$\Phi$ maps $S_1 L^2$ to $\Ng_\infty$ with 
the inverse $\z=-w\ph$ on its image, 
in particular $\Ng_\infty\not=\{0\}$.  \\[3pt]
(i) Let first $|x|\leq 10$. Let $p=2q/(q-1)$ we have 
\[
\int_{|y|\leq 20} |\log |x-y| v(y) \z(y)| dy 
\leq 
\|\log|y|\|_{L^p(|y|\leq 30)}\|v\|_{L^{2q}(|y|\leq 20)}\|\z\|_{L^2(\R^4)}; 
\]
if $|y|>20$, we have 
$\log |x-y|\leq \log (2 |y|) \leq 2 \log |y|$  and 
\[
\int_{|y|>20}\log |x-y| |v(y) \z(y)| dy \leq 2
\|(\log |y|) v\|_{L^2(|y|>20)}) \|\z\|_2. 
\]
Thus, $|N_0 (v\z)(x)|\leq C$ for $|x|\leq 10$. \\[3pt]
(ii) Let next $|x|>10$. Since $P\z=0$ or $\int_{\R^4} v(y)\z(y) dy=0$, 
we have  
\bqn \lbeq(log-)
(N_0v\z)(x) = -\frac1{8\pi^2}
\int_{\R^4}(\log |x-y| -\log |x|)v(y) \z(y) dy.
\eqn 
If $|y|>2|x|$, then  $|x| < |x-y| < |x||y|$. Hence,   
$0<\log |x-y| -\log |x| <\log |y|$ and 
\[
\int_{|y|>2|x|} (\log |x-y| -\log |x|)|v(y) \z(y)| dy \\
\leq  \|\la \log |y|\ra v\|_2 \|\z\|_2 
\]
If $|y|<|x|/2$, then 
$|\log |x-y| -\log |x|| \leq |y||x-\th{y}|^{-1}\leq 1$ for a $0<\th<1$ and 
\[
\int_{|y|<|x|/2} |(\log |x-y| -\log |x|)v(y) \z(y)| dy 
\leq \int_{\R^4} |(v\z)(y)| dy \leq \|v\|_2 \|\z\|_2.
\]
Finally let $|x|/2<|y|<2|x|$. Then, $|\log |x||\leq \log 2|y|$ and 
\[
\int_{|x|/2<|y|<2|x|} |(\log |x|)v(y) \z(y)| dy  
\leq 2 \|\la \log |y|\ra v\|_2 \|\z\|. 
\]
Let $p=2q/(q-1)$ again. Then
\[
\int_{|x-y|<2} |(\log |x-y|)v(y) \z(y)| dy \leq 
\|\log|y|\|_{L^p(|y|\leq 2)}\|v\|_{2q} \|\z\|_2 . 
\]
If $|x-y|>2$ and $|x|/2<|y|<2|x|$, then 
$\log |x-y|\leq \log |x|+ \log |y| \leq 2 \la \log |y|\ra$ and 
\[
\int_{|x-y|>2,|x|/2<|y|<2|x|} |(\log |x-y|)v(y) \z(y)| dy 
\leq 2\|\la \log |y|\ra v\|_2 \|\z\|_2. 
\]
Thus, $|N_0 v\z(x)|\leq C$ also for $|x|\geq 10$. 

To finish the proof we show that $\z=-w\ph \in S_1 L^2 \setminus\{0\}$ 
if $\ph \in \Ng_\infty\setminus \{0\}$. Then, 
$\lap^2(\ph+ N_0 V\ph )=(\lap^2 + V)\ph=0$, 
$|\xi|^4 \Fg(\ph+ N_0 V\ph)(\xi)=0$ and 
$\Fg(\ph+ N_0 V\ph)\in \Sg'(\R^4)$ vanishes outside $\{0\}$. Hence  
\[
\Fg(\ph+ N_0 V\ph)(\xi)= \sum_{{\rm finite}} c_\a D^\a \delta (\xi) 
\]
for constants $c_\a$, or $(\ph+ N_0 V\ph)(x)$ is a polynomial. 
But, $\ph \in L^\infty$ and $\la \log|x| \ra^2 V \in (L^1\cap L^q)(\R^4)$ imply 
that 
\[
(N_0 V\ph)(x)= -\frac1{8\pi^2}\left(\int_{|x-y|<2} + \int_{|x-y|\geq 2}\right)
\log |x-y| V(y) \ph(y)dy  
\]
is bounded by $C(1+\log\ax)$ for a constant $C>0$. Hence it must be that  
$\ph+ N_0 V\ph=c$ for a constant $c$ and $N_0 V\ph(x)\in L^\infty$. 
It follows that $\int V \ph dx =0$ 
because otherwise $|N_0 V\ph(x)|\geq C |\log |x||$ for large $|x|$ 
for a $C>0$. Hence $P(w\ph)=0$ or $w\ph=Q(w\ph)$,  
\[
cv=(v+ vN_0V)\ph= (U + N_0^{(v)})(w\ph) = T_0 Q(w\ph)
\]
and $QT_0 Q(w\ph) = 0$. Moreover  
$w\ph\not=0$ because $w\ph=0$ would implies $N_0 V\ph=0$, 
$0\not= \ph=c$, $w=0$ and $V=0$, 
which is a contradiction. This proves statement (1) of the lemma. \qed \\[3pt]

\paragraph{\bf Proof of \reflmb(reso-intro)} 
Here we assume that $\la \log |x| \ra^2 \ax^3 V \in 
(L^1 \cap L^q)(\R^4)$ for a $q>1$. Let $q'=q/(q-1)$. 
Let $\ph \in \Ng_\infty(H)$ and $|x|\geq 10^{10}$. 

Since $\ph=\Phi(\z)$ is the isomorphism from $S_1$ to $\Ng_\infty(H)$ 
with the inverse $\z=-w\ph$, we have from \refeq(z-ph) and 
\refeq(log-) that  
\bqn \lbeq(ph-log)
\ph(x)= -c_0 + \frac1{8\pi^2}\int_{\R^4} (\log |x-y|-\log|x|) V(y) \ph (y) dy,
\eqn 
where $c_0$ is given by \refeq(c-0). Let 
\begin{gather*}
\D_1= \{y \colon |y|>|x|/4, |x-y|<10\}, \\  
\D_2= \{y \colon |y|>|x|/4, |x-y|\geq 10\}, \quad 
\D_3= \{y \colon |y|\leq |x|/4\},
\end{gather*}
and split the integral on the right of \refeq(ph-log) as 
\[
\left(\int_{\D_1}+\int_{\D_2} +\int_{\D_3}\right)
(\log |x-y|-\log|x|) V(y) \ph (y) dy 
= \sum_{j=1}^3 I_j. 
\]
(1) For $y \in \D_1$, we have $\log |x|\leq \log 4|y|$ and 
$C^{-1} \ax \leq \ay \leq C\ax$ for a large $C>1$. 
Then, H\"older's inequality implies 
\begin{align*}
|I_1(x)|& \leq (|\log|y|\|_{L^p(|y|<10)} \|V(y)\|_{L^q(\D_1)}
+ \|\la \log |y|\ra V(y)\|_{L^1(\D_1)}) \|\ph\|_\infty \\
& \leq C\ax^{-3}(\|\ay^{3}V\|_{L^q}+
\|\ay^{3}\la \log|y|\ra V\|_{L^1})\|\ph\|_\infty \leq C_1 \ax^{-3}. 
\end{align*}
(2) For $y \in \D_2$, we have $|x-y|>10$, $\log |x-y| \leq \log (|x|+|y|)
\leq 2\log |y|$ and 
\[
|I_2(x)|\leq C\ax^{-3}\|\ay^{3}\la \log |y| \ra V\|_1 \|\ph\|_\infty
\leq C_2 \ax^{-3}\,.
\]
Thus, $I_1(x)$ and $I_2(x)$ may be put into the remainder $O(|x|^{-3})$ 
of \refeq(reso-intro).

\noindent 
(3) For $y \in \D_3$, $|x-y|\geq 3|x|/4 >10^9$. 
Let $f(\th)= \log |x-\th{y}|-\log |x|$ for $0\leq \th \leq 1$. Then, 
Taylor's formula implies 
\begin{align} \lbeq(Taylor)
& f(1) = f(0)+ f'(0) + \frac12 f''(0) + 
\int_0^1 \frac{(1-\th)^2}{2}f'''(\th) d\th\, ,   \\
& f'(0)= - \sum_{j=1}^4 \frac{x_j y_j}{|x|^2},  \quad 
f''(0)= \frac{|y|^2}{|x|^2}
- \sum_{j,k=1}^4 \frac{2x_j x_k y_j y_k}{|x|^4}\,, \lbeq(Tay-1)   \\
& f'''(\th) =
- \frac{6(y \cdot (\th y -x))|y|^2}{|x-\th y|^4}
+ \frac{8(y \cdot (\th y -x))^3}{|x-\th y|^6}. \lbeq(Tay-3)
\end{align}
We substitute \refeq(Taylor) for $\log |x-y|-\log |x|$ in the 
integral for $I_3(x)$. 
Since $|x-\th{y}|\geq 3|x|/4$ for $|y|\leq |x|/4$ and $0<\th<1$,
\[
R_3 (x,y) \colon= \int_0^1 \frac{(1-\th)^2}{2}f'''(\th) d\th 
\absleq (7/3)(|y|/|x|)^3. 
\]
and the contribution of $R_3(x,y)$ to $I_3(x)$ is bounded in modulus by 
\[
C\ax^{-3} \int_{\D_3} |y|^3 |V(y) \ph (y)| dy\leq C \ax^{-3}\|\ay^3 V\|_1\,.
\]
Since $|f'(0)|\leq (|y|/|x|)$, $|f''(0)|\leq C(|y|/|x|)^2$ and 
$|x|/4\leq |y|$ in $\R^4 \setminus \D_3$,  
\[
\int_{\R^4\setminus\D_3} (f'(0)+ \frac12f''(0))V(y) \ph(y) dy \\
\absleq C \ax^{-3}\|\ay^3 V\|_1\,.
\]
(4) Combining estimates in (1), (2) and (3), we obtain  
\[
\ph(x)= \int_{\R^3}(f'(0)+ \frac12f''(0))V(y) \ph(y) dy + \Og(\ax^{-3}).
\]
Restoring $V(y)\ph(y)= - v(y)\z(y)$, we obtain  
\begin{gather}
\ph(x)= - c_0 
- \sum_{j=1}^4 \frac{a_j x_j}{|x|^2} 
+ \sum_{j,k=1}^4 \frac{(2a_{jk}-b\d_{jk}) x_j x_k}{|x|^4} 
+ O(|x|^{-3}), \lbeq(phi-exp)  \\
a_j=\frac1{8\pi^2}\int_{\R^4} y_j v(y) \z(y) dy, \quad 
b= \frac1{8\pi^2}\int_{\R^4} |y|^2 v(y) \z(y) dy ,\lbeq(c-a) \\ 
a_{jk}=\frac1{8\pi^2}\int_{\R^4} y_j y_k v(y) \z(y) dy, \lbeq(Ajk)
\end{gather}
where $\d_{jk}$ is the Kronecker delta. This completes the proof. 
\qed 

\bglm \lblm(characterization)
\ben
\item[{\rm(1)}] Let $\z \in S_1 \HL$. Then,   
$\z \in S_2 \HL$ iff $T_0\z=0$. 
\item[{\rm(2)}] Let $\z \in S_2 \HL$. Then, 
$\z \in S_3 \HL$ iff $(x^\a v,\z)=0$ 
for $|\a|\leq 1$. 
\item[{\rm(3)}] Let $\z \in S_3 \HL$. Then, 
$\z \in S_4 \HL$ iff $(x^\a v, \z)=0$ 
for  $|\a|\leq 2$. 
\een
\edlm
\bgpf (1) If $\z \in S_1\HL$ and $T_0\z=0$, then 
$T_1 \z=S_1T_0 P T_0 S_1\z=0 $ and $\z\in S_2\HL$. 
Conversely, if $\z\in S_2\HL$, then $PT_0 S_1 \z = 0$ since 
$\Ker T_1= \Ker PT_0 S_1$, But  $\z \in S_1\HL$ implies 
$PT_0\z = 0$ and simultaneously $QT_0 Q\z= QT_0\z=0$. Hence $T_0\z=0$.  

\noindent 
(2)  If  $\z \in S_3 \HL$, then $(v, \z)=0$ and   
\bqn \lbeq(T2-d)
0=(T_2\z, \z)= \frac{2i}{4^4\pi} 
\sum_{j=1}^4 \left|\int_{\R^4}x_j v(x) \z(x)dx \right|^2 . 
\eqn 
It follows that $\la x_j v,  \z\ra=0$ for $1\leq j \leq 4$. 
Conversely if $\la x^\a v,  \z\ra=0$ for all $|\a|\leq 1$, then   
$T_2\z(x)$ is equal to $-i(4^4\pi)^{-1}$ times  
\[
S_2 M_v \int_{\R^4} |x-y|^2 v(y) \z(y) dy 
= (S_2 v)(x) \int_{\R^4} y^2 v(y) \z(y) dy.  
\]
But $S_2 v =0$ and, hence, $T_2\z(x)=0$. Since $\z\in S_2\HL$, $\z \in S_3\HL$.

\noindent 
(3) Let $\z \in S_4\HL \subset S_3 \HL$. Then, 
$(x^\a v,\z)=0$ for $|\a|\leq 1$ by (2) and  
\begin{multline*} 
0= (T_3 \z, \z)
= \frac{2}{3\cdot 4^3} \int_{\R^4}(|x|^2|y|^2 + 2(x\cdot y)^2)v(x)v(y) \z(x)
\overline{\z(y)} dx dy \\
= \frac{2}{3\cdot 4^3} \left|\int_{\R^4}|x|^2v(x)\z(x) dx \right|^2 
+ \frac{1}{3\cdot 4^2}\sum_{j,k=1}^4 
\left|\int_{\R^4}x_j x_k v(x) \z(x) dx \right|^2.
\end{multline*}
It follows that 
$(x^\a v, \z)=0$ also for $|\a|=2$. 
Conversely, if $\z\in S_2\HL$ satisfies 
$( x^\a v, \z )=0$ for $|\a|\leq 2$, then (2) implies $\z \in S_3\HL$ and 
\[
T_3 \z(x)= S_3 \left(v(x)\int_{\R^4} 
\Big(|y|^4 -4 \sum_{j=1}^4 x_j  y_j \cdot |y|^2 \Big) v(y) \z(y)dy\right). 
\]
Since $S_3 v=0$ and $S_3(x_j v)=0$ for $j=1, \dots, 4$ by (2), 
$T_3\z=0$. Hence, $\z\in S_4\HL$. This completes the proof. 
\edpf  

\paragraph{\bf Proof of \reflmb(resonance) (2) and (3).} 
For $\z \in S_1 L^2$, let $\ph =\Phi(\z)\in \Ng_\infty$ and 
$c_0, \ab$ and $A$ be the coefficients 
of the expansion \refeq(reso-intro) for $\ph(x)$. 

\noindent 
(2) Since $P$ is one-dimensional, $\rank T_1\leq 1$. It follows that 
$T_1\vert_{S_1L^2}$ is invertible, then $\rank S_1=1$. We have  
$\Ker T_1= \Ker (PT_0 S_1)^\ast P T_0 S_1= \Ker PT_0S_1=\{0\}$.  
Thus, $c_0 =\|v\|_2^{-2}\la T_0\z, v\ra\not=0$ for all 
$\z\in S_1 L^2\setminus \{0\}$ 
and $H$ has only $s$-wave resonances.

\noindent 
(3) (i) If $\z \in S_1L^2 \ominus S_2 L^2$, 
$c_0 = \|v\|_2^{-2}\la PT_0\z, v\ra\not=0$ and $\ph$ 
is $s$-wave resonance. 

\noindent 
(ii) Remark that $i T_2 $ is selfadjoint in $\HL$. 
Let $\z\in S_2 L^2\ominus S_3L^2$. 
Then $\z\in S_2 L^2$ implies $T_0\z=0$ by \reflm(characterization) and $c_0=0$. 
We have 
\begin{multline} \lbeq(T2)
i \la T_2 \z, \z \ra= \frac1{4^4\pi}
\int_{\R^4\times \R^4} |x-y|^2 (v\z)(x)\overline{(v\z)(y)}dx dy \\
= \frac{-1}{2^7\pi} \sum_{j=1}^4 \left|\int_{\R^4}x_j v(x)\z(x)dx\right|^2 
\leq 0
\end{multline}
and $i T_2\leq 0$ on $S_2L^2$. It follows that  
$\la T_2 \z, \z \ra=0$ implies $\z \in \Ker T_2\vert_{S_2 \HL}$ 
and $\z\in S_3\HL$. Hence,  
$i \la T_2 \z, \z\ra <0$ for non-trivial $\z \in S_2 L^2\ominus S_3L^2$,
which implies $\ab\not=0$ in \refeq(reso-intro) by \refeq(c-a)  
and $\ph$ is $p$-wave resonance. 

\noindent 
(iii) 
Suppose next $\z\in S_3 L^2\ominus S_4L^2$. 
Then $c_0=0$ as previously and $\z\in S_3 L^2$ implies 
$\ab=0$ by virtue of \reflm(characterization). 
For $\z \in S_3L^2$ we have 
\[
( T_3 \z, \z)= \frac1{48}\sum_{j,k=1}^4 
\left|\int_{\R^4}x_j x_k (v\z)(x)dx \right|^2 + 
\frac1{96} \left|\int_{\R^4}x^2 (v\z)(x)dx \right|^2  
\]
as previously and $T_3\geq 0$ on $S_3\HL$. It follows 
$\la T_3 \z, \z \ra >0$ for non-trivial $\z\in S_3 L^2\ominus S_4L^2$. 
Suppose $A=0$. Then  
$a_{jk}=0$ for $j\not=k$ and $2a_{jj}-b=0$ for $j=1, \dots, 4$ in 
\refeqss(phi-exp,c-a,Ajk). But $\sum_{j=1}^4 a_{jj}=b$ by \refeq(Ajk) and 
$0=\sum_{j=1}^4 (2a_{jj}-b)=-2b$. Hence $a_{jk}=0$ for all $1\leq j,k\leq 4$  
which contradicts to $\la T_3 \z, \z\ra>0$. Thus $A\not=0$ and 
$\ph$ is $d$-wave resonance. 

\noindent 
(iv) Finally let $\z\in S_4 L^2 \setminus\{0\}$. Then, we already have shown 
that $c_0=0$ and $\ab=0$.
Moreover $\la T_3\z, \z\ra=0$ and \refeq(Ajk) implies $A=0$. 
Thus, $\ph$ must be zero energy eigenfunction of $H$. 
\qed

\section{Singularity of the first kind.} \lbsec(first-kind)

In this section we prove $W_{-}\chi_{\leq{a}}(|D|)$ is GOP for 
sufficiently small $a>0$ when $H$ has singularity of the first 
kind at zero. 
We assume $\la \log |x|\ra^2 \ax^8 V \in (L^1\cap L^q)(\R^4)$ for 
a $q>1$ unless otherwise stated explicily. 
We shall in what follows repeatedly and inductively use the following 
lemma due to Jensen and Nenciu \cite{JN}.  

\begin{lemma}[\cite{JN}] \lblm(JN) 
Let $A$ be a closed operator in a Hilbert space $\Hg$ 
and $S$ a projection. Suppose $A+S$ has bounded inverse. 
Then, $A$ has bounded inverse if and only if 
\bqn \lbeq(JN-B)
B= S - S(A+S)^{-1}S 
\eqn 
has bounded inverse in $S\Hg$ and, in this case, 
\bqn \lbeq(JN-A)
A^{-1}= (A+S)^{-1}+ (A+S)^{-1}SB^{-1}S (A+S)^{-1}.  
\eqn 
\end{lemma}

\subsection{Threshold analysis 1}
We begin with the lemma which generally holds when $H$ is singular at zero. 
Let $\{\z_1, \dots, \z_n\}$ be the orthonormal basis of $S_1\HL$ so that  
\bqn \lbeq(extofS1) 
S_1 u = (\z_1 \otimes \z_1 + \cdots+ \z_n \otimes \z_n)u, \quad u \in Q\HL. 
\eqn 
We use the same letter $S_1$ to denote the extension of $S_1$ to 
$L^2(\R^4)$. Let 
\bqn 
T_{4,l}(\lam)=\tg_2(\lam) G^{(v)}_{4} + G^{(v)}_{4,l}\,. \lbeq(T4-l)
\eqn 

\bglm \lblm(MS1-good) 
Suppose $H$ is singular at zero. Then: \\[3pt]
(1) $\Mg(\lam^4)+S_1$ is 
invertible in $\HL$ for small $\lam>0$,  
\begin{gather}  
(\Mg(\lam^4)+S_1)^{-1}= \tD_{0}(\lam) - \lam^2
\tD_{0}(\lam)G_2^{(v)}\tD_{0}(\lam) + Y_1(\lam) , \lbeq(MS1) \\
Y_1(\lam)=\Rg_{em}\in \Og_{\Hg_2}^{(4)}(\lam^4 \log\lam) \lbeq(Y1)
\end{gather}
and $M_v (\Mg(\lam^4)+S_1)^{-1}M_v=\Gg\Vg\Sg + \Rg_{em}(\lam)$ 
is GPR for small $\lam>0$. \\[3pt]
(2) If $V$ satisfies $\la \log |x|\ra^2 \ax^{12} V \in (L^1\cap L^q)(\R^4)$ for a 
$q>1$, then 
\bqn \lbeq(Y1-remark)
Y_1(\lam)= -\lam^4 \tD_0(\lam) 
\{T_{4,l}(\lam)\tD_0(\lam)- (G_2^{(v)}\tD_0(\lam))^2 \} + 
\Og^{(6)}_{\Hg_2}(\lam^6\log\lam).
\eqn 
\edlm 
\bgpf (1) We have 
\bqn 
\Mg(\lam^4)+S_1 = g_0(\lam)P + T_0 +S_1 + M^{(v)}_2(\lam). \lbeq(s-first-1) 
\eqn 
Repeating the proof of \reflm(reg-M) by using Feshbach formula 
with $QT_0 Q+ S_1$ replacing $QT_0 Q$, we see that 
$g_0(\lam)P + T_0 +S_1$ is invertible for small $\lam>0$ in $\HL$ and  
\bqn 
(g_0(\lam)P + T_0 +S_1)^{-1}= h_1 (\lam)L_0 + D_{0} = \colon \tD_{0}(\lam)
\lbeq(gPT0-inv-a) 
\eqn 
where $D_{0}=Q(QT_0Q +S_1)^{-1}Q$, 
$h_1(\lam)= (g_0(\lam)+ c_1)^{-1}$, $c_1$ being the same constant as in 
\reflm(t0-inv) with new $D_{0}$ and
\bqn
L_0  = \begin{pmatrix} P & -PT_0Q D_{0} \\
-D_{0} QT_0P  & D_{0} QT_0PT_0Q D_{0}
\end{pmatrix}\,
\lbeq(gPT0-inv-b) 
\eqn  
in the decomposition $\HL= P \HL \oplus Q\HL$. 
(Here we used the same notation as in \reflm(t0-inv) since they become 
the same when $S_1=0$ and the notation of \reflm(t0-inv) will not be used 
in what follows. This is the convention used  in \cite{GT-1}.) 
Then, as in the proof of \reflm(reg-M),     
\bqn \lbeq(MS1a)
(\Mg(\lam^4)+S_1)^{-1} 
= \tD_{0}(\lam)(1+ M^{(v)}_2(\lam) \tD_{0}(\lam))^{-1}  
\eqn 
and by expanding the right side 
and applying \refeq(remainder-est) for $n=1$ and $n=2$ to the remainder 
we obtain \refeq(MS1).  

\noindent 
(2) Expand \refeq(MS1a) by using  
$(1+X)= 1- X + X^2- X^3(1+X)^{-1}$ with $X=M^{(v)}_2(\lam) \tD_{0}(\lam)$ 
and estimate the remainder by using \refeq(remainder-est) 
for $n=1,2$ and $n=3$. We obtain \refeq(Y1-remark).
\edpf 

Let $B_1(\lam)= S_1 - S_1 (\Mg(\lam^4)+ S_1)^{-1} S_1$. 
Since $S_1 \tD_{0}(\lam) S_1= h_1(\lam)T_1+ S_1$, \refeq(MS1) implies 
\bqn 
B_1(\lam)= -h_1(\lam)T_1 + 
\lam^2 S_1 \tD_{0}(\lam) G_2^{(v)}\tD_{0}(\lam)S_1 - 
S_1 Y_1(\lam)S_1.  \lbeq(B-1) 
\eqn
Since $T_1$ is invertible, $B_1(\lam)$ is also invertible in $S_1\HL$ 
and  
\reflm(JN) for the pair $(\Mg(\lam^4), S_1)$ implies 
\begin{gather} \lbeq(Mlam4inv)
\Mg(\lam^4)^{-1}= (\Mg(\lam^4)+S_1)^{-1}+ M_{ess}^{(1)}(\lam) , \\
M_{ess}^{(1)}(\lam)= 
(\Mg(\lam^4)+S_1)^{-1}S_1 B_1(\lam)^{-1}S_1 (\Mg(\lam^4)+S_1)^{-1}. 
\lbeq(M1ess-def)
\end{gather} 
In what follows $A(\lam)\equiv B(\lam)$ means 
$A(\lam)- B(\lam) = \Gg\Vg\Sg + \Rg_{em}(\lam)$ unless otherwise 
stated explicitly.

\bglm \lblm(s-mod) Suppose that $H$ has singularity of the first 
kind at zero. Then, $\rank S_1=1$ and 
\bqn \lbeq(main-sl)
\Qg_v(\lam) \equiv (a\log \lam + b)(v\z) \otimes (v\z)\,,
\eqn 
where $a\in \R \setminus\{0\}$ and $b\in \C$ are constants 
and $\z$ is the normalized basis vector of $S_1\HL$.    
\edlm
\bgpf We have proved that $\rank S_1=1$ in \reflm(resonance) (2). 
Then, $T_1 = d_0^{-1} (\z \otimes \z)$ with $d_0= c_0^{-2} \|V\|_1^{-1}>0$  
and $\lam^2 S_1 \tD_{0}(\lam) G_2^{(v)}\tD_{0}(\lam)S_1 - 
S_1 Y_1(\lam)S_1= \Og_{\C}^{(4)}(\lam^2\log\lam)(\z\otimes \z)$.
Hence, from \refeq(B-1) we have      
\bqn \lbeq(B1-inv)
B_1(\lam)^{-1}= d(\lam)(\z \otimes \z),\  
d(\lam)= - d_0 h_1(\lam)^{-1}(1 + \Og_{\C}^{(4)}(\lam^2(\log\lam)^2))
\eqn 
and $h_1(\lam)^{-1}=g_0(\lam)+c_1=\Og_{\C}(\log\lam)$. 

By \reflm(MS1-good)  
$M_v(\Mg(\lam^4)+S_1)^{-1}M_v={\Gg\Vg\Sg}+ \Rg_{em}(\lam)$. 
Let $F_0(\lam)=\tD_{0}(\lam) - \lam^2 \tD_{0}(\lam)G_2^{(v)} \tD_{0}(\lam)$. 
Then, $F_0(\lam)$ is $\Gg\Vg\Sg$ and 
$(\Mg(\lam^4)+S_1)^{-1}= F_0(\lam) + \Rg_{em}(\lam)$ by \refeq(MS1). 
Then, \refeq(M1ess-def) and \refeq(B1-inv) imply that 
\bqn \lbeq(Mess-e)
M_{ess}^{(1)}(\lam)= d(\lam) (F_0(\lam) + \Rg_{em}(\lam))(\z \otimes \z)
(F_0(\lam) + \Rg_{em}(\lam)).
\eqn
Expanding the right of \refeq(Mess-e) and using 
$d(\lam)= - d_0 h_1(\lam)^{-1}+ {\rm GMU}$ and $D_0 \z=D_0 S_1 \z= \z$, 
we obtain  
\begin{align}
M_{ess}^{(1)}(\lam) & = d(\lam)F_0(\lam)(\z\otimes \z)F_0(\lam) + \Rg_{em}(\lam)
\notag \\
& = -d_0 h_1(\lam)^{-1}(\z\otimes \z) + \Gg\Vg\Sg.  \lbeq(M1ess-2)
\end{align} 
We sandwich \refeq(M1ess-2) by $M_v$ and obtain \refeq(main-sl) .
\edpf

\paragraph{\bf Proof of \refthb(main-rephrase) 
when $H$ has singularity of first kind}  
By virtue of \reflm(s-mod), $W_{-}\chi_{\leq {a}}(|D|)u(x)$ 
is equal to \refeq(33) modulo GOP. We have 
$v\z\in \ax^{-4}\la \log |x|\ra^{-1}L^1(\R^4)$ by the assumption 
and $\int_{\R^4} v(x)\z(x) dx=0$. Thus, the following lemma 
implies \refth(main-rephrase) for this case. 
The lemma is more than necessary for this purpose and it 
is stated in this fashion for the later purpose. 

\bglm \lblm(s-low)
Suppose that $f\in L^1(\R^4)$, $\ax g \in L^1(\R^4)$ and 
\bqn \lbeq(S1-cancel) 
\int_{\R^4} g(x) dx=0. 
\eqn 
Then, the operators $\tW_k$ defined for $k=0,1,2, \dots$ by 
\bqn \lbeq(sta0-s)
\tW_k{u}(x) = 
\int_0^\infty \big(R^{+}_0 (\lam^4)(f\otimes g) \Pi(\lam)u \big)(x)
\lam^3 (\log\lam)^k \chi_{\leq{a}}(\lam)d\lam 
\eqn  
for $u \in \Dg_\ast$ are GOP. %
\edlm 
\bgpf Let $\m_k(\lam) = \lam (\log\lam)^k \chi_{\leq {a}}(\lam)$ 
for $k=0,1, \dots$; $\m_k$ are GMU. By virtue of \refeq(S1-cancel)
\bqn \lbeq(after-cancel)
\la g, \Pi(\lam)u \ra 
= \int_{\R^4} g(z)(\Pi(\lam)u(z) - \Pi(\lam)u(0))dz
\eqn 
and $(\Pi(\lam)u(z) - \Pi(\lam)u(0))$ may be expressed as in \refeq(reason). 
Then, as in \refeq(intro-35), $\tW_k{u}(x)$ becomes 
the $\sum_{j=1}^4 \int_0^1 d\th$ of 
\bqn \lbeq(7-16)
i \int_{0}^\infty \left(\int_{\R^4} (R_0^{+}(\lam^4)f)(x) 
z_j g(z)\Pi(\lam)R_ju(\th{z}) dz \right) 
\lam^3 \m_k(\lam)d\lam ,
\eqn 
where $\la \cdot, \cdot \ra_z$ means the coupling of functions of $z\in \R^4$.
Let 
\[
T_j(y,z)= i z_j f(y)g(z) \in L^1(\R^4\times \R^4) 
\]
and, by using \refeq(spect-proj) and \refeq(mult) express as 
\[
\mu_k (\lam)\Pi(\lam)R_ju(\th{z})
=\Pi(\lam)(\tau_{-{\th}z}R_j\mu_k(|D|)u)(0).
\] 
Then, \refeq(7-16) is equal by \refeq(T-K) to 
\begin{align*}
& \int_{0}^\infty R_0^{+}(\lam^4) T_j \Pi(\lam)(\tau_{-{\th}z}R_j \m(|D)u)(0)
\lam^3 d\lam \\
&= \int_{\R^8} T_j(y,z) \tau_{y} K \tau_{-\th{z}}R_j \m(|D|)u dy dz 
\end{align*}
and \reflm(T-K) and Minkowski's inequality imply that
\[
\|\refeq(7-16)\|_p \leq C \|T_j\|_{\Lg^1}\|\tau_{-{\th}z}R_j \m(|D)u\|_p 
\leq C \|u\|_p.
\]
This proves that $\tW_k$, $k=0,1, \dots$ are GOP.   
\edpf

\section{Singularity of the second kind} \lbsec(2nd)

In this section we assume 
$\ax^{12}\la \log |x| \ra^2 V \in (L^1\cap L^q)(\R^4)$ 
for a $q>1$ and prove \refth(main-rephrase) (2). Thus, we assume 
$T_1$ is singular in $S_1\HL$ and $T_2= S_2 G_2^{(v)}S_2$ is invertible 
in $S_2\HL$. In what follow we abuse  notation and write 
\bqn 
\Og^{(\ell)}_{S_j\HL}(f(\lam)) = \Og^{(\ell)}_{\Bb(S_j\HL)}(f(\lam)), \quad 
j=1,\dots, 4.
\eqn

\subsection{Threshold analysis 2}  
Recall \refeqss(Mlam4inv,M1ess-def,B-1).  
We apply \reflm(JN) to study $B_1(\lam)^{-1}$ for small $\lam>0$.   
In view of \refeq(B-1) we let    
\begin{gather} \lbeq(tB-1)
\tB_1(\lam) \colon = -h_1(\lam)^{-1} B_1(\lam)= 
T_1 - \lam^2 h_1(\lam)^{-1} \tT_1(\lam) + \tT_4(\lam); \\
\tT_1(\lam)=S_1 \tD_{0}(\lam) G_2^{(v)}\tD_{0}(\lam) S_1 
\in \Og_{S_1\HL}^{(4)}(1) \,, \lbeq(gather-tT1) \\
\tT_4(\lam)= S_1h_1(\lam)^{-1}Y_1(\lam)S_1 
\in \Og_{S_1\HL}^{(4)}(\lam^4 (\log\lam)^2). \lbeq(tQ4)
\end{gather} 

We begin with two lemma which do not assume that $T_2$ is non-singular 
in $S_2\HL$. Before \reflm(second-Mlam) we need only 
$\ax^{8}\la \log |x| \ra^2 V \in (L^1\cap L^q)(\R^4)$.  

\bglm \lblm(identities)
We have the following identities: 
\begin{align}
& S_2 D_{0} = S_2 = D_{0} S_2.  \lbeq(S2D0) \\
& S_2 T_0 =T_0 S_2=0, \quad L_0 S_2= S_2 L_0=0. \lbeq(S2T0)  \\
& S_2 \tD_{0}(\lam)= \tD_{0}(\lam) S_2 = S_2. \lbeq(S2tD1) \\
& S_2 \tT_1(\lam)S_2 = S_2 G_2^{(v)}S_2 = T_2.  \lbeq(S2tTS_2=T_2)
\end{align} 
\edlm 
\bgpf (1) Since $S_1 D_{0} = S_1 = D_{0} S_1$ and $S_2 \subset S_1$, 
we have \refeq(S2D0). 

\noindent 
(2) $0=QT_0 Q S_1=QT_0 S_1$ is equivalent to $T_0 S_1= PT_0 S_1$; 
$\Ker_{S_1L^2} T_1= \Ker_{S_1L^2} PT_0 S_1= \Ker_{S_1L^2} T_0 S_1$. 
Hence $T_0 S_2= T_0 S_1 S_2=0$ and by the duality $S_2 T_0 =0$ . 
This implies the first of \refeq(S2T0). Then,     
\[
S_2 L_0= 
S_2(P -PT_0Q D_{0} Q - QD_{0} QT_0P + D_{0} QT_0PT_0Q D_{0})=0.
\]
We likewise have $L_0 S_2=0$ and the second of \refeq(S2T0) follows. 

\noindent 
(3) Eqns. \refeq(S2D0) and \refeq(S2T0) imply   
$S_2\tD_{0}(\lam)= S_2(h_1(\lam)L_0+ D_{0}) = S_2$ and likewise 
$\tD_{0}(\lam)S_2= S_2$. 
Eqn. \refeq(S2tTS_2=T_2) is obvious from \refeq(S2tD1). 
\edpf  

\bglm \lblm(first-line) For small $\lam>0$, 
$\tB_1(\lam)+S_2$ is invertible in $S_1\HL$ and 
\begin{gather} 
(\tB_1(\lam)+S_2)^{-1}=D_1 + 
D_1 \lam^2 h_1(\lam)^{-1} \tT_1 (\lam)D_1+
\Og^{(4)}_{S_1\HL}(\lam^4(\log\lam)^2)\,, \notag \\
(T_1+ S_2)^{-1}=D_1\,. \lbeq(tB1S_2inv)
\end{gather}
\edlm 
\bgpf  From \refeq(tB-1) 
$\tB_1(\lam)+S_2=  ({\bf 1}_{S_1L^2} - L_1(\lam))(T_1+S_2)$ 
with  
\bqn \lbeq(D1-L1)
L_1(\lam)= \lam^2 h_1(\lam)^{-1} \tT_1 (\lam)D_1 - \tT_4(\lam) D_1 
\in \Og^{(4)}_{\Hg_2}(\lam^2\log\lam).
\eqn  
It follows that $\tB_1(\lam)+S_2$ is invertible in $S_1\HL$ and 
\bqn  
(\tB_1(\lam)+S_2)^{-1}
=D_1 + D_1 L_1(\lam) + D_1 L_1(\lam)^2({\bf 1}_{S_1L^2} - L_1(\lam))^{-1}.
\lbeq(first-line) 
\eqn  
Substituting \refeq(D1-L1) and using also \refeq(tQ4),  
we obtain \refeq(tB1S_2inv). 
\edpf

\bglm \lblm(B2)
Let $B_2(\lam)= S_2 - S_2(\tB_1(\lam)+S_2)^{-1}S_2$. Then:
\bqn 
B_2(\lam) 
=- \lam^2 h_1(\lam)^{-1} (T_2 + \Og^{(4)}_{S_2\HL}(\lam^2 \log\lam)), 
\lbeq(B2-def) 
\eqn 
$B_2(\lam)$ is invertible in $S_2 \HL$ for small $\lam>0$ and 
\bqn 
B_2(\lam)^{-1}=- \lam^{-2} h_1(\lam) T_2^{-1}+ \Og^{(4)}_{S_2\HL} (1)\,.  
\lbeq(B2-inv) 
\eqn  
\edlm 
\bgpf 
Multiply \refeq(tB1S_2inv) by $S_2$ from both sides. 
Since $D_1 S_2 = S_2 D_1 = S_2$ and 
$S_2 \tT_1(\lam) S_2 = T_2$ by \refeq(S2tTS_2=T_2),
\bqn 
S_2(\tB_1(\lam)+S_2)^{-1}S_2= S_2 + \lam^2 h_1(\lam)^{-1} T_2 + 
\Og^{(4)}_{S_2\HL}(\lam^4(\log\lam)^2) \lbeq(BS2inv) 
\eqn 
and we obtain \refeq(B2-def). Since $T_2$ is invertible in $S_2 \HL$, 
the rest of the statements is obvious.  
\edpf 

Since $B_2(\lam)^{-1}$ exists in $S_2\HL$ 
and $B_1(\lam)^{-1} = -h_1(\lam)^{-1}\tB_1(\lam)^{-1}$, 
we obtain by using \reflm(JN) that    
\begin{align}
& B_1(\lam)^{-1}= -h_1(\lam)^{-1}(\tB_1(\lam)+S_2)^{-1}  
-h_1(\lam)^{-1}J_2(\lam), \lbeq(tB1-inv) \\
& J_2(\lam) = 
(\tB_1(\lam)+S_2)^{-1}S_2 B_2(\lam)^{-1}S_2(\tB_1(\lam)+S_2)^{-1}.  
\lbeq(J2-def)  
\end{align}
Substituting \refeq(tB1-inv) in \refeq(M1ess-def) yields  
\begin{gather} 
M_{{\rm ess}}^{(1)}(\lam)= \Ng_{1}(\lam)+ \Ng_{2}(\lam), 
\lbeq(M1ess-expression) \\
\Ng_{1}(\lam)= -h_1(\lam)^{-1}
(\Mg(\lam^4)+S_1)^{-1}S_1 (\tB_1(\lam)+S_2)^{-1}S_1 (\Mg(\lam^4)+S_1)^{-1} 
\notag \\
\Ng_{2}(\lam) = -h_1(\lam)^{-1}
(\Mg(\lam^4)+S_1)^{-1}S_1 J_2(\lam) S_1 (\Mg(\lam^4)+S_1)^{-1}. \lbeq(N2-def)
\end{gather} 

\bglm \lblm(GPR-part1) Suppose that 
$S_1 D_1 S_1= \sum_{j,k=1}^n c_{jk}(\z_j \otimes \z_k)$.  Then, 
\bqn 
M_v \Ng_{1}(\lam) M_v \equiv 
-\sum_{j,k=1}^n h_1(\lam)^{-1}c_{jk} (v \z_j)\otimes (v\z_k)  \lbeq(except-a)
\eqn 
and $M_v \Ng_{1}(\lam) M_v$ is GPR. 
\edlm 
\bgpf Substitute \refeq(MS1) for $(\Mg(\lam^4)+S_1)^{-1}$ 
and \refeq(tB1S_2inv) for $(\tB_1(\lam)+S_2)^{-1}$ 
and expand the result. Then, the terms which contain  
$Y_1(\lam)\in \Og_{\Hg_2}^{(4)}(\lam^4\log\lam)$ of \refeq(MS1) or 
$\Og_{S_1L^2}^{(4)}(\lam^4(\log \lam)^2)$ of \refeq(tB1S_2inv) 
are $\Rg_{em}(\lam)$. What remains is $\Gg\Vg\Sg$ except   
$-h_1(\lam)^{-1}M_v \tD_{0}(\lam)S_1 D_1 S_1 \tD_{0}(\lam) M_v $ 
which is equal to $-h_1(\lam)M_v S_1 D_1S_1 M_v$ 
modulo $\Gg\Vg\Sg$. This implies \refeq(except-a). Since 
$\int_{\R^4} v(x) \z_j(x) dx=0$, for $j=1, \dots, n$, $M_v \Ng_1(\lam) M_v$ 
is GPR by \reflm(s-low). \edpf

The following lemma is the clue to the proof of \refth(main-rephrase). 
Let $\{\z_1, \dots, \z_m\}$ be the orthonormal basis of  $S_2\HL$ such that 
\bqn \lbeq(eig-T2)
T_2\z_j= i a_j^{2}\z_j, \quad a_j>0, \ j=1, \dots, m
\eqn 
(recall \refeq(T2)) and expand it to the basis 
$\{\z_1, \dots, \z_n\}$ of $S_1\HL$, $m \leq n$. 

\bglm \lblm(second-Mlam) Suppose that $H$ has singularity of the 
second kind at zero, Then, modulo 
$\Gg\Vg\Sg+ \Rg_{em}(\lam)$, $\Qg_v(\lam)$ is equal to  
\bqn \lbeq(second-limit)
-i \lam^{-2} \sum_{j=1}^m a_j^{-2} (v\z_j)\otimes (v\z_j) 
+ h_1(\lam)^{-1} \sum_{j,k=1}^n a_{jk}(\lam) (v\z_j)\otimes (v\z_k)  
\eqn 
where $a_{jk}(\lam)$, $j,k=1, \dots, n$ are GMU. 
\edlm  
\bgpf 
It suffices by \reflm(GPR-part1) to prove 
\refeq(second-limit) for $M_v \Ng_2 (\lam) M_v$ in place of 
$\Qg_v(\lam)$. 
We substitute \refeq(tB1S_2inv) for $(\tB_1(\lam)+S_2)^{-1} $ 
and \refeq(B2-inv) for $B_2(\lam)^{-1}$ in \refeq(J2-def) 
and express the result via the basis $\{\z_j\}$. We obtain  
\bqn   \lbeq(inner)
S_1 J_2(\lam) S_1 = 
i\sum_{j=1}^m a_j^{-2} \lam^{-2}h_1(\lam) \z_j \otimes \z_j 
+ \sum_{i,k=1}^n a_{ik}(\lam) \z_i \otimes \z_k  
\eqn 
where $a_{ik}(\lam)$ are GMU. We then substitute 
\refeq(MS1) with $Y_1(\lam)$ of \refeq(Y1-remark) 
for $(\Mg(\lam^4)+S_1)^{-1}$ and 
\refeq(inner) for $S_1 J_2(\lam) S_1$ in \refeq(N2-def) 
and expand the result. Then,  
the part 
$-\lam^2 \tD_{0}(\lam)G_2^{(v)}\tD_{0}(\lam)  
 -\lam^4 \tD_0(\lam) 
\{T_{4,l}(\lam)\tD_0(\lam)- (G_2^{(v)}\tD_0(\lam))^2 \}$ 
of $(\Mg(\lam^4)+S_1)^{-1}$ cancels $\lam^{-2}$ of \refeq(inner) 
and produces $\Gg\Vg\Sg$ and the part 
$\Og^{(6)}_{\Hg_2}(\lam^6\log\lam)$ does $\Rg_{em}(\lam)$.
Thus, we obtain that 
\begin{align*}
& M_v \Ng_2 (\lam)M_v \equiv 
-h_1(\lam)^{-1}M_v \tD_{0}(\lam)\times \refeq(inner)\times \tD_{0}(\lam)M_v 
\notag \\ 
&  \equiv -\sum_{j=1}^m ia_j^{-2} \lam^{-2}
(v\z_j) \otimes (v\z_j)+
\sum_{i,k=1}^n a_{ik}(\lam)h_1(\lam)^{-1}(v\z_i) \otimes (v\z_k) 
\end{align*}
where we used $D_{0}=D_{0}^\ast$, 
$S_1D_{0}= D_{0} S_1= S_1$, $S_2D_{0}= D_{0} S_2= S_2$ and $S_2 L_0=L_0S_2=0$ 
in the second step. \reflm(second-Mlam) follows. 
\edpf  

\subsection{\bf Proof of \refthb(main-rephrase) (2)} 
We follow the argument which is outlined in the introduction 
and which patterns after that of Theorem 5.13 of \cite{Ya-2dim-new}. 
We shall, however, need some new estimates towards the end of the proof. 

By virtue of \reflms(s-low,second-Mlam), we need study only 
\bqn  
\W_{{\rm red}}\,u = \sum_{j=1}^m ia_j^{-2} 
\int_0^\infty R^{+}_0 (\lam^4) 
(v\z_j) \otimes (v\z_j)
\Pi(\lam) u \lam  
\chi_{\leq{a}}(\lam)d\lam \,.
\lbeq(sta0-second) 
\eqn 
We shall deal typically with  the operator 
\bqn 
\W u = \int_0^\infty (R^{+}_0 (\lam^4)(v\z)\otimes 
(v\z)\Pi(\lam) u ) \lam \chi_{\leq{a}}(\lam)d\lam , \quad 
\z\in S_2\HL.
\lbeq(sta0-second-a)
\eqn   
\paragraph{\bf Good and bad parts} 
Since $\int_{\R^4} v(x)\z(x) dx =0$, we may replace 
$\Pi(\lam) u(z)$ by $\Pi(\lam)u(z) -\Pi(\lam) u (0)$   
which is equal to \refeq(Pi-sec), that is, to      
\bqn 
\sum_{l=1}^4 i \lam z_l (\Pi(\lam)R_l u)(0) 
+ \sum_{m,l=1}^4 {z_m z_l \lam^2}\int_0^1 (1-\th)
(\Pi(\lam)\t_{-\th{z}}R_m R_l u)(0)d\th, \lbeq(Pi-sec-a)
\eqn 
and which we substitute in \refeq(sta0-second-a) to produce 
\bqn 
\W u = \W_{B}u + \W_{G}u, 
\eqn  
where $\W_{B}$ and $\W_{G}$ are the operator produced by the first 
and the second terms respectively. 
We call $\W_G$ the {\it good part} and $\W_B$ the {\it bad part} of $\W$.  \\

\paragraph{\bf Good part is GOP}
\bglm \lblm(Good)
The good part $\W_{G}$ is a GOP.
\edlm 
\bgpf Let $T_{m,l}(x,y)= (v\z)(x)y_m y_l (v\z)(y)$ 
and $R_m R_l u=u_{m,l}$. Then, $T_{m,l}\in \Lg^1$ and by virtue 
of \refeq(OmegaT), we may express $\W_G u$ in the following form:    
\begin{align*}
& \sum_{m,l=1}^4 \int_0^1  (1-\th) 
\left(\int_0^\infty  R^{+}_0 (\lam^4)T_{m,l}\Pi(\lam) 
(\tau_{-\th{z}}\chi_{\leq{a}}(|D|)u_{m,l})(0) \lam^3 d\lam \right)d\th \\
& = \sum_{m,l=1}^4 \int_0^1  (1-\th) 
\W(T_{m,l})(\tau_{-\th{z}}\chi_{\leq{a}}(|D|)u_{m,l})d\th .
\end{align*} 
\reflm(T-K) and Minkowski's inequality imply 
that $\W_{G}$ is GOP. 
\edpf 

\bgrm The proof shows that \reflm(Good) holds if 
$\z \otimes \z$ is replaced by $a \otimes \z$ such that  
$a(x) v(x) \in L^1(\R^4)$ and $\z\in Q \HL$.
\edrm 

\paragraph{\bf High energy part of bad part } 
Since $\sum_{l=1}^4 i \lam z_l (\Pi(\lam)R_l u)(0) $ is $\Vg\Sg$,
$\W_B u(x)$ becomes the product   
\begin{align} 
& \W_{B}u(x)= \sum_{l=1}^4 i \la z_l v, \z\ra \W_{B,l} u(x), \lbeq(B-l) \\
& \W_{B,l} u(x) = 
\int_0^\infty R^{+}_0 (\lam^4)(v\z)(x) 
(\Pi(\lam)R_l u)(0)\lam^2 \chi_{\leq{a}}(\lam)d\lam. \lbeq(Wjk2-red)  
\end{align}
Ignoring the harmless constant $i\la z_l v, \z\ra $ 
and Riesz transforms,  we consider typically the operator $W_B$ defined by 
\bqn \lbeq(Bdef)
W_B u(x) = \int_0^\infty (R^{+}_0 (\lam^4)\ph)(x)
(\Pi(\lam)u)(0)\lam^2 \chi_{\leq{a}}(\lam)d\lam\,,
\eqn 
where $\ph(x)  = v(x)\z(x)$, $\z \in S_2\HL$ (do not confuse this $\ph$ 
with $\ph=\Phi(\z)$ of \reflm(resonance)). 
Difficulty here is of course that \refeq(Bdef) has only $\lam^2$ 
instead of $\lam^3$. We decompose  
\[
W_B u = \chi_{\geq 4a}(|D|)W_B u + \chi_{\leq 4a}(|D|)W_B u 
=: W_{B,\geq }u + W_{B,\leq }u 
\]
as in \refeq(B-decomp) to which we refer for more details.
Let $\m(\xi)= \chi_{\geq 4a}(\xi)|\xi|^{-4}$. We have 
$\m \in L^p(\R^4)$ for all $1<p\leq \infty$. 

\bglm \lblm(mu) We have $\hat{\m}(x)\in L^p(\R^4)$ for $1\leq p <\infty$. 
For all $1\leq p \leq \infty$, 
${\m}(|D|)\in \Bb(L^p(\R^4))$ and $\mu(|D|)\ph \in L^p(\R^4)$ 
\edlm 
\bgpf Since $\m \in C^\infty(\R^4)$ and 
$|\pa^\a \m(\xi)|\leq C_\a \la\xi\ra^{-4-|\a|}$, 
integration by parts shows that 
$\hat{\m}\in C^\infty(\R^4\setminus \{0\})$ and 
and rapidly decreasing at infinity along with derivatives; 
for the small $|x|$ behavior, we observe that $\hat{\m}$ is equal 
modulo a smooth function to 
\[
\frac1{(2\pi)^2}\int_a^\infty 
\left(\int_{{\mathbb S}^3}e^{irx\w}d\w\right)\frac{dr}{r} 
=\int_a^\infty  \frac{J_1(r|x|)}{r^2|x|} dr
\]
and the well-known property of the Bessel function implies 
the right side is equl to $C \log |x| + O(|x|^2)$ as $|x|\to 0$. 
Thus, $\hat{\m}(x)\in L^p(\R^4)$ for all $1\leq p <\infty$ 
and $\m(|D|)$ is bounded in $L^p(\R^4)$ 
for all $1\leq p \leq \infty$. Since 
$\ax^{6}\la \log |x|\ra v \in (L^2 \cap L^{2q})(\R^4)$, 
$\ax^{6}\la \log |x|\ra \ph \in (L^1 \cap L^{\frac{2q}{q+1}})(\R^4)$ 
and $\mu(|D|)\ph(x) = (2\pi)^{-2}(\hat\m \ast \ph)(x)\in L^p(\R^4)$ 
for all $1\leq p \leq \infty$. 
\edpf 

\bglm \lblm(Bad) The operator $W_{B,\geq}$ is bounded in $L^p(\R^4)$ 
for $1<p<4$ and, if $a>0$ is sufficiently small, it is unbounded 
in $L^p(\R^4)$  for $4 \leq p \leq \infty$. 
\edlm 
\bgpf By Fourier transform we have 
\begin{multline}  \lbeq(chge2a)
\chi_{\geq{4a}}(|D|) R^{+}_0 (\lam^4)\ph(x) 
= \Fg^\ast\left( \m(\xi)\hat{\ph}(\xi)+   
\lam^4 \frac{\m(\xi)\hat{\ph}(\xi)}{|\xi|^4-\lam^{4}-i0}\right) \\
= \m(|D|)\ph(x) + \m(|D|)\lam^4 R^{+}_0 (\lam^4)\ph(x),
\end{multline}
which we substitute in \refeq(B-decomp) and split it 
as $W_{B,\geq} u(x)= W^{(1)}_{B,\geq} u(x) +W^{(2)}_{B,\geq} u(x)$ 
by defining 
\begin{align}  
& W^{(1)}_{B,\geq} u(x) = \m(|D|)\ph(x) \int_0^\infty 
\Pi(\lam)u(0)\lam^2 \chi_{\leq{a}}(\lam)d\lam\,,  \lbeq(B1) \\ 
& W^{(2)}_{B,\geq} u(x)= 
\int_0^\infty  \m(|D|) R^{+}_0 (\lam^4)\ph(x) 
\Pi(\lam)u(0)\lam^6 \chi_{\leq{a}}(\lam)d\lam. \lbeq(B2)
\end{align}
(1) Let $\n(\lam)= \lam^3 \chi_{\leq{a}}(\lam)$. Then, $\n(\lam)$ is GMU and 
\refeq(K-first) implies 
\begin{align*}
W^{(2)}_{B,\geq}u(x)& = \m(|D|)\int_{\R^4}\ph(y) \t_y 
\left(\int_0^\infty \Rg_\lam (x) \Pi(\lam)\n(|D|)u (0)\lam^3 d\lam \right) dy \\
& = \m(|D|)\int_{\R^4}\ph(y) (\t_y K \n(|D|)u)(x)  dy\,. 
\end{align*}
Minkowski's inequality and \reflm(T-K) imply that 
$W^{(2)}_{B,\geq}$ is GOP. \\[3pt]
(2) Define the linear functional $\ell(u)$ by 
\[
\ell(u)= \int_0^\infty 
\Pi(\lam)u(0)\lam^2 \chi_{\leq{a}}(\lam)d\lam
\]
so that $W^{(1)}_{B,\geq}u(x)=\m(|D|)\ph(x) \ell(u)$. 
We have $\m(|D|)\ph(x)\in L^p(\R^4)$ for all $1\leq p \leq \infty$ 
by \reflm(mu); polar coordinates $\xi=\lam\w$ and  
the Parseval identity imply that $\ell(u)$   
\begin{align}
& \frac1{(2\pi)^2}\int_0^\infty \int_{{\mathbb S}^3} \widehat{u}(\lam\w)
\chi_{\leq{a}}(\lam)\lam^2 d\w d\lam 
=\frac1{(2\pi)^2}\int_{\R^4} \widehat{u}(\xi)
\frac{\chi_{\leq{a}}(|\xi|)}{|\xi|}d\xi \notag \\
& =\frac1{(2\pi)^2}\int_{\R^4} u(x)f(x)dx, \quad  
f(x)= \Fg \left(\frac{\chi_{\leq{a}}(|\xi|)}{|\xi|}\right). \lbeq(Parseval)
\end{align}
It is evident that 
$f \in L^q(\R^4)$ if and only if $4/3<q\leq \infty$. 
Hence $W^{(1)}_{B,\geq}$ is bounded in $L^p(\R^4)$ for $1\leq p <4$ 
and it is unbounded for $4 \leq p \leq \infty$ unless $\m(|D|)\ph=0$.  
However, if $\m(|D|)\ph=0$ for all $a>0$, then it must be that 
$\ph(=v\z)=0$ and, \refeq(inverse) implies that 
there must exist constant solution $u= -\|v\|^{-2}(PT_0 \z, v)$ 
of $(\lap^2 + V)u=0$. However, this implies $V=0$ which is a contradiction. 
This proves the lemma. 
\edpf 

\paragraph{\bf Proof of the negative part of \refthb(main-rephrase)(2)}
\bglm  If $H$ has singularity of the second kind at zero, then 
$W_{-}$ is unbounded in $L^p(\R^4)$ for $4\leq p \leq \infty$ 
\edlm 
\bgpf We prove the lemma when $\rank S_2=1$. Modification 
for general case by using Hahn-Banach theorem can be done 
by following the argument in the part (iv) of the proof of 
Theorem 1.4 (2b) of \cite{Ya-4dim-new} which we omit here. 

We prove the lemma by reductio ad absurdum. Suppose 
$W_{-}$ is bounded in $L^p(\R^4)$ for 
a $4\leq p \leq \infty$. Then, so must be $\W$ 
of \refeq(sta0-second-a) for all $0<a<\infty$ and, 
since $\W_G$ is GOP by \reflm(Good), 
so must be $\W_{B,\geq}=\chi_{\geq{4a}}(|D|)\W_B$.  
Then, since $W^{(2)}_{B,\geq}$ is GOP by part (1) of the proof 
of \reflm(Bad), we conclude that  
\bqn \lbeq(36a) 
\W^{(1)}_{B,\geq}= \sum_{l=1}^4 \la v z_l, \z \ra W^{(1),l}_{B,\geq}u(x) 
= \m(|D|)(v\z)(x) \tilde{\ell}(u), 
\eqn 
must also be bounded in $L^p(\R^4)$ for that $p$, 
where  $W^{(1),l}_{B,\geq}u=W^{(1)}_{B,\geq}R_l u$ (see \refeq(B1)) and 
\bqn 
\tilde{\ell}(u) = 
\left\la u, \sum_{l=1}^4 \la v z_l, \z \ra f_l(x) \right\ra, 
\quad 
f_l(x)= \Fg \left(\frac{\xi_l\chi_{\leq{a}}(|\xi|)}{|\xi|^2} \right)(x).
\lbeq(36)
\eqn 
For sufficiently small $a>0$, we have $\m(|D|)v\z\not=0$. 
By virtue of \refeq(T2)   
$\a\colon =(\la v z_1, \z \ra, \dots, \la v z_4, \z\ra)$ is a non-zero vector 
and, hence, $\a \cdot \xi$ is non-trivial linear function of $\xi$. 
It follows that  
\[
\sum_{l=1}^4 \la v z_l, \z \ra f_l(x)= 
\Fg\left(\a \cdot \xi |\xi|^{-2}\chi_{\leq a}(|\xi|) \right)\not\in L^q(\R^4)
\]
for any $1\leq q \leq 4/3$, because otherwise 
$(\a \cdot \xi)|\xi|^{-2}\chi_{\leq a}(|\xi|) \in L^p(\R^4)$ 
for some $4<p<\infty$ by Hausdorff-Young's inequality which is 
clearly impossible. Thus, $\tilde{\ell}$ 
is unbounded on $L^p(\R^4)$ for any $4\leq p \leq \infty$ 
by the Riesz theorem.  This is a contradiction. 
\edpf 

\paragraph{\bf Low energy part of the bad part} 

The following lemma completes the proof of \refth(main-rephrase) (2).

\bglm \lblm(Bl-p) Let $\z\in S_2\HL$. Then,  
$\chi_{\leq 4a}(|D|){W}_{B}$ is bounded in $L^p(\R^4)$ 
for $1<p<4$. 
\edlm 
\bgpf Let $u \in \Dg_\ast$. It suffices to prove the lemma for   
\[
W_{B,\leq } u(x) = \int_0^\infty   
\chi_{\leq 4a}(|D|) R^{+}_0 (\lam^4)(v\z)(x) 
\Pi(\lam)u(0)\lam^2 \chi_{\leq{a}}(\lam)d\lam \,. 
\]
We denote $\r(\lam) = \Pi(\lam)u(0)\lam^2 \chi_{\leq{a}}(\lam)$.  
Then, $\r\in C_0^\infty((0,\infty))$. Let $I$ be its support. 
Denote $\ph(x)=(v\z)(x)$. It is well known that 
that $W_{B,\leq }u = \lim_{\ep \downarrow{0}} W_{B,\leq }^{\ep} u$ in 
$\HL$, where for $\ep>0$ 
\bqn \lbeq(wb-ep) 
W_{B,\leq }^{\ep} u(x) = \int_0^\infty   
\chi_{\leq 4a}(|D|) R_0 (\lam^4+ i\ep)\ph(x) \r(\lam) d\lam\,. 
\eqn 
We have $\ph \in \ax^{-6}(L^2\cap L^{2q})(\R^4)$ for a 
$q>1$. Then,  
\bqn \lbeq(dx-1ch)
\chi_{\leq 4a}(|D|) R_0 (\lam^4+i\ep)\ph(x) 
= \frac1{(2\pi)^2}\int_{\R^4} \frac{e^{ix\xi}
\chi_{\leq 4a}(|\xi|)\hat{\ph}(\xi)}{(|\xi|^4-\lam^4-i\ep)}d\xi
\eqn 
is bounded by $C \ax^{-\frac32}$ uniformly for $\lam\in I$ and $\ep>0$  
and converges as $\ep \to 0$ uniformly on compact sets of $I \times \R^4$ 
to $\chi_{\leq 4a}(|D|)R_0^{+}(\lam^4) \ph(x)$ along with the derivatives. 

Let $f \in C_0^\infty(\R^4)$. It follows by changing the order of integrations 
by $dx$ and $d\lam$ and, the order of the integral by $d\lam$  
and $\lim_{\ep \to {0}}$ that
\begin{align} 
& \la f, W_{B,\leq }u \ra  =
\lim_{\ep \to 0} \la f, W_{B,\leq }^{\ep}u \ra \notag \\
& =  \int_0^\infty 
\lim_{\ep \to 0} \left(
\int_{\R^4} f(x)
(\chi_{\leq 4a}(|D|)R_0 (\lam^4+i\ep)\ph)(x)dx\right) \r(\lam) d\lam .
\lbeq(chdx)
\end{align} 
Since $\hat{\ph}(0)=0$, we have 
\bqn \lbeq(hatph)
\hat{\ph}(\xi)= 
\frac1{(2\pi)^2}\sum_{m=1}^4 \int_0^1 
\int_{\R^4} (-iz_m \xi_m )e^{-i{\th}z\xi} \ph(z) dz d\th.
\eqn 
We substitute \refeq(hatph) in \refeq(dx-1ch) and, then, substitute 
the result in \refeq(chdx). The inner integral of \refeq(chdx) becomes 
$\sum_{m=1}^4 \int_0^1 d\th $ of 
\begin{align}  
& \left\la f(x), 
\frac{-i}{(2\pi)^4}\iint_{\R^8} \frac{\xi_m e^{i(x-{\th}z)\xi} 
\chi_{\leq 4a}(|\xi|)z_m {\ph}(z)}{(|\xi|^4-\lam^4-i\ep)}d\xi dz 
\right\ra_{\HL(\R^4)}
\lbeq(dx-2ch)
\\
&= \frac{-i}{(2\pi)^4}
\int_{\R^4} z_m {\ph}(z) \left\la R_m f, 
\tau_{{\th}z} \int_{\R^4}e^{ix\xi} \frac{|\xi| \chi_{\leq 4a}(|\xi|)}
{(|\xi|^4-\lam^4-i\ep)}d\xi
\right\ra  dz \,, \notag 
\end{align} 
where we have used $|R_m f(x)|\leq C\la x\ra^{-4}$ 
(see \cite{Stein-old}, page 59) to change 
the order of integrations by $dx$ and $dz$. 
We split the inner most integral on the right by setting      
\bqn 
\frac{|\xi| }{(|\xi|^4-\lam^4-i\ep)}   
=\frac{\lam}{(|\xi|^4-\lam^4-i\ep)}
+ \frac{(|\xi|-\lam) }{(|\xi|^4-\lam^4-i\ep)}\,.
\lbeq(mult-1)
\eqn 
It is obvious that the remark  
below \refeq(dx-1ch) applies to each part with the obvious modifications. 
 Thus, the $\lim_{\ep \to 0}$ 
may be put in front the $\xi$-integral and 
$\la f, W_{B,\leq }u \ra$ becomes $Z_1+ Z_2$, where 
$Z_1$ and $Z_2$ are those produced respectively 
by the first and the second on the right side of \refeq(mult-1). 

We first study $Z_1$ which is equal to 
$\sum_{m=1}^4 \int_0^1 d\th \int_{\R^4}z_m {\ph}(z)dz $ of 
\bqn \lbeq(44p)
\int_0^\infty \la R_m f, 
\chi_{\leq 4a}(|D|)\Rg_\lam(x-{\th}z)\ra \r(\lam)\lam d\lam 
\eqn 
where we used the fact that  
$|\la R_m f, \chi_{\leq 4a}(|D|)
\Rg_\lam(x-{\th}z)\ra |\leq C\la \th{z}\ra^{-\frac32}$ for $\lam \in I$ to 
integrate by $d\lam$ before doing by $d{\th}dz$. Changing the order of 
integrations once more and restoring the definition of $\r(\lam)$, 
we obtain 
\begin{align*}
\refeq(44p) & = \left\la R_m f,
\chi_{\leq 4a}(|D|) \tau_{\th{z}} \int_0^\infty 
\Rg_\lam(x)\r(\lam)\lam d\lam \right\ra \\
& = \left\la f, R_m \chi_{\leq 4a}(|D|) \tau_{\th{z}}K \chi_{\leq {a}}(|D|)u 
\right\ra.
\end{align*} 
Changing the order of integrations, we obtain 
$Z_1 = \la f, \Zg_1 u \ra$, where  
\[
\Zg_1u(x)= \sum_{m=1}^4 \int_0^1 \left( 
\int_{\R^4} z_m\ph(z) R_m \chi_{\leq 4a}(|D|)
\tau_{\th{z}} K \chi_{\leq{a}}(|D|)u dz \right) d\th 
\] 
and it satisfies 
$\|\Zg_1u \|_p \leq C \|\ax \ph \|_1 \|u\|_p$ for all $1<p<\infty$. 
by \reflm(allp) and Minkowski's inequality.   

We next study $Z_2$. For $\lam>0$ and $\ep>0$ we have 
\[
\frac{|\xi|-\lam}{(|\xi|^4-\lam^4-i\ep)}\absleq 
\frac{1}{(|\xi|+\lam)(|\xi|^2+\lam^2)}
\]
and, as $\ep \to 0$, the left-hand side converges to the right-hand side 
for all $(\xi,\lam)\in \R^4 \times I$. Then, proceeding as previously 
we obtain $Z_2 = \la f, \Zg_2 u\ra$, where   
\bqn \lbeq(Z2def)
\Zg_2 u(x)= \sum_{m=1}^4 \int_0^1 \left( 
\frac{i}{(2\pi)^4} 
\int_{\R^4} z_m {\ph} (z) (\tau_{{\th}z} R_m L u)(x)dz \right) d\th. 
\eqn 
where $Lu(x)$ is defined for $u \in \Dg_\ast$ by 
\[ 
Lu(x)= \int_0^\infty \left( \int_{\R^4}
e^{ix\xi} \frac{\chi_{\leq 4a}(|\xi|)}
{(|\xi|^2+\lam^2)(|\xi|+\lam)}d\xi\right)
\Pi(\lam)u(0)\lam^2 \chi_{\leq{a}}(\lam)d\lam.
\] 
Then, we have the following lemma whose proof is postponed to Appendix 

\bglm \lblm(intL) The $L$ is bounded in $L^p(\R^4)$ for $1<p<4$. 
\edlm 

We apply \reflm(intL) and Minkowskii's inequality to \refeq(Z2def) 
and obtain 
$\|\Zg_2 u\|_p \leq C \|u\|_p$ for $1<p<4$. It follows that 
\[
|\la f, W_{B,\leq} u\ra| = |\la f, \Zg_1u + \Zg_2 u \ra |\leq
C \|f\|_{p'} \|u\|_p ,\ p'=p/(p-1)
\] 
and $\|W_{B,\leq} u\|_p \leq C \|u\|_p$. This completes the proof. 
\edpf

\section{Singularities of third and fourth kind }  \lbsec(3rd-4th)

In this section we prove statements (3) and (4) of 
\refth(main-rephrase) when $H$ has singularities of the third or 
of the fourth kind at zero assuming that 
$\ax^{16}\la \log |x|\ra^2 V \in (L^1 \cap L^q)(\R^4)$ for a $q>1$.
Thus, 
$T_2=S_2 G_2^{(v)}S_2$ is singular 
on $S_2\HL$, $S_3$ is the projection to $\Ker T_2\vert_{S_2\HL}$  
and we have a sequence of projections 
$Q \supset S_1 \supset S_2 \supset S_3 \not=\{0\}$. 
We choose the basis  $\{\z_1, \dots, \z_n\}$ of $S_1\HL$ such that 
$\{\z_1, \dots, \z_m\}$ is the one of $S_2\HL$ such that 
$\z_1, \dots, \z_r$, $r \leq m$ are eigenvectors 
of $T_2$ with non-zero eigenvalues $-ia_1^{2}, \dots, -ia_r^{2}$ 
and $\z_{r+1}, \dots, \z_m \in S_3 \HL$.  

\subsection{\bf Threshold analysis 3. First step}
\reflms(MS1-good,GPR-part1) imply that    
\bqn \lbeq(Qgv=N2v)
\Qg_v(\lam)\equiv M_v \Ng_2(\lam)M_v = \colon \Ng_2^{(v)}(\lam) 
\quad \mbox{modulo GPR}. 
\eqn 
and $\Ng_2(\lam)$ is given by \refeq(N2-def) and \refeq(J2-def). 

Let 
$\tB_2(\lam) \colon = - \lam^{-2} h_1(\lam) B_2(\lam)$. 
We apply \reflm(JN) to the pair $(\tB_2(\lam),S_3)$. 
\refeq(B2-def) implies  
\bqn \lbeq(B2-deff)
\tB_2(\lam)= T_2+ \Og^{(4)}_{S_2\HL}(\lam^2 \log\lam); 
\eqn 
since $(T_2+S_3)^{-1}$ exists in $S_2\HL$, so does 
$(\tB_2(\lam)+ S_3)^{-1}$  for small $\lam>0$ and 
\bqn \lbeq(tB2S3-rough)
(\tB_2(\lam)+ S_3)^{-1}= D_2 + \Og^{(4)}_{S_2\HL}(\lam^2 \log\lam), 
\quad D_2= (T_2+S_3)^{-1}. 
\eqn 
Assume for the moment that  
\bqn 
B_3(\lam) = S_3 - S_3 (\tB_2(\lam)+ S_3)^{-1} S_3 \lbeq(B3-def)
\eqn 
is invertible in $S_3\HL$. 
Then, \reflm(JN) implies $\tB_2(\lam)^{-1}$ is equal to 
\bqn 
(\tB_2(\lam)+ S_3)^{-1} + 
(\tB_2(\lam)+ S_3)^{-1}S_3 B_3(\lam)^{-1}S_3 
(\tB_2(\lam)+ S_3)^{-1}. \lbeq(tB2-inv)
\eqn 
If we substitute $\refeq(tB2-inv) \times (- \lam^{-2} h_1(\lam))$  
for $B_2(\lam)^{-1}$ in \refeq(J2-def)  
and the result for $J_2(\lam)$ in \refeq(N2-def), 
then $J_2(\lam)$ becomes $J_2(\lam) = J_{2,1}(\lam)+ J_{2,2}(\lam)$ 
with  
\bqn 
J_{2,1}(\lam)=S_1 (\tB_1(\lam)+S_2)^{-1}
S_2 (\tB_2(\lam)+S_3)^{-1}S_2(\tB_1(\lam)+S_2)^{-1}S_1
\lbeq(E1-def)
\eqn 
and $J_{2,2}(\lam)$ being obtained from \refeq(E1-def) by 
replacing $(\tB_2(\lam)+S_3)^{-1}$ by 
$(\tB_2(\lam)+ S_3)^{-1}S_3 B_3(\lam)^{-1}S_3 
(\tB_2(\lam)+ S_3)^{-1}$ 
and, subsequently $\Ng_2^{(v)}(\lam)$ becomes the sum 
and 
\bqn \lbeq(N2-sum)
\Ng_2^{(v)}(\lam)=\Ng_{2,1}^{(v)}(\lam)+ \Ng_{2,2}^{(v)}(\lam)\,.
\eqn 
where, for $j=1,2$ 
\bqn 
\Ng_{2,j}^{(v)}(\lam)= \lam^{-2} M_v (\Mg(\lam^4)+S_1)^{-1}
J_{2,j} (\lam)(\Mg(\lam^4)+S_1)^{-1}M_v\,. \lbeq(A1-a) 
\eqn

In this subsection we prove the following lemma. 

\bglm \lblm(tB2S3) {\rm (1)} With $\b_{jk}(\lam) \in \Og_{\C}^{(4)}(1)$, 
$1\leq j,k\leq n$,
\bqn \lbeq(N21-def)
\Ng_{2,1}^{(v)}(\lam)\equiv 
- i\sum_{j=1}^m a_j^{-2} \lam^{-2} 
(v\z_j) \otimes(v\z_j) + \sum_{j,k=1}^n h_1(\lam)^{-1} \b_{jk}(\lam) 
(v\z_j) \otimes(v\z_k)\,.
\eqn 
{\rm (2)} The operator produced by \refeq(sta0-low) with 
$\Ng_{2,1}^{(v)}(\lam)$ in place of   
$\Qg_v(\lam)$ is bounded in $L^p(\R^4)$ for $1<p<4$ 
and unbounded for $4\leq p \leq \infty$. 
\edlm 
\bgpf \reflm(resonance) implies that $\z \in S_1\HL$ 
is given by $\z=-w\ph$ by a $\ph \in \Ng_{\infty}(H)$ and 
$\z\in \ax^{-6}\la \log |x| \ra^{-1} (L^2 \cap L^{2q})(\R^4)\subset L^1(\R^4)$. 
Hence $J_{2,1}(\lam)$ is $\Gg\Vg\Sg$. 
Substitute \refeq(MS1) with \refeq(Y1-remark)  
for each of two $(\Mg(\lam^4)+S_1)^{-1}$'s in \refeq(A1-a) and expand the result. 
Then, the terms on the right of \refeq(MS1) except $\tD_{0}(\lam)$ 
produces $\Gg\Vg\Sg+ \Rg_{em}(\lam)$ and we have  
\[
\Ng_{2,1}^{(v)}(\lam)\equiv 
\lam^{-2} M_v\tD_{0}(\lam) J_{2,1}(\lam) \tD_{0}(\lam)M_v \,.
\]
Then, since 
$\tD_{0}(\lam)= D_0 + h_1(\lam) L_0$ and $S_1D_0= D_0 S_1= S_1$, 
\begin{align} \lbeq(E11)
& \Ng_{2,1}^{(v)}(\lam)\equiv  \lam^{-2} M_v J_{2,1}(\lam) M_v   \\
& + \lam^{-2}h_1(\lam)
M_v(L_0J_{2,1}(\lam)+J_{2,1}(\lam)L_0)M_v   
+ \lam^{-2}h_1(\lam)^2 M_v L_0 J_{2,1}(\lam)L_0.  \notag 
\end{align}
Note that the right side of \refeq(E11) is $\Vg\Sg$. 
We then substitute \refeq(tB1S_2inv) 
for each of two $(\tB_1(\lam)+S_2)^{-1}$'s in 
$J_{2,1}(\lam)$ of \refeq(E11) and expand the result. 
Then, the term 
$\Og^{(4)}_{S_2\HL}(\lam^4 (\log\lam)^2)$ of \refeq(tB1S_2inv) produces 
$\Gg\Vg\Sg$ for $\Ng_{2,1}^{(v)}(\lam)$ ; 
$\lam^2 h_1(\lam)^{-1} D_1 \tT_1(\lam) D_1$ does so for the second line 
of \refeq(E11) and for the first produces   
\bqn \lbeq(bjk-l)
\sum_{j,k=1}^n h_1(\lam)^{-1} b_{jk}(\lam)(v \z_j) \otimes (v\z_k), 
\quad b_{jk}(\lam)\in \Og^{(4)}_{\C}(1)\,.
\eqn 
Thus, we may replace $(\tB_1(\lam)+S_2)^{-1}$ by $D_1$ in 
$J_{2,1}(\lam)$ and, subsequently 
$J_{2,1}(\lam)$ by $S_2 (\tB_2(\lam) + S_3)^{-1} S_2$ 
in \refeq(E11), which, since $D_1S_2= S_2 D_1=S_2$, yields 
that   
\bqn 
\Ng_{2,1}^{(v)}(\lam)\equiv  
\lam^{-2} M_v S_2 (\tB_2(\lam) + S_3)^{-1} S_2 M_v + \refeq(bjk-l). 
\eqn 
We next substitute \refeq(tB2S3-rough) for $(\tB_2(\lam) + S_3)^{-1}$.
Then $\Og_{S_2\HL}^{(4)}(\lam^2\log\lam)$ in \refeq(tB2S3-rough)
yields the operator of the form \refeq(bjk-l) again for 
$\Ng_{2,1}^{(v)}(\lam)$ and 
\bqn \lbeq(47)
\Ng_{2,1}^{(v)}(\lam) \equiv \lam^{-2} M_v S_2 D_2 S_2  M_v+ 
\sum_{j,k=1}^m h_1(\lam)^{-1}\b_{jk}(\lam) (v \z_j)\otimes (v\z_k)
\eqn 
where $\b_{jk}(\lam)\in \Og_{\C}^{(4)}(1)$. Since $D_2=(T_2+ S_3)^{-1}$, 
the choice of the basis of $S_2\HL$ produces  
\bqn \lbeq(9-6)
\lam^{-2} M_v S_2 D_2 S_2M_v = - i\sum_{j=1}^m a_j^{-2} \lam^{-2} 
(v\z_j) \otimes(v\z_j)
\eqn 
where $a_{r+1}^{2}= \dots =a_m^{2}=i$. Combining \refeq(47) with \refeq(9-6), 
we obtain (1). \\[3pt] 
(2) The second term on \refeq(N21-def) is GPR by \reflm(s-low) and 
the first is of the form 
which appeared in \refeq(sta0-second) and, hence, produces the operator which 
is bounded $L^p(\R^4)$ for $1<p<4$ and unbounded for $4\leq p$. 
\edpf 

Eqn. \refeq(Qgv=N2v) and \reflm(tB2S3) imply the following. 
\bgcor \lbcor(Q=Z) If $H$ has singularities of the third or fourth kind 
at zero, then, modulo the operator 
which is bounded in $L^p(\R^4)$ for $1<p<4$, $W_{-}\chi_{\leq {a}}(|D|)$ 
is equal to the operator $\Zg$ defined by 
\bqn \lbeq(Q=Z)
\Zg u = \int_0^\infty R_0(\lam^4)\Ng_{2,2}^{(v)}(\lam) \Pi(\lam)u \lam^3 
\chi_{\leq {a}}(\lam)u d\lam \,.
\eqn 
\edcor

\subsection{\bf Key lemma} 
For studying $\Ng_{2,2}^{(v)}(\lam)$ we need the following lemma, of which 
we mostly use in this section only the information on the size that 
$B_3(\lam)^{-1}= \Og_{S_3\HL}^{(4)}(\lam^{-2}h_1(\lam))$ 
or $B_3(\lam)^{-1}= \Og_{S_3\HL}^{(4)}(\lam^{-2})$
when singularities are of the third kind or of the fourth kind 
respectively.

\bglm \lblm(b3-inv)
\ben 
\item[{\rm (1)}]  If $H$ has singularity of the third kind, then, 
\bqn \lbeq(B3-inv-remainder)
B_3(\lam)^{-1}
= \lam^{-2} \tg_2(\lam)^{-1} S_3 T_3^{-1} S_3+ 
\Og^{(4)}_{S_3\HL}(\lam^{-2}(\log\lam)^{-2}).
\eqn 
\item[{\rm (2)}]  If the singularity is of the fourth kind, then
\bqn \lbeq(B3-inv-for4)
B_3(\lam)^{-1}
= \lam^{-2} S_4 T_4^{-1} S_4+ 
\Og^{(4)}_{S_3\HL}(\lam^{-2}(\log\lam)^{-1}).
\eqn 
\item[{\rm (3)}] If the singularity is of the fourth kind but 
$d$-wave resonances are absent from $H$, then    
\bqn \lbeq(B3-inv-for4-suppl)
B_3(\lam)^{-1}
= \lam^{-2} S_4 T_4^{-1} S_4+ \Og^{(4)}_{S_3\HL}((\log\lam)^2).
\eqn 
\een
\edlm 

For proving \reflm(b3-inv), we prepare a few lemmas. 
\bglm \lblm(S34-cn) {\rm (1)} 
Following identities are satisfied by $S_3$:  
\begin{gather}\lbeq(MvG)
G_2^{(v)}S_3= i(4^4\pi)^{-1} v \otimes S_3(x^2 v), \  
S_3 G_2^{(v)} = i(4^4\pi)^{-1}S_3 (x^2 v)\otimes v. \\
S_j  G_2^{(v)} S_3 =S_3  G_2^{(v)} S_j =0\,, \ 
j=0,1,2,3. \lbeq(3rd-cancel) \\
\tT_1(\lam)D_1 S_3 = - h_1(\lam)S_1T_0P  G_2^{(v)} S_3. \lbeq(tT1S_2) \\
S_2 \tT_1(\lam)D_1 S_3= 0.  \lbeq(S2tT1S_3)
\end{gather}
{\rm (2)} We have following identities for $S_4$:  
\begin{gather} 
G_2 M_v S_4= S_4 M_v G_2=0, \lbeq(S4G2)  \\
\tT_1(\lam)S_4= S_4 \tT_1(\lam)=0.    \lbeq(S4tT)
\end{gather}
\edlm 
\bgpf (1) \reflm(characterization) (2) evidently 
implies \refeq(MvG).   
Then, \refeq(3rd-cancel) follows since $S_j v=0$, $j=0, \dots, 3$. Recall that 
$\tT_1(\lam)= S_1 \tD_{0}(\lam) G_2^{(v)}\tD_{0}(\lam)S_1$.  
Then, $D_1S_2=S_2$, $L_0S_2=0$ and \refeq(3rd-cancel) imply   
\[
\tT_1(\lam)D_1 S_3 = 
(S_1+ h_1(\lam)S_1 L_0) G_2^{(v)}S_3
= h_1(\lam)S_1 L_0  G_2^{(v)} S_3.
\] 
Substitute \refeq(gPT0-inv-b) for $L_0$. Then, 
$Q  G_2^{(v)} S_2=0$ and $S_1 D_{0} Q= S_1$ imply \refeq(tT1S_2). 
Since $S_2T_0=0$ by \refeq(S2T0), \refeq(S2tT1S_3) follows from 
\refeq(tT1S_2). \\[3pt]
(2) \reflm(characterization) (3) implies \refeq(S4G2). 
Since $S_4\tD_0(\lam)= S_4(D_0+ h_1(\lam)L_0)=S_4$,  
\refeq(S4tT) follows from \refeq(S4G2).  
\edpf
The following lemma is the precise version of \refeq(tB2S3-rough). 
The lemma is more 
than necessary for the proof for \reflm(b3-inv), however, we need 
it in this form for the proof of \reflm(upto-2). 

\bglm \lblm(coll) Modulo $\Og^{(4)}_{S_2\HL}(\lam^4 \log\lam)$ we have 
that 
\bqn \lbeq(coll)
(\tB_2(\lam)+S_3)^{-1}\equiv  
D_2- D_2 F_3(\lam)D_2 + F_{3,sq}(\lam),
\eqn 
where $F_3(\lam)$ and $F_{3,sq}(\lam)$ are given by 
\begin{align}
F_3(\lam)& \equiv \lam^2 S_2 \{T_{4,l}(\lam)- G_2^{(v)}\tD_0(\lam) G_2^{(v)}
+ h_1(\lam)^{-1}(\tT_1(\lam)D_1)^2\}S_2  \notag \\
& + \lam^4 h_1^{-1}(\lam) \tg_2(\lam) S_2 
\{G_4^{(v)}S_1 G_2^{(v)} D_1 
+ G_2^{(v)}S_1 D_1 S_1 G_4^{(v)}\}S_2 \notag \\
& + \lam^4 h_1(\lam)^{-2} S_2 (\tT_1(\lam)D_1)^3) S_2, \lbeq(F3-detail) \\
F_{3,sq}(\lam)&   
\equiv 
\lam^4 D_2 \{S_2 (\tg_2(\lam)G_{4}^{(v)}
+ h_1(\lam)^{-1}(\tT_1(\lam)D_1)^2)S_2 D_2\}^2. \lbeq(Fsq-detail) 
\end{align}
\edlm 
\bgpf Expanding \refeq(first-line) to the third order, we have 
by \refeq(D1-L1) that 
\bqn \lbeq(tB1-f) 
(\tB_1(\lam)+S_2)^{-1}= \sum_{j=0}^3 D_1 L_1(\lam)^j 
+\Og^{(4)}(\lam^8(\log\lam)^4).
\eqn
Then, $B_2(\lam) = S_2 -S_2 (\tB_1(\lam)+S_2)^{-1}S_2$ becomes 
\[
B_2(\lam)= - S_2L_1(\lam)S_2 - S_2 \big(\sum_{j=2}^3 D_1 L_1(\lam)^j 
+\Og^{(4)}(\lam^8(\log\lam)^4)\big) S_2.
\]
Recall \refeq(D1-L1): 
$L_1(\lam)=\lam^2 h_1(\lam)^{-1} \tT_1 (\lam)D_1 - \tT_4(\lam)D_1$ 
and \refeq(tQ4) and \refeq(Y1-remark) for $\tT_4(\lam)$ which yields    
\bqn 
\tT_4(\lam) \equiv - \lam^{4}h_1^{-1}(\lam)S_1 \tD_0(\lam)
\{T_{4,l}(\lam)\tD_0(\lam)-  (G_2^{(v)}\tD_0(\lam))^2\}S_1 
\lbeq(T4-recall)
\eqn 
modulo $\Og_{S_1\HL}(\lam^6(\log\lam)^2)$. Then, 
$\tB_2(\lam)=-\lam^{-2}h_1(\lam)B_2(\lam)$ becomes, since 
$S_2\tT_1 (\lam)D_1 S_2 = T_2$, 
\begin{gather} 
\tB_2(\lam) = T_2 + F_3(\lam)+ 
\Og_{S_2\HL}^{(4)}(\lam^6(\log\lam)^{3}),  \lbeq(B2-N)  \\
F_3(\lam)=\lam^{-2}h_1(\lam)S_2 \Big(
- \tT_4(\lam) + \sum_{j=2}^{3} L_1(\lam)^j \Big)S_2 \lbeq(F3) 
\end{gather}
and $S_2L_0=L_0S_2=0$ implies that $F_3(\lam)$ is given by \refeq(F3-detail) . 
From \refeq(B2-N) we deduce that modulo $\Og^{(4)}_{S_2\HL}(\lam^6(\log\lam)^3)$
\begin{align}
(\tB_2(\lam)+S_3)^{-1} 
& \equiv  D_2 ({\bf 1}_{S_2\HL}+ F_3(\lam)D_2)^{-1} \notag \\
& \equiv   D_2- D_2 F_3(\lam)D_2 + D_2(F_3(\lam) D_2)^2 
\lbeq(tB2S3-inv) 
\end{align}
and $F_{3,sq}(\lam)=D_2(F_3(\lam) D_2)^2$ is given by \refeq(Fsq-detail). 
\edpf

\bglm  \lblm(pre-B3-inv) 
Let $\tL = S_3 G_2^{(v)}(- L_0+ L_0S_1D_1S_1L_0)G_2^{(v)} S_3$ and 
\bqn \lbeq(Cgdef)
\Cg(\lam)\colon = T_3 + \tg_2(\lam)^{-1}S_3 G_{4,l}^{(v)}S_3 
+ \tg_2(\lam)^{-1}h_1(\lam)\tL.
\eqn 
Then, $B_3(\lam)=S_3 - S_3(\tB_2(\lam)+ S_3)^{-1}S_3$  is equal to 
\bqn   \lbeq(9-B3-e) 
B_3(\lam)= \lam^2 \tg_2(\lam) \Cg(\lam) - \lam^4 \tg_2(\lam)^2 F(\lam) 
\eqn 
where 
\bqn 
F(\lam)\colon 
=S_3(G_4^{(v)}S_2 D_2 S_2G_4^{(v)})S_3 + \Og^{(4)}_{S_3\HL}((\log\lam)^{-1}).
\lbeq(F-def)
\eqn 
\edlm 
\bgpf By virtue of \refeq(coll) we have   
\bqn \lbeq(9-B3)
B_3(\lam)= S_3 F_{3}(\lam) S_3 -  S_3F_{3,sq}(\lam)S_3 
+ \Og^{(4)}_{S_2\HL}(\lam^4 \log\lam).
\eqn 
Since $Q G_2^{(v)}S_3= S_3 G_2^{(v)}Q=0$ by \refeq(3rd-cancel), 
the second and the third lines of \refeq(F3-detail) vanish when 
sandwiched by $S_3$ and, modulo $\Og^{(4)}_{S_3\HL}(\lam^4\log\lam)$  
\begin{align} \lbeq(9-F3)
& S_3 F_{3}(\lam) S_3 \equiv \lam^2 S_3 (T_{4,l}(\lam)+ h_1(\lam)\tL)S_3,  \\
\lbeq(9-F3sq)
& S_3F_{3,sq}(\lam)S_3 
\equiv \lam^4 \tg_2(\lam)^2 S_3(G_4^{(v)}S_2 D_2 S_2G_4^{(v)})S_3 .
\end{align} 
Recalling that $S_3 G_4^{(v)} S_3= T_3$, we obtain the lemma. 
\edpf 

\paragraph{\bf Proof of \reflmb(b3-inv) (1)} 
If $H$ has singularity of the third kind, then $T_3$ is invertible in 
$S_3\HL$ and so is 
$\Cg(\lam)= \big(1_{S_3\HL} + 
(\tg_2(\lam)^{-1}S_3 G_{4,l}^{(v)}S_3 + 
\tg_2(\lam)^{-1}h_1(\lam)\tL)T_3^{-1}\big)T_3$ 
for small $\lam>0$ and 
\[
\Cg(\lam)^{-1}= T_3^{-1}- 
\tg_2(\lam)^{-1} T_3^{-1} S_3 G_{4,l}^{(v)}S_3 T_3^{-1}
+ \Og_{S_3\HL}((\log\lam)^{-2}).
\]
Then, $B_3(\lam)= 
\lam^2 \tg_2(\lam) (1 -  \lam^2 \tg_2(\lam)F(\lam)\Cg(\lam)^{-1})\Cg(\lam)$ 
is also invertible in $S_3\HL$ for small $\lam>0$ 
and    
\[ 
B_3(\lam)^{-1}
= \lam^{-2} \tg_2(\lam)^{-1}\Cg(\lam)^{-1} 
+   \Cg(\lam)^{-1}F(\lam)\Cg(\lam)^{-1} + 
\Og_{S_3\HL}(\lam^2\tg_2(\lam)).
\] 
This implies the lemma. 
\qed \\

\paragraph{\bf Proof of \reflmb(b3-inv)(2)} 
Let $H$ have singularity of the fourth kind and $S_4$ be the projection 
to ${\rm Ker}\,T_3\vert_{S_3\HL}$. Then, it is well-known (cf. \cite{GT-1}) that 
$T_4= S_4G_{4,l}^{(v)}S_4$  is non-singular in $S_4\HL$. 
Let $S_4^\perp=S_3\ominus S_4$. 

\bglm \lblm(B3-inv-fourth) 
For small $\lam>0$,  $\Cg(\lam)^{-1}$ exists in $S_3\HL$ and 
\bqn \lbeq(C1-structure)
\Cg(\lam)^{-1} = \tg_2(\lam)S_4 T_4^{-1} S_4 + Z(\lam),
\eqn 
where $Z(\lam)=\Og^{(4)}_{S_3\HL}(1)$ and 
in the decomposition $S_3\HL = S_4^\perp\HL \oplus S_4\HL$ 
\begin{gather*}
Z(\lam)= 
\begin{pmatrix} 
d(\lam) & -d(\lam)S_4^\perp G_{4,l}^{(v)}S_4 T_4^{-1} \\
- T_4^{-1}S_4 G_{4,l}^{(v)}S_4^\perp {d(\lam)} & 
T_4^{-1}S_4 G_{4,l}^{(v)}S_4^\perp {d(\lam)} 
S_4^\perp G_{4,l}^{(v)}S_4 T_4^{-1}
\end{pmatrix},  \\
d(\lam)= (S_4^\perp T_3 S_4^\perp)^{-1}+  
\Og^{(4)}_{S_4^\perp\HL}((\log\lam)^{-1}).
\end{gather*} 
\edlm  
\bgpf Since $G_2^{(v)}S_4=S_4 G_2^{(v)}=0$  by \refeq(S4G2), we have 
$\tL S_4= S_4 \tL=0$ and, in the 
decomposition $S_3\HL = S_4^\perp \oplus S_4\HL$, 
\bqn \lbeq(Cgdef-a)
\Cg(\lam) 
= \begin{pmatrix}
S_4^\perp \Cg(\lam)S_4^\perp  & 
\tg_2(\lam)^{-1}S_4^\perp G_{4,l}^{(v)}S_4 \\
\tg_2(\lam)^{-1}S_4 G_{4,l}^{(v)}S_4^\perp & \tg_2(\lam)^{-1}T_4 
\end{pmatrix}\,. 
\eqn 
We apply \reflm(FS) to $\Cg(\lam)$. Then,  
$a_{22}=\tg_2(\lam)^{-1}T_4$ is invertible in $S_4\HL$;  since 
$S_4^\perp T_3 S_4^\perp $ is invertibe in $S_4^\perp \HL$, 
\begin{align*}
a_{11}-a_{12}a_{22}^{-1}a_{21}
& = S_4^\perp \Cg(\lam)S_4^\perp - 
\tg_2(\lam)^{-1}S_4^\perp G_{4,l}^{(v)}
S_4 T_4^{-1} S_4 G_{4,l}^{(v)}S_4^\perp \\
& = S_4^\perp T_3 S_4^\perp  + \Og^{(4)}_{S_4^\perp \HL}((\log\lam)^{-1})
\end{align*}
is also invertible for small $\lam>0$ and 
\bqn \lbeq(dlam)
d(\lam)= (a_{11}-a_{12}a_{22}^{-1}a_{21})^{-1}
= 
(S_4^\perp T_3 S_4^\perp)^{-1}(1+ \Og^{(4)}_{S_4^\perp}(\log\lam)^{-1})\,. 
\eqn 
It follows by \reflm(FS) that $\Cg_1(\lam)^{-1}$ exists for small $\lam>0$ and  
\bqn 
\Cg_1(\lam)^{-1} = \tg_2(\lam)S_4 T_4^{-1} S_4  + Z(\lam).
\eqn 
This proves the lemma. 
\edpf 

Since $\Cg(\lam)^{-1}$ exists and 
$\Cg(\lam)^{-1}=\Og_{S_3\HL}^{(4)}(\log\lam)$ by \refeq(C1-structure), 
we still have 
$B_3(\lam)= 
\lam^2 \tg_2(\lam) (1 -  \lam^2 \tg_2(\lam)F(\lam)\Cg(\lam)^{-1})\Cg(\lam)$ 
and, repeating the proof of part (1),   
\begin{align}
& B_3(\lam)^{-1}= 
\lam^{-2} \tg_2(\lam)^{-1}\Cg(\lam)^{-1} 
+   \Cg(\lam)^{-1}F(\lam)\Cg(\lam)^{-1} + 
\Og_{S_3\HL}^{(4)}(\lam^2\tg_2(\lam)^3) \notag \\
& = \lam^{-2} \tg_2(\lam)^{-1} 
(\tg_2(\lam)S_4 T_4^{-1} S_4 + Z(\lam)) + \Og_{S_3\HL}^{(4)}((\log\lam)^2). 
\lbeq(B3-4)
\end{align}
This implies \refeq(B3-inv-for4). 

\paragraph{\bf Proof of \reflmb(b3-inv) (3)}
\reflm(resonance) (3) implies that 
$d$-resonances are absent from $H$ if and only if 
$S_3\HL \ominus S_4\HL=\{0\}$ or $T_3=0$ on $S_3\HL$. 
Then, $S_4= S_3$, $S_4^\perp=0$ and \refeq(Cgdef-a) becomes 
$\Cg(\lam)= \tg_2(\lam)^{-1} S_4 T_4 S_4$. It follows that $Z=0$ 
in \refeq(C1-structure). Then, \refeq(B3-4) implies 
\refeq(B3-inv-for4-suppl). 
\qed

\subsection{Simplification} 
In this subsection we shall simplify $\Ng_{2,2}^{(v)}$ of \refeq(N2-sum) 
modulo GPR. For shortening formulae we introduce     
\begin{align}
E_{2,l}(\lam)& =  S_1 (\tB_1(\lam)+S_2)^{-1}
S_2 (\tB_2(\lam)+S_3)^{-1}S_3,  \lbeq(Mess-E2l)\\
E_{2,r}(\lam)& = S_3 (\tB_2(\lam)+S_3)^{-1}S_2(\tB_1(\lam)+S_2)^{-1}S_1, 
 \lbeq(Mess-E2r)\\
E_2(\lam) & = E_{2,l}(\lam) B_3(\lam)^{-1} E_{2,r}(\lam) 
 \lbeq(Mess-E2c)
\end{align}
and express $\Ng_{2,2}^{(v)}(\lam) $ in the form  
\bqn \Ng_{2,2}^{(v)}(\lam) 
= \lam^{-2} M_v(\Mg(\lam^4)+S_1)^{-1}E_2(\lam)(\Mg(\lam^4)+S_1)^{-1}M_v\,.
 \lbeq(Mess-E2)
\eqn 
Note that $E_2(\lam)$ is sandwiched by $S_1$ and, hence, is $\Vg\Sg$  
but not $\Gg\Vg\Sg$ since, by 
\refeqss(B3-inv-remainder,B3-inv-for4,B3-inv-for4-suppl) 
$E_2(\lam)\in \Og_{S_1\HL}^{(4)}(\lam^{-2}\tg_2(\lam)^{-1})$ 
if the singularity are of the third kind and 
$E_2(\lam)\in \Og_{S_1\HL}^{(4)}(\lam^{-2})$ 
if they are of the fourth kind. 

We remark that the proof of the following lemma will use the 
information only on the size of $B_3(\lam)^{-1}$ of \reflm(b3-inv)

\bglm \lblm(upto-2) 
{\rm (1)} If $H$ has singularity of the third kind, then 
$\Ng_{2,2}^{(v)}(\lam)$ is unchanged modulo GPR when the left sides of 
$\approx$ if the following equations are replaced by the right sides:  
\begin{align}
& (\Mg(\lam^4)+S_1)^{-1} \approx 
\tD_{0}(\lam)- \tD_{0}(\lam)\lam^2 G_2^{(v)}\tD_{0}(\lam) .  
\lbeq(resd-a-suppl-third) \\
& (\tB_1(\lam)+S_2)^{-1} \approx 
D_1+ \lam^2 h_1(\lam)^{-1}\tT_1(\lam)D_1. \lbeq(11S22-suppl-third) \\
& (\tB_2(\lam)+S_3)^{-1}\approx 
D_2 - \lam^2 D_2 S_2 
(T_{4,l}{ -G_2^{(v)}\tD_0(\lam)G_2^{(v)}})
S_2 D_2 \lbeq(tB2S3-inv-det-suppl-third) \\
& \hspace{3cm} - \lam^2 h_1(\lam)^{-1} D_2 S_2 (\tT_1(\lam)D_1)^2 S_2 D_2.
\notag 
\end{align}

\noindent 
{\rm (2)} If $H$ has singularity of the fourth kind, then 
the same as {\rm (1)} holds if the right sides of 
\refeqss(resd-a-suppl-third,11S22-suppl-third,tB2S3-inv-det-suppl-third) 
are replaced as follows:  
\bqn 
(\Mg(\lam^4)+S_1)^{-1} \approx 
\tD_{0}- 
\lam^2 \tD_{0}G_2^{(v)}\tD_{0} - 
\lam^4 \tg_2(\lam)D_0 G_4^{(v)}D_{0},   
\lbeq(resd-a-suppl) 
\eqn
where the variable $\lam$ is omitted from $\tD_0(\lam)$; 
\begin{multline}%
(\tB_1(\lam)+S_2)^{-1} \approx 
D_1+ \lam^2 h_1(\lam)^{-1}D_1 \tT_1(\lam)D_1 \lbeq(11S22-suppl) \\ 
+  \lam^4   h_1(\lam)^{-1} \tg_2(\lam) D_1 G_4^{(v)}D_1 
+ \lam^4 h_1(\lam)^{-2} D_1 (G_2^{(v)} D_1)^2, 
\end{multline}
where we wrote $D_1S_1= S_1 D_1 = D_1$ for simplicity;  
\bqn 
(\tB_2(\lam)+S_3)^{-1} 
\approx D_2 - D_2 F_{3}(\lam) D_2 + F_{3,sq}(\lam) \lbeq(tB2S3-inv-det-suppl)
\eqn 
where $F_{3}(\lam)$ and $F_{3,sq}(\lam)$ are as in \reflm(coll).
\edlm 
\bgpf We prove (2) and explain how to obtain (1) from (2) at the end of 
the proof. The proof is divided into several steps. 
Observe that $E_2(\lam) \in \Og^{(4)}_{S_1\HL}(\lam^{-2})$ by 
\reflm(b3-inv) (2).
\\[5pt]
(i)  Denote the right side of \refeq(resd-a-suppl) by $A_1(\lam)$ and let 
\[ 
\tA_1(\lam)= \tD_{0}(\lam)- \lam^2 \tD_{0}(\lam)G_2^{(v)} \tD_{0}(\lam) 
- \lam^4 \tg_2(\lam)\tD_{0}(\lam) G_4^{(v)}\tD_{0}(\lam)\,.
\]  
We write \refeq(MS1a) for $(\Mg(\lam^4)+S_1)^{-1}$ in the form  
\[
\tD_{0}(\lam)(1+ M^{(v)}_{2\to 6}(\lam)\tD_{0}(\lam))^{-1}
(1+ M^{(v)}_8(\lam)(1+ M^{(v)}_{2\to 6}(\lam)\tD_{0}(\lam))^{-1})^{-1} .
\]
Since 
$M^{(v)}_8(\lam)(1+ M^{(v)}_{2\to 6}(\lam)\tD_{0}(\lam))^{-1}
=\Og^{(4)}_{\Hg_2}(\lam^8 \log\lam)$ by \refeq(remainder-est)
\begin{align}
& 
(\Mg(\lam^4)+S_1)^{-1}= 
\tD_{0}(\lam)(1+ M^{(v)}_{2\to 6}(\lam)\tD_{0}(\lam))^{-1} + 
\Og^{(4)}_{\Hg_2}(\lam^8\log\lam) \notag \\
& = \sum_{j=0}^3 \tD_{0}(\lam)(-M^{(v)}_{2\to 6}(\lam)\tD_{0}(\lam))^j  
+\Og^{(4)}_{\Hg_2}(\lam^8\log\lam).\lbeq(m1-inv)
\end{align}
On substituting \refeq(m1-inv) in \refeq(Mess-E2) 
$\Og^{(4)}_{\Hg_2}(\lam^8\log\lam)$ produces 
$\Og^{(4)}_{\Lg^1 }(\lam^4)$ for $\Ng_{2,2}^{(v)}(\lam)$ which is GPR 
by \refprop(R-theo) and we may ignore it from \refeq(m1-inv).  
Then, the terms of order $\Og^{(4)}_{\Hg_2}(\lam^4)$ which appear 
in the sum on the right  of \refeq(m1-inv)  
produce $\Gg\Vg\Sg$ for $\Ng_{2,2}^{(v)}(\lam)$ and they may also be ignored.  
Thus, $(\Mg(\lam^4)+S_1)^{-1} \approx \tA_1(\lam)$. 
Since $\lam^4 \tg_2(\lam)\tD_{0}(\lam) G_4^{(v)}\tD_{0}(\lam)= 
\lam^4 \tg_2(\lam)D_0 G_4^{(v)}D_{0}+ \Og_{\Hg_2}(\lam^4)$, 
we may further replace $\tA_1(\lam)$ by $A_1(\lam)$ and 
$(\Mg(\lam^4)+S_1)^{-1} \approx A_1(\lam)$. 
This proves \refeq(resd-a-suppl-third). \\[3pt]
(ii) Let $\Ng_{2,{\rm red}}= \lam^{-2} M_v A_1(\lam)E_2(\lam) A_1(\lam) M_v$,  
which is equal to $\Ng_{2,2}^{(v)}(\lam)$ modulo GPR. Let   
$F_1(\lam)= A_1(\lam)S_1-S_1A_1(\lam)$. Then, 
\[
F_1(\lam)= h_1(\lam)[L,S_1] + 
\Og^{(4)}_{\Hg_2}(\lam^2) \in \Og_{\Hg_2}^{(4)}(h_1(\lam)). 
\] 
On replacing $A_1(\lam)S_1$ on the left by 
$S_1 A_1(\lam)+ F_1(\lam)$ and $S_1 A_1(\lam)$ on the right 
by $A_1(\lam)S_1 - F_1(\lam)$, $\Ng_{2,{\rm red}}(\lam)$ becomes 
\begin{multline*} 
\lam^{-2} S_1 M_v A_1(\lam)E_2(\lam)A_1(\lam)M_v S_1
- \lam^{-2} M_vS_1 A_1(\lam)E_2(\lam)F_1(\lam) M_v \\
+ \lam^{-2} M_vF_1(\lam)E_2(\lam)A_1(\lam)S_1M_v 
- \lam^{-2} M_vF_1(\lam)E_2(\lam) F_1(\lam)M_v. 
\end{multline*}
The point here is that 
the first term is sandwiched by $S_1M_v$ and $M_v S_1$  and 
and other terms carry at least one 
$F_1(\lam)\in \Og^{(4)}_{\Hg_2}(h_1(\lam))$ and, hence,  
by virtue of \reflm(s-low) and by \reflm(T-K), 
terms of order $\Og^{(4)}(\lam^4\log\lam)$ in the 
formulae which will appear 
for $(\tB_1(\lam)+S_2)^{-1}$  and $(\tB_2(\lam)+S_3)^{-1}$ 
in the following step (iii) produce GPR for $\Ng_{2,2}^{(v)}(\lam)$. \\[3pt]
(iii) We show \refeq(11S22-suppl). 
Since $L_1(\lam)\in \Og^{(4)}_{\Hg_2}(\lam^{2}\log\lam)$, \refeq(tB1-f) 
implies 
\bqn \lbeq(line-2)
(\tB_1(\lam)+S_2)^{-1} = D_1 + D_1 L_1(\lam) + D_1 L_1(\lam)^2 + 
\Og^{(4)}_{\Hg_2}(\lam^{6}(\log\lam)^3).  
\eqn 
Then \refeq(D1-L1) and \refeq(T4-recall) imply that modulo 
$\Og_{\Hg_2}^{(4)}(\lam^4\log\lam)$ 
\begin{gather*}
D_1 L_1(\lam)\equiv 
\lam^2 h_1(\lam)^{-1}D_1 \tT_1(\lam)D_1  
+  \lam^4   h_1(\lam)^{-1} \tg_2(\lam) D_1 G_4^{(v)}D_1, \\
D_1 L_1(\lam)^2 \equiv \lam^4 h_1(\lam)^{-2} D_1 (G_2^{(v)}D_1)^2,
\end{gather*} 
where we used $D_1S_1= S_1D_1= D_1$ and, hence, $D_1 D_0 = D_0 D_1 = D_1$.
Since $\Og_{\Hg_2}^{(4)}(\lam^4\log\lam)$ may be ignored from the right 
side of \refeq(line-2) by the result of (ii) above, 
we obtain \refeq(11S22-suppl).  

(iv) Since the term $\Og^{(4)}_{S_2\HL}(\lam^4 \log\lam)$ in 
$(\tB_2(\lam)+S_3)^{-1}$ may be ignored by (ii) above, 
\refeq(coll) implies \refeq(tB2S3-inv-det-suppl). 
This completes the proof of statement (2). 

If $B_3(\lam)^{-1}= \Og_{S_3\HL}^{(4)}(\lam^{-2}\tg_2(\lam)^{-1})$, then 
the proof of (2) implies that terms in the class $\Og_{\Hg_2}(\lam^4\log\lam)$ 
may be ignored from \refeq(m1-inv) and, hence, from \refeq(resd-a-suppl) 
and that those in  $\Og_{\Hg_2}(\lam^4(\log\lam)^2)$ from 
\refeq(11S22-suppl) and  \refeq(tB2S3-inv-det-suppl). The statement (1) 
follows.    
\edpf 

\subsection{\bf Proof of \refthb(main-rephrase) (3). Singularity 
of the third kind} 
We have $B_3(\lam)^{-1} \in \Og_{S_3\HL}^{(4)}(\lam^{-2}(\log\lam)^{-1})$. 
We apply (1) of \reflm(upto-2) to $\Ng_{2,2}^{(v)}(\lam)$. Denote the right of 
\refeqss(resd-a-suppl-third,11S22-suppl-third,tB2S3-inv-det-suppl-third) 
by $\tD_0(\lam)+a$, $D_1+ b$ and $D_2+c$ respectively so that 
$a\in \Og_{\HL}^{(4)}(\lam^2)$, 
$b$, $c \in \Og_{S_1\HL}^{(4)}(\lam^2\log\lam)$ are $\Gg\Vg\Sg$ and 
\begin{align} 
& \Ng_{2,2}^{(v)}(\lam) \equiv 
\lam^{-2}M_v (\tD_0(\lam)+a_l)S_1(D_1+ b_l)S_2 (D_2+c_l)S_3 \notag \\ 
& \hspace{1cm} \times B_3(\lam)^{-1}
S_3 (D_2+ c_r)S_2 (D_1+ b_r) S_1 (\tD_0(\lam)+a_r)M_v\,, \lbeq(Mess-3) 
\end{align} 
where we have added the indices $l$ and $r$ to 
distinguish the ones on the left  
and the right of $B_3(\lam)^{-1}$. 
We expand the right of \refeq(Mess-3). The result is $\Vg\Sg$; 
the terms which contain 
more than two of $\{a_l,a_r, b_l, b_r,c_l,c_r\}$ are in the class 
$\Og_{\Lg^1}^{(4)}(\lam^2(\log\lam)^2)$ and they are $\Gg\Vg\Sg$; 
those which contain two of them are also GPR because they 
are $\Gg\Vg\Sg$ if they contain $a_l$ or $a_r$ and, if not, 
they are of the form 
\[
M_v \tD_0(\lam)\Og_{S_1\HL}^{(4)}({ \log}\lam)\tD_0(\lam)M_v=  
M_v S_1\Og_{S_1\HL}^{(4)}({\log}\lam)S_1M_v +\Gg\Vg\Sg \,.
\]
Thus, modulo GPR, $\Ng_{2,2}^{(v)}(\lam)$ is the 
sum of the terms which contain at most one of $\{a_l,a_r, b_l, b_r,c_l,c_r\}$.
We denote the term which contains none of by $\Og(\emptyset)$ 
and those which contain $a_l$, etc. by $\Og(a_l)(\lam)$, etc. respectively and 
we estimate the operators produced by 
\refeq(sta0-low) with $\Og(\emptyset)(\lam), \Og(a_{l})(\lam), \dots$ 
in place of $\Qg_v(\lam)$ individually. 

Recall that the basis  of $S_3\HL$ is given by 
$\{\z_{r+1}, \dots, \z_m\}$, $r < m$. 
By virtue of \refeq(B3-inv-remainder) we have   
\bqn \lbeq(B3-inv)
B_3(\lam)^{-1}= \sum_{j,k=r+1}^{m}\lam^{-2}(\log\lam)^{-1}
c_{jk}(\lam)\z_j \otimes \z_k 
\eqn 
with $c_{jk}(\lam)\in \Og_{\C}^{(4)}(1)$ for small $\lam>0$, 
$j,k=r+1, \dots, m$.  

\bglm \lblm(L-good) Operators 
$\Og(a_{l})(\lam), \Og(b_l)(\lam)$ and $\Og(c_l)(\lam)$ are GPRs. 
\edlm 
\bgpf The proof uses the fact that 
\begin{align*}
\Og(a_l)(\lam) & = - h_1(\lam) M_v L_0 G_2^{(v)} S_3 B_3(\lam)^{-1}S_3 M_v, \\
\Og(a_l)(\lam) & = - h_1(\lam) M_v L_0 G_2^{(v)} S_3 B_3(\lam)^{-1}S_3 M_v, \\
\Og(c_l)(\lam) & = M_vS_2 D_2 S_2 \{ 
T_{4,l}- G_2^{(v)}(S_1 D_1S_1 L_0)G_2^{(v)} \\ 
& + h_1(\lam)G_2^{(v)}(L_0S_1D_1S_1 L_0 -L_0) G_2^{(v)} \}S_3 
B_3(\lam)^{-1}S_3 M_v 
\end{align*}
all have $S_3 M_v$ 
on the right ends which cancels the singularity $\lam^{-2}$ 
and it will be similar for all of them. Thus, we 
prove the lemma only for $\Og(a_{l})(\lam)$ and comment on how to modify 
the argument for others at the end of the proof. 

We set, for $j,k=r+1, \dots, m$, 
$\m_{jk}(\lam)=h_1(\lam)(\log\lam)^{-1}c_{j,k}(\lam)$, 
$\r_j(x)= (M_vL_0 G_2^{(v)}\z_j)(x)$ and  
$\ph_k(x)= (v\z_k)(x)$.  
Then, $\m_{jk}(\lam)$ are GMU,   
$\r_j, \ \ph_k \in \ax^{-8}\la \log |x|\ra^{-1} L^1(\R^4)$,    
$\la x^\a, \ph_k \ra=0$ for $|\a|\leq 1$ and 
\[
\Og(a_{l})(\lam)
= -\lam^{-2} \sum_{j,k=r+1}^m \m_{jk}(\lam) (\r_j \otimes \ph_k).  
\]
Thus, with $u_{jk}= \chi_{\leq{a}}(|D|)\m_{jk}(|D|)u$, 
operator produced by $\Og(a_{l})(\lam)$ becomes 
\bqn \lbeq(9-Ijk)
-\sum_{j,k=r+1}^m I_{jk}u=
-\sum_{j,k=r+1}^m 
\int_0^\infty R^{+}_0 (\lam^4)(\r_j \otimes \ph_k) \Pi(\lam) u_{jk} \lam  d\lam. 
\eqn 
We prove that all $I_{jk}$ are GOP. Since the proof is similar, we prove this 
only for $I_{(r+1)(r+1)}$ omitting the index $j=k=r+1$ but writing $\tilde{u}$ 
for $u_{(r+1)(r+1)}$ with $u \in \Dg_\ast$. Note that 
the integral by $d\lam$ is then over a compact subset of $(0,\infty)$. 

Since $\la x^\a, \ph \ra=0$ for $|\a|\leq 1$, 
$\Pi(\lam) \tu(z)$  may be replaced by 
\refeq(Pi-sec-a) and its the first term does not contribute. Hence, we have  
\bqn \lbeq(se-inner)
\la \ph, \Pi(\lam) \tu \ra= 
\sum_{l,m=1}^4 \lam^2 \int_0^1 (1-\th) 
\left(\int_{\R^4}\ph_{lm}(z)(\Pi(\lam)\t_{-\th{z}}\tu_{lm})(0) dz\right) 
d\th, 
\eqn 
where $\ph_{lm}(z)=z_lz_m \ph(z)$ and $\tu_{lm}= R_l R_m \tu$, 
$l,m=1, \dots, 4$, which we substitute in  
\refeq(9-Ijk) ($j=k=r+1$ which we will omit). Then, after the change 
of order of integrations, $Iu(x) $ becomes  
$\sum_{l,m=1}^4 \int^1_0(1-\th) I_{(lm)}(\th)u(x) d\th$ , where 
$I_{(lm)}(\th)u(x)$ is equal to 
\[
\iint_{\R^8} \r(y)\ph_{lm}(z) \tau_y \left(\int_0^\infty\Rg_\lam(x)
\Pi(\lam)\t_{-\th{z}}\tu_{lm}(0) \lam ^3  d\lam\right)dydz.
\] 
The integral inside the parenthesis is equal to $K \t_{-\th{z}}\tu_{lm}$ 
(recall \refeq(K-first)) and \refeq(T-K) implies that for any $1<p<\infty$
\[
\|I_{(lm)}(\th)u\|_p \leq C \|\r\|_1 \|\ph_{lm}\|_1 
\|\tu_{lm}\|_p  \leq C\|u\|_p\,. 
\]
This proves that $I_{jk}$, $j,k=1, \dots,m$ are GOP and
$\Og(a_{l})(\lam)$ is a GPR. 

Since $h_1(\lam)$ does not play any role except that it is a GMU, 
the entire argument for $\Og(a_{l})(\lam)$ applies for proving that 
$\Og(b_{l})(\lam)$ 
is GPR. The argument for $\Og(c_{l})(\lam)$ is similar. 
The only point we have to note is that, instead of $h_1(\lam)$ 
in $\Og(a_{l})(\lam)$, $\Og(c_{l})(\lam)$ contains the singularity 
$\tg_2(\lam)$  which is hidden in  
$T_{4,l}= G_4^{(v)}\tg_2(\lam) + G_{4,l}^{(v)}$, however, this is 
harmless since $\tg_2(\lam) (\log \lam)^{-1} c_{jk}(\lam)$ is still a GMU. 
\edpf

We next prove the followng lemma for 
\begin{align*}
\Og(a_r)(\lam) & = - h_1(\lam) M_v S_3 B_3(\lam)^{-1}S_3 G_2^{(v)} L_0 M_v, \\
\Og(b_r)(\lam)& = M_v S_3 B_3(\lam)^{-1}
S_3  G_2^{(v)} L_0 S_1 (D_0+ h_1 (\lam)L_0)M_v, \\
\Og(c_r)(\lam) & = M_v S_3 B_3(\lam)^{-1}S_3
\{T_{4,l}- G_2^{(v)}(L_0 S_1 D_1 S_1 ) G_2^{(v)} \\ 
& + h_1(\lam) G_2^{(v)}(L_0 S_1 D_1 S_1 L_0 - L_0)G_2^{(v)}\} S_2 D_2 S_2 M_v.
\end{align*}

\bglm \lblm(R-1to2) Operators produced by 
$\Og(a_r)(\lam), \Og(b_r)(\lam)$ and $\Og(c_r)(\lam)$ are bounded 
in $L^p(\R^4)$ for $1<p<2$. 
\edlm 

For the proof of \reflm(R-1to2) we use the following lemma.  
\bglm \lblm(tZg-4)
Let $\z\in S_3\HL$ and $\r \in L^1(\R^4)$. Then $\Zg^{(r)}$ defined by 
\bqn \lbeq(Zg-4-simple)
\Zg^{(r)} u= \int_0^\infty R^{+}_0 (\lam^4)(v\z \otimes \r) \Pi(\lam) u 
\lam \chi_{\leq{a}}(\lam)d\lam, \ u \in \Dg.
\eqn  
is bounded in $L^p(\R^4)$ for $1<p<2$. Moreover, $\Zg^{(r)}$ is unbounded for 
$2\leq {p}\leq \infty$ unless $\int_{\R^4} \r(x) dx=0$
\edlm 
\bgpf  The proof patterns after that of \reflm(Bad). We split 
\bqn \lbeq(ZXl)
\Zg^{(r)}u = \chi_{\leq 4a}(|D|)\Zg^{(r)} u + 
\chi_{>4a}(|D|)\Zg^{(r)} u =\colon \Zg^{(r)}_{\leq 4a}u+ \Zg^{(r)}_{>4a}u. 
\eqn 
(1) We first show that $\Zg^{(r)}_{>4a}$ 
is bounded in $L^p(\R^4)$ for $1<p <2$ and is unbounded for $2\leq p\leq \infty$ 
which implies in particular that $\Zg^{(r)}$ is unbounded 
for $2\leq p\leq \infty$  
since $\chi_{>4a}(|D|)$ is bounded in $L^p(\R^4)$ for all $1\leq p \leq \infty$.  
\[
\Zg^{(r)}_{>4a}u(x)= \int_0^\infty \chi_{>4a}(|D|) 
R^{+}_0 (\lam^4)(v\z \otimes \r) \Pi(\lam) u 
\lam \chi_{\leq{a}}(\lam)d\lam \,. 
\] 
Let $\ph= v\z$ and $\m(\xi)= \chi_{>4a}(\xi)|\xi|^{-4}$. Then, 
$\mu(|D|)\ph \in L^p(\R^4)$ for all $1\leq p \leq \infty$, 
$\m(|D|)$ is GOP (cf. \reflm(mu)) and by \refeq(chge2a) 
\[
\chi_{>4a}(|D|) R^{+}_0 (\lam^4)\ph(x)
= \m(|D|)\ph(x) + \m(|D|)\lam^4 R^{+}_0 (\lam^4)\ph(x)\,. 
\]
It follows that 
$\Zg^{(r)}_{>4a}u= \Zg^{(r,1)}_{>4a}u + \Zg^{(r,2)}_{>4a}u$, 
where 
\begin{align*}
& \Zg^{(r,1)}_{> 4a} u
= \m(|D|)\ph(x) \ell(u), \ \ \ell(u) \colon = \int_0^\infty \la \r, 
\Pi(\lam)u\ra\lam \chi_{\leq{a}}(\lam)d\lam,
\\
& \Zg^{(r,2)}_{>4a} u= 
\m(|D|)\int_0^\infty R^{+}_0 (\lam^4)(v\z \otimes \r) \Pi(\lam) u 
\lam ^5\chi_{\leq{a}}(\lam)d\lam.
\end{align*}
Then, \reflm(T-K) implies 
$\|\Zg^{(r,2)}_{>4a}u\|_p \leq C_p \|\ph\|_1\|\r\|_1\|u\|_p$ for $1<p<\infty$. 
Assume $\r \in C_0^\infty(\R^4)$ first. Then, changing the order of integrations, 
letting $(\lam,\w)$ be the polar coordinates of $\xi \in \R^4$ 
and using Parseval's identity, we obtain  as in \refeq(Parseval) that 
\[
\ell(u)
=\frac1{(2\pi)^2}\int_{\R^8} \r(y) u(x) J_a(x-y) dx dy, 
J_a (x)= \Fg\left(\frac{\chi_{\leq{a}}(|\xi|)}{|\xi|^{-2}}\right)(x),
\] 
Here $J_a(x)$ is smooth 
and $|J_a (x)|\leq C \ax^{-2}$. It follows by Young's inequality 
that $|\ell(u)|\leq C \|\r\|_1 \|J\|_{p'}\|u\|_p$ 
for $1\leq p <2$ and $p'=p/(p-1)$ and   
$\|\Zg^{(r,1)}_{>4a} u\|_p \leq C \|\r\|_1 \|u\|_p$. Thus, 
$\|\Zg^{(r)}_{>4a} u\|_p \leq C \|\r\|_1 \|u\|_p$. \\[3pt]
(2) We note that $v \z \not=0$ as otherwise $\z=0$ by virtue of 
\reflm(resonance). 
Then, the repetition of the argument of \reflm(Bad) implies  
$\Zg^{(r)}$ is unbounded in $L^p(\R^4)$ for all $2\leq p <\infty$ 
since 
\[
\int_{\R^4} J_a (x-y) \r(y) dy \not\in L^{p'}(\R^4), \quad 1\leq p'=p/(p-1)<2   
\]
unless $\check{\r}(0)=0$. We omit the details. \\[3pt]
(3) Denote $\ph(x)= v(x) \z(x)$.  We next show that 
\[
\Zg^{(r)}_{\leq 4a}u(x) = 
\int_0^\infty \chi_{\leq {4a}}(|D|)
R^{+}_0 (\lam^4)(\ph \otimes \r) \Pi(\lam) u 
\lam \chi_{\leq{a}}(\lam)d\lam
\]
satisfies for $1<p<2$ that 
$\|\Zg^{(r)}_{\leq 4a}u\|_p \leq C \|\ax^2 \ph\|_1 \|\r\|_1 \|u\|_p$ 
which will finish the proof of the lemma.  
We may assume $u \in \Dg_\ast$ and $\ph, \r \in C_0^\infty(\R^4)$. 
Let $\ph_{kl}(y)=y_k y_l \ph(y)$ for $k,l=1, \dots, 4$.  
Then, since $\la x^\a, \ph \ra=0 $ for $|\a|\leq 1$,   
\[
\hat{\ph}(\xi)= \sum_{k,l=1}^4 \frac{-\xi_k \xi_l}{(2\pi)^2} \int_0^1 (1-\th)
\left(\int_{\R^4} e^{-iy\xi\th} \ph_{kl}(y) dy \right)d\th 
\]
and $\chi_{\leq 4a}(|D|)R_0^{+}(\lam^4)\ph(x) $ is equal to  
\[ 
\lim_{\ep \downarrow 0}\sum_{k,l=1}^4 \int_0^1 (1-\th)
\left(\frac{-1}{(2\pi)^4} 
\iint_{\R^8} 
\frac{e^{i(x-\th y)\xi}\xi_k \xi_l\chi_{\leq 4a}(|\xi|)}
{\xi^4-(\lam+i\ep)^4} \ph_{kl}(y) dy d\xi\right) d\th.
\]
After this, we repeat the argument in the proof of \reflm(Bl-p) after 
\refeq(hatph) with $\xi_k\xi_l$ replacing $\xi_m$ and 
\bqn \lbeq(K1+K2)
\frac{|\xi|^2}{\xi^4-(\lam+i\ep)^4}= \frac{1}{2(\xi^2+(\lam+i\ep)^2)} + 
\frac{1}{2(\xi^2-(\lam+i\ep)^2)},
\eqn 
in place of \refeq(mult-1).  Eqn. \refeq(K1+K2) and 
the definition of $\Gg_{z}(x)$  (cf. \refeq(Green)) yield 
\[
\frac{1}{(2\pi)^4} \int_{\R^4} 
\frac{e^{ix\xi}|\xi|^2}{\xi^4-(\lam+ i\ep)^4} d\xi 
= \frac{1}{2}\left(\Gg_{i\lam-\ep}(x)+ \Gg_{\lam+i\ep}(x)\right) 
\]
and, defining $\c_{kl,a}(D)= R_j R_k \chi_{\leq 4a}(|D|)$, we conclude that  
$\Zg^{(r)}_{\leq 4a}u$ is equal to $\sum_{j,k}^4 \int_0^1 (1-\th) d\th$ of
\[
\int_{\R^4} \ph_{kl}(y)\t_{\th{y}}
\left(\c_{kl,a}(D)\int_0^\infty (\Gg_{i\lam}(x)+ \Gg_{\lam}(x))
\la \r , \Pi(\lam) u \ra \lam \chi_{\leq a}(\lam)d\lam\right) dy.
\]
Here the inner most integral is equal to 
\bqn \lbeq(F-Klas)
Fu(x) = \int_{\R^4}\r(z)(K_1 + K_2)(\tau_{-z}\chi_{\leq a}(|D|)u)(x)dz  
\eqn  
and, by \reflms(3-1,3-2), $\|Fu\|_ p \leq C \|\r\|_1 \|u\|_p$ for $1<p<2$.
Then, Minkowski's inequality implies the desired estimate. 
This completes the proof of lemma.  
\edpf

\paragraph{\bf Proof of \reflmb(R-1to2)}
Operators $\Og(a_r)(\lam), \Og(b_r)(\lam)$ and $\Og(c_r)(\lam)$ all 
have $M_v S_3$ 
on the left ends. The proof is similar and we prove the lemma only 
for $\Og(a_r)(\lam)$. The modification for others is obvious.  
We use the notation in the proof of \reflm(L-good). Let $u \in \Dg_\ast$. 
Substitute $\Og(a_r)(\lam)$ for $\Qg_v(\lam)$ in \refeq(sta0-low) 
and use \refeq(mult). Then, \refeq(sta0-low) becomes 
\bqn \lbeq(9-Jjk)
\sum_{j,k=r+1}^m J_{jk}u, \quad J_{jk}u = \int_0^\infty R^{+}_0 (\lam^4)
(v\z_j \otimes \r_k) 
\Pi(\lam) u_{a,jk} \lam  d\lam\,,
\eqn 
where $\r_k(x)= (M_vL_0 G_2^{(v)}\z_k)(x)$ and 
$u_{a,jk}= \chi_{\leq{a}}(|D|)\m_{jk}(|D|)u$.  
Then $\r_k \in L^1(\R^4)$ for $k=r+1, \dots, m$ and \reflm(tZg-4) 
implies that $\|J_{jk} u\|_p \leq C_p \|\ax^2 v\z_j\|_1 
\|\r_k\|_1 \|u\|_p$ for $1<p<2$. 
The lemma follows. 
\qed  

\paragraph{\bf Completion of the proof of \refthb(main-rephrase)(3)} 
The following lemma completes the proof of \refth(main-rephrase).

\bglm The operator produced by $\Og(\emptyset)(\lam)$ 
is bounded in $L^p(\R^4)$ for $1<p<2$.
\edlm  
\bgpf  
We have $ \Og(\emptyset)(\lam) = \lam^{-2}M_v S_3 B_3(\lam)^{-1}S_3 M_v$ 
and the operator in question is equal to 
\bqn \lbeq(Zg-1)
\Zg_0 u= \int_0^\infty R^{+}_0 (\lam^4) M_v S_3 B_3(\lam)^{-1}S_3M_v \Pi(\lam) u 
\lam \chi_{\leq{a}}(\lam)d\lam 
\eqn  
and \refeq(B3-inv) implies  $\Zg_0 u = \sum_{j,k=r+1}^m J_{jk}u$ with  
\bqn
\lbeq(Z0sum)
J_{jk}u= 
\int_0^\infty R^{+}_0 (\lam^4)|v\z_j\ra \la v\z_k, \Pi(\lam)\m_{jk}(|D|)u\ra  
\lam ^{-1}\chi_{\leq a}(\lam)d\lam, 
\eqn 
and $\m_{jk}(\lam)\colon =(\log\lam)^{-1}c_{jk}(\lam)$ are GMU. 
We prove $\|J_{(r+1)(r+1)}u\|_p\leq C\|u\|_p$ for $1<p<2$. 
We omit indices and denote $\tu= \mu_{(r+1)(r+1)}(|D|)u$. 
The proof for others is similar. 
Let $\ph= v\z$, $\ph_{lm}(x)=x_lx_m \ph(x)$ and $\tu_{lm}= R_lR_m \tu$,  
$1\leq l, m\leq 4$.
We substitute \refeq(se-inner) for 
$\la \ph, \Pi(\lam)\tu\ra$. Then, $Ju$ becomes 
$\sum_{l,m=1}^4 \int_0^1 (1-\th) d\th$ of 
\begin{align*}
J_{(lm),\th}(x)& = \int_0^\infty R^{+}_0 (\lam^4)|v\z \ra 
\la \ph_{lm}(z), \Pi(\lam)\t_{-\th{z}} \tu_{lm}(0)\ra \lam \chi_{\leq{a}}(\lam)d\lam \\
& = \int_0^\infty R^{+}_0 (\lam^4)|v\z \ra 
\la \ph_{lm,\th}(z), \Pi(\lam)\tu_{lm}(z)\ra \lam \chi_{\leq{a}}(\lam)d\lam ,
\end{align*}
where $\ph_{lm,\th}(z)= \th^{-4}\ph_{lm}(\th^{-1}z)$. Then, applying 
\reflm(tZg-4) for $\r=\ph_{lm,\th}(z)$ and 
$u=R_l R_m \tu$, we obtain 
that for $1<p<2$
\[
\|J_{(lm),\th} u\|_p \leq C \|\ph_{lm,\th}\|_1 \|\tu\|_p 
\leq C_1 \|\ph_{lm}\|_1 \|u\|_p 
\]
with constants $C$ and $C_1$ which bare independent of $0<\th<1$. 
Hence, $J$ is bounded in $L^p(\R^4)$ for $1<p<2$ 
and the lemma follows. 
\edpf

\subsection{\bf Proof of \refthb(main-rephrase)(4)} 
By virtue of \refcor(Q=Z), it suffices to study $\Zg$ 
of \refeq(Q=Z) with $\Ng_{2,2}^{(v)}(\lam)$ of 
\refeq(Mess-E2) and 
$B_3(\lam)^{-1}$ is given by \refeq(B3-inv-for4) 
if $T_3\not=0$ and by \refeq(B3-inv-for4-suppl) if $T_3=0$. 
Then, the following lemma implies \refth(main-rephrase) (4) 
because $\Zg$  is bounded in $L^p(\R^4)$ for $1<p<2$ if 
$B_3(\lam)^{-1}= \Og^{(4)}_{S_3\HL}(\lam^{-2}(\log\lam)^{-1})$ 
by the result of the previous subsection and, hence, 
the second term in the right of \refeq(B3-inv-for4) produces 
an operator bounded in $L^p(\R^4)$ for $1<p<2$. 

\bglm \lblm(last-lemma)
Suppose that $B_3(\lam)^{-1}
= - \lam^{-2} S_4 T_4^{-1}S_4+\Og_{S_4\HL}^{(4)}((\log\lam)^2)$ 
as in \refeq(B3-inv-for4-suppl). 
Then, $\Zg$ is bounded in $L^p(\R^4)$ for $1<p<4$. 
\edlm  

We prove \reflm(last-lemma) by a series of lemma. 
The strategy is similar to but is more complicated than 
that of the previous subsection. 
Denote the right sides of 
\refeqss(resd-a-suppl,11S22-suppl,tB2S3-inv-det-suppl) 
by $\tD_0(\lam)+\ta$, $D_1+ \tb$ and $D_2+\tc$ respectively and 
$B_3(\lam)^{-1}= - \lam^{-2} S_4 \tB_3(\lam)^{-1}S_4$ with 
\bqn \lbeq(tB3) 
\tB_3(\lam)^{-1} = T_4^{-1} +\Og_{S_4\HL}^{(4)}(\lam^2(\log\lam)^2).
\eqn 
We recall 
\bqn \lbeq(order)
\ta\in \Og^{(4)}_{\Hg}(\lam^2), \quad
\tb\in \Og^{(4)}_{\Hg}(\lam^2 \log \lam), \quad
\tc\in \Og^{(4)}_{\Hg}(\lam^2 \log \lam). 
\eqn 
Then, \reflm(upto-2) implies that modulo GPR 
\begin{align} 
& \Ng_{2,2}^{(v)}(\lam)\equiv \lam^{-4} 
J_l(\lam) \tB_3(\lam)^{-1}J_r(\lam), \lbeq(M42) \\ 
& J_l(\lam)= 
M_v (\tD_0(\lam)+\ta_l)S_1(D_1+ \tb_l)S_2 (D_2+\tc_l)S_4, 
\lbeq(M42-l) \\ 
& J_r(\lam)= S_4 (D_2+ \tc_r)S_2 (D_1+ \tb_r) S_1 
(\tD_0(\lam)+\ta_r)M_v.
\lbeq(M42-r)
\end{align}
where we have added the indices $l$ and $r$ as previously. 
Since $G_2 M_v S_4= S_4 M_v G_2=0$ and 
$\tT_1(\lam)S_4= S_4 \tT_1(\lam)=0$ 
by \refeqs(S4G2,S4tT), we have 
\begin{align}
& S_4(D_2 F_3(\lam)D_2) \equiv  S_4 
(\lam^2 T_{4,l}(\lam)+ \lam^4 (h_1^{-1}\tg_2)(\lam) 
G_4^{(v)}S_1 G_2^{(v)})S_2 D_2, \notag  \\ 
& S_4 F_{3,sq}(\lam) \equiv 
\lam^4 \tg_2(\lam)S_4 G^{(v)}_{4}\tilde{D}_2 
(\tg_2(\lam)G_{4}^{(v)}+
h_1(\lam)^{-1}(\tT_1(\lam)D_1)^2)S_2 D_2 \lbeq(F3-simp-r)
\end{align} 
modulo $S_4 \Og^{(4)}_{\Hg_2}(\lam^4 \log\lam)$ and 
\begin{align} 
& (D_2 F_3(\lam)D_2)S_4 
\equiv D_2 S_2 (\lam^2 T_{4,l}(\lam) 
+ \lam^4 (h_1^{-1}\tg_2)(\lam) 
G_2^{(v)}\tilde{D}_1 G_4^{(v)})S_4,  \notag \\
&  F_{3,sq}(\lam)S_4 \equiv 
\lam^4 \tg_2(\lam)D_2 S_2 (\tg_2(\lam)G^{(v)}_{4}+ 
h_1(\lam)^{-1}(\tT_1(\lam)D_1)^2)\tilde{D}_2 G^{(v)}_{4}S_4 
\lbeq(F3-simp-l) 
\end{align}
modulo $\Og^{(4)}_{\Hg_2}(\lam^4 \log\lam)S_4$, 
where $\tilde{D}_1= S_1 D_1 S_1$ and $\tilde{D}_2= S_2 D_2 S_2$. 
Since $S_4 \Og^{(4)}_{\Hg_2}(\lam^4 \log\lam)$ in $\tc_l$ and 
$\Og^{(4)}_{\Hg_2}(\lam^4 \log\lam)S_4$ in $\tc_r$ produce 
GPR for $\Ng_{2,2}^{(v)}(\lam)$ by (ii) 
of the proof of \reflm(upto-2), 
we may and do replace $S_4(D_2 F_3(\lam)D_2)$, $(D_2 F_3(\lam)D_2)S_4$, 
etc. by the right sides of the corresponding formulas above 
without changing $\Ng_{2,2}^{(v)}(\lam)$ modulo GPR. 
We expand the right of \refeq(M42). The result is a $\Vg\Sg$. 

\bglm \lblm(three)
The sum of the terms in the expansion of \refeq(M42) 
which contain three or more of 
$\{\ta_l,\ta_r, \tb_l, \tb_r,\tc_l, \tc_r\}$ is GPR. 
\edlm 
\bgpf The sum is $\Gg\Vg\Sg$ 
since it is $\Vg\Sg$ and belongs to 
$\Og_{\Lg^1}^{(4)}(\lam^2(\log\lam)^3)$ by 
virtue of \refeq(order). Hence, it is  GPR by \refcor(vari-sep).
\edpf 

Thus, we consider only the terms which 
contain {\it at most two} 
from $\{\ta_l,\ta_r, \tb_l, \tb_r,\tc_l, \tc_r\}$ and we use 
the notation of \S\S 9.4, viz. 
$\Og(\emptyset)(\lam)$ is the term which contains none of them, 
$\Og(\ta_r)(\lam), \Og(\ta_l)(\lam)$, etc. are those 
which contain $\ta_r, \ta_l$, etc only and 
$\Og(\ta_r, \tb_l)(\lam)$, etc. are those which contain $\ta_r$ and $\tb_l$, 
etc. respectively.

\bglm \lblm(none) 
The operator  produced $\Og(\emptyset)(\lam)$ 
is bounded in $L^p(\R^4)$ for $1<p<4$. 
\edlm 
\bgpf We have $\Og(\emptyset)(\lam)=\lam^{-4}S_4 \tB_3(\lam)^{-1} S_4$ 
and it produces   
\[
Z_{4-4}u= \int_0^\infty R^{+}_0 (\lam^4)M_v
S_4 \tB_3(\lam)^{-1}S_4 M_v \Pi(\lam) u \lam^{-1} 
\chi_{\leq{a}}(\lam)d\lam. 
\]
Let 
$\{\z_{r'}, \dots, \z_m\}$, $r'\geq r+1$ be the orthonormal basis of 
$S_4 \HL$ and   
\bqn \lbeq(Ejdef)
E_j u(x)= \int_0^\infty (R^{+}_0 (\lam^4)(v\z_j)(x) 
\Pi(\lam)u(0) \lam^2 \chi_{\leq{a}}(\lam)d\lam, \ 
u \in \Dg_\ast 
\eqn 
for $j=r', \dots, m$. Then, $E_j$ is equal to 
$W_B u$ of \refeq(Bdef) with $\z_j$ in place of $\z$ 
and $\{E_j\}_{j=r', \dots, m}$ are bounded in $L^p(\R^4)$ 
for $1<p<4$ by \reflms(Bad,Bl-p). 
Via $\{\z_{r'}, \dots, \z_l\}$, we express 
\bqn \lbeq(matrix-T4)
S_4 \tB_3(\lam)^{-1}S_4= \sum_{j,k=r'}^m \tilde{t}_{jk}(\lam) 
\z_j \otimes \z_k, 
\quad 
\tilde{t}_{jk}(\lam) =t_{jk}+\Og_{\C}^{(4)}(\lam^2(\log\lam)^2), 
\eqn 
where $t_{jk}= \la \z_j, T_{4}^{-1}\z_k\ra$ and $\{\tilde{t}_{jk}(\lam)\}$ 
are GMU. 
Since $\z\in S_4\HL$ satisfies $\la x^\a, v\z\ra=0$ for 
$|\a|\leq 2$ (cf. \reflm(characterization)), 
Taylor expansion of $e^{i\lam{z}\w}$ 
to the third order in \refeq(reason) or \refeq(Pi-sec-a) 
implies that $\la v\z, \Pi(\lam) u \ra $ for 
$\z\in S_4\HL$ may be expressed as 
\bqn \lbeq(9-linear-funct)
\lam^3 
\sum_{|\a|=3}C_\a \int_0^1(1-\th)^2  \left(\int_{\R^4} z^\a (v\z)(z) \Pi(\lam)
(R^\a \t_{-\th{z}} u)(0) dz\right) d\th  
\eqn 
where $C_\a$ are unimportant constants and $R^\a= R_1^{\a_1}\cdots R_4^{\a_4}$ 
for $\a=(\a_1, \dots, \a_4)$. 
It follows that $Z_{4-4}u(x)$ is equal to 
\bqn \lbeq(z44)
\sum_{j,k=r'}^m \sum_{|\a|=3}C_\a \int_0^1(1-\th)^2
\left(\int_{\R^4}
z^\a (v\z_k)(z) E_j (R^\a \t_{-\th{z}}\tilde{t}_{jk}(|D|))
u(x) dz\right) d\th  
\eqn 
and the Minkowski inequality implies that for $1<p<4$  
\bqn \lbeq(z44-result)
\|Z_{4-4}u(x) \|_p \leq C 
\sum_{j,k=r'}^m \|\la z \ra^3(v\z_k)\|_1 \|u\|_p \|E_j\|_{\Bb(L^p)}\,.
\eqn 
This proves the lemma. 
\edpf 

\bglm \lblm(Jl-S4) 
The operator produced by the sum of the terms 
which contain none of $\{\ta_r, \tb_r, \tc_r\}$ 
is bounded in $L^p(\R^4)$ for $1<p<4$. 
\edlm 
\bgpf The sum of the terms,  
$\lam^{-4}J_l(\lam) \tB_3(\lam)^{-1}S_4 M_v$, 
produces 
\[
Z_{l-4}u= \int_0^\infty R^{+}_0 (\lam^4)
J_l(\lam) \tB_3(\lam)^{-1}S_4 M_v \Pi(\lam) u \lam^{-1} 
\chi_{\leq{a}}(\lam)d\lam
\]
We expand $J_l(\lam)$ of \refeq(M42-l). \\[3pt]
(i) The term which contains none of 
$\{\ta_l, \tb_l, \tc_l\}$ produces $Z_{4-4}$ which is bounded in 
$L^p(\R^4)$ for $1<p<4$. \\[3pt]
(ii) The terms which contain at least one of $\{\ta_l, \tb_l, \tc_l\}$ 
are GPR. To see this we argue as follows. We first note that 
$\la v\z, \Pi(\lam)u\ra$ for $\z\in S_4$ is equal 
to \refeq(9-linear-funct). We then define 
$\tE_j u(x)$ by \refeq(Ejdef) with $J_l(\lam)\z_j$ in place of $v\z_j$. 
Then, $Z_{l-4}u(x)$ becomes \refeq(z44) with $\tE_j$ replacing $E_j$. 
For $j=r', \dots, m$, $J_l(\lam)\z_j= \sum_{n=1}^8 \lam \s_n(\lam) f_{jn}$ 
with $f_{jn}\in L^1(\R^4)$ and $\s_n \in \Og^{(4)}_{\C}(\lam(\log\lam)^2)$ 
by virtue of \refeq(order). It follows that $\tE_j$ becomes
\[
\tE_ju(x)= \sum_{n=1}^8 \int_0^\infty (R_0^{+}(\lam^4)f_{jn})(x) 
(\Pi(\lam)\s_j(|D|)u)(0)
\lam^3\chi_{\leq{a}}(\lam) d\lam 
\]
for $j=r', \dots, m$ and they are GOP by \reflm(T-K). 
Hence, we see that $Z_{l-4}$ is GOP as in \refeq(z44-result) but 
replacing $\|E_j\|_{\Bb(L^p)}$ by $\|\tE_j\|_{\Bb(L^p)}$ and 
$1<p<4$ by $1<p<\infty$.
\edpf  

Next we consider the operators produced by the terms which contain two of 
$\{\ta_r, \tb_r,\tc_r\}$ and none of $\{\ta_l, \tb_l,\tc_l\}$. 

\bglm \lblm(ar-brcr) 
The operator $\Og(\ta_r, \tb_r)(\lam)$ and $\Og(\ta_r, \tc_r)(\lam)$ 
are GPR.
\edlm 
\bgpf (1) We have $\Og(\ta_r, \tb_r)(\lam)
=\lam^{-4} M_v S_4 \tB_3(\lam)^{-1}S_4 \tb_r S_1 \ta_r M_v$.  
By \refeqs(S4G2,S4tT),   
$S_4 \tb_r= \lam^4 h_1(\lam)^{-1} \tg_2(\lam) S_4 G_4^{(v)} D_1$ 
and $\ta_r \in \Og_{\Hg_2}^{(4)}(\lam^2)$.
Hence, $\Og(\ta_r, \tb_r)(\lam)\in \Og_{\Lg^1}^{(4)}(\lam^6(\log\lam)^2)$ 
and it is GPR by \refprop(R-theo). \\[3pt]
(2) We have 
$\Og(\ta_r, \tc_r)(\lam)
=\lam^{-4} M_v S_4 \tB_3(\lam)^{-1} S_4 \tc_r S_2 \ta_r M_v $.  
Then, we apply \refeq(F3-simp-l) for $S_4\tc_r$, use that    
$\ta_r \in \Og_{\Hg_2}^{(4)}(\lam^2)$ and the matrix representation 
\refeq(matrix-T4) of $S_4 \tB_3(\lam)^{-1} S_4$. Then,  
we have modulo $\Gg\Vg\Sg$ that 
\begin{align*}
& \Og(\ta_r, \tc_r)(\lam) \equiv 
\tg_2(\lam) M_v S_4 \tB_3(\lam)^{-1} S_4 G_{4}^{(v)}S_2 G_2^{(v)}D_0 M_v \\
& =   \tg_2(\lam) \sum_{j,k=r'}^m \tilde{t}_{jk}(\lam) (v\z_j) \otimes \r_k, 
\quad \r_k= M_v D_0 G_2^{(v)}S_2 G_4^{(v)}\z_k. 
\end{align*}
We evidently have 
$\az \r_k \in L^1(\R^4)$ for $k=1, \dots, 4$ and, 
since $D_0= QD_0 $ and $Qv=0$, that $\int_{\R^4} \r_k(z) dz =0$ 
Then, 
\reflm(s-low) implies that $\Og(\ta_r, \tc_r)(\lam)$ is GPR 
since $\tilde{t}_{jk}(\lam)$ are GMU. 
\edpf 

\bglm \lblm(oturi-lp4)
The operator produced by $\Og(\tb_r, \tc_r)(\lam)$ is bounded in 
$L^p(\R^4)$ for $1<p<4$.
\edlm 
\bgpf By virtue of \refeq(F3-simp-l), we have 
\[ 
\Og(\tb_r, \tc_r)(\lam)=
\lam^{-4} M_v S_4 \tB_3(\lam)^{-1} S_4  \tc_r S_2 \tb_r S_1 \tD_0(\lam)M_v  
\]
which modulo GPR is equal to 
\[
-h_1(\lam)^{-1} M_v S_4 \tB_3(\lam)^{-1} S_4 (\tg_2(\lam)G_4^{(v)}+ 
G_{4,l}^{(v)}) \tilde{D}_2 G_2^{(v)}\tD_0(\lam)\tilde{D}_1 \tD_0(\lam)M_v. 
\]
We further simplify this modulo GPR: We 
first replace $\tB_3(\lam)^{-1} $ 
by $T_4^{-1}$ since $\Og^{(4)}_{S_4\HL}(\lam^2(\log\lam)^2)$ 
produces GPR; secondly, we replace the right most 
$\tD_0(\lam)=D_0 + h_1(\lam) L_0$ by $h_1(\lam)L_0$, which is possible 
since $D_0 M_v \Pi(\lam)= D_0 M_v (\Pi(\lam)-\Pi(0))$ can be written 
in the form \refeq(reason) and produces GPR; this $h_1(\lam)$ 
cancels $h_1(\lam)^{-1}$ in the front and $G_{4,l}^{(v)}$ may be removed; 
finally, since $\tg_2(\lam)h_1(\lam)$ is GMU, another 
$\tD_0(\lam)$ may be replaced by $D_0$. In this way we have obtained 
that modulo GPR  
\begin{align}  \lbeq(tbr-tcr-simple)
& \Og(\tb_r, \tc_r)(\lam)\equiv 
\tg_2(\lam) M_v S_4 T_4^{-1} S_4 G_4^{(v)}\tilde{D}_2 G_2^{(v)}
\tilde{D}_1 L_0 M_v \\
&  = \sum_{j,k=1}^l \tg_2(\lam) {t}_{jk} (v\z_j)\otimes {\r}_{k}, 
\quad {\r}_{k}=  M_v L_0 \tilde{D}_1 G_2^{(v)}\tilde{D}_2 G_4^{(v)}\z_k 
\notag
\end{align}
and $\Og(\tb_r, \tc_r)(\lam)$ produces modulo GOP 
the linear combination of 
\begin{align} \lbeq(brcr)
&\int_0^\infty (R^{+}_0 (\lam^4)((v\z_j)\otimes \r_{k})\Pi(\lam)u 
\lam^3 \tg_2(\lam) \chi_{\leq{a}}(\lam)d\lam \\
& = \int_{\R^4} \r_{k}(z)
\left(\int_0^\infty (R^{+}_0 (\lam^4)(v\z_j)(x) 
(\Pi(\lam)\tau_{-z} u)(0) 
\lam^2 \m(\lam)\chi_{\leq{a}}(\lam)d\lam\right) dz, \notag 
\end{align}
where $\m(\lam)\colon = \lam \tg_2(\lam)$ is GMU.  The function inside 
the brackets is equal to $W_B u$ of \refeq(Bdef) with $\ph$ and $u$ 
being replaced $v\z_j$ and $\tau_{-z}\m(|D|)u$ respectively. 
Then \reflm(Bad) and \reflm(Bl-p) imply the lemma. 
\edpf 

\bgrm We believe that \reflm(oturi-lp4) holds for $1<p<\infty$ 
since $\int_{|R^4}v(x) \z_j(x) dx=0$. We do not, however, 
pursue this here. 
\edrm   

We next consider the terms which contain one  
from $\{\ta_r, \tb_r,\tc_r\}$ and another from $\{\ta_l, \tb_l,\tc_l\}$. 

\bglm If $\tf_l\in \{\ta_l, \tb_l,\tc_l\}$ and 
$\tg_r\in \{\ta_r, \tb_r,\tc_r\}$, then $\Og(\tf_l,\tg_r)(\lam)$ is GPR. 
\edlm 
\bgpf (i) Operators $\Og(\ta_l, \tg_r)(\lam) $ and 
$\Og(\tb_l, g_r)(\lam)$ are 
$\Gg\Vg\Sg$ since
\[
\ta_l S_4= -\lam^4 \tg_2(\lam)D_0G_4^{(v)}S_4, \quad 
\tb_l S_4= \lam^4 h_1(\lam)^{-1}\tg_2(\lam)D_1G_4^{(v)}S_4
\]
by virtue of the cancellation properties 
\refeqs(S4G2,S4tT) of  $S_4$, 
$\tg_r\in \Og_{\Hg_2 }^{(4)}(\lam^2\log\lam)$ 
and $M_v$ sandwiches them. 

(ii) The operators 
$\Og(\tc_l, \ta_r)(\lam)$ and 
$\Og(\tc_l, \tb_r)(\lam) $ are also $\Gg\Vg\Sg$ 
since $\tc_l=\Og_{S_2\HL} (\lam^2 \log\lam)$ 
and $S_4\ta_r, S_4 \tb_r \in \Og_{\Hg_2} (\lam^4 (\log\lam)^2)$. 

(iii) Let $\{\z_1, \dots, \z_m\}$ be the basis of $S_2\HL(\R^4)$. 
Then, \refeq(F3-simp-l), \refeq(F3-simp-r), 
the cancellation properties \refeqs(S4G2,S4tT) of $S_4$ 
and the matrix representation \refeq(matrix-T4) for $\tB_3(\lam)^{-1}$ 
jointly imply that  $\Og(\tc_l, \tc_r)(\lam)$ is equal to
\[
\lam^{-4}M_v S_2 \tc_l S_4 \tB_3(\lam)^{-1} S_4 \tc_r S_2 M_v 
= \sum_{j,k=1}^m a_{jk}(\lam)(v\z_j) \otimes (v\z_k)
\]
with $a_{jk}(\lam)\in \Og^{4}_{\C}((\log\lam)^2)$, 
$j,k=1, \dots, m$. Then, \reflm(s-low) implies that 
$\Og(\tc_l, \tc_r)$ is GPR.  
\edpf 

Finally we consider the terms which contain 
one from $\{\ta_r, \tb_r, \tc_r\}$ but none of 
$\{\ta_l, \tb_l, \tc_l\}$, viz. $\Og(\ta_r)(\lam)$, 
$\Og(\tb_r)(\lam)$ and $\Og(\tc_r)(\lam)$. 

\bglm Operator $\Og(\ta_r)(\lam)$ is GPR 
and $\Og(\tb_r)$ produces an operator bounded in $L^p(\R^4)$ for $1<p<4$. 
\edlm 
\bgpf (i) We have  
$\Og(\ta_r)= - \tg_2(\lam) M_v S_4 \tB_3(\lam)^{-1}S_4 G_4^{(v)}D_0 M_v$ 
and, in terms of the basis $\{\z_{r'}, \dots, \z_m\}$ of $S_4\HL$,
\[
\Og(\ta_r)(\lam)= - \sum_{j,k=r'}^m 
\tg_2(\lam)\tilde{t}_{jk}(\lam) (v\z_j)\otimes 
(M_v D_0 G_4^{(v)}\z_k).
\]
Here $g(x)=M_v D_0 G_4^{(v)}\z_k(x)$ satisfies $\int_{\R^4}g(x) dx=0$ 
as previously, since $D_0=QD_0$ and $Q v=0$. Thus,  
\reflm(s-low) implies $\Og(\ta_r)$ is GPR.

(ii) By virtue of \refeq(11S22-suppl) and \refeq(S4G2), 
$\Og(\tb_r)$ is equal modulo GPR to 
\[
h_1(\lam)^{-1} \tg_2(\lam) 
M_v S_4 T_4^{-1}S_4 G_4^{(v)}D_1 S_1 (D_0 + h_1(\lam) L_0)M_v.
\]
Expanding the sum, we have two terms. The term which arises from  
$D_0$ is GPR as in (i) above, since 
$\int_{\R^4} (M_v D_0 S_1D_1 G_4^{(v)} \z_k)(x)dx=0 $;  
the term which contains $h_1(\lam) L_0$ is equal to the
\[
\sum_{j,k=1}^l \tg_2(\lam){t}_{jk}(v\z_j)\otimes \tilde{\r}_k, \quad 
\tilde{\r}_k= M_v L_0 S_1 D_1G_4^{(v)}\z_k \in L^1(\R^4)
\]
which is the same as \refeq(tbr-tcr-simple) with ${\r}_k$ 
being replaced by $\tilde{\r}_k$. Since $\ax \tilde{\r}_k\in L^1(\R^4)$,
the proof of \reflm(oturi-lp4) implies that the second term 
also produces a bounded operator in $L^p(\R^4)$ for $1<p<4$. 
This proves the lemma. 
\edpf

The next lemma for $\Og(\tc_r)(\lam)$ concludes the proof of 
\refth(main-rephrase)(4). 

\bglm \lblm(tcr)
The operator $\Og(\tc_r)(\lam)$ produces a bounded operator 
in $L^p(\R^4)$ for $1<p<4$.
\edlm 
\bgpf We have 
$\Og(\tc_r)(\lam)= \lam^{-4}M_v S_4 \tB_3(\lam)^{-1}S_4 \tc_r S_2 M_v$. 
Since it has $S_2 M_v$ on the right end, 
$\Og_{\Hg}^{(4)}(\lam^4(\log\lam)^2)$ in \refeq(F3-simp-l) 
for $S_4 \tc_l$ produces GPR by \reflm(s-low) and,  
$\Og(\tc_r)(\lam) \equiv 
\lam^{-2}M_v S_4 \tB_3(\lam)^{-1}S_4 T_{4,l}(\lam)S_2 D_2 S_2 M_v$ 
modulo GPR. 
Then, 
$\Og^{(4)}_{S_4\HL}(\lam^2(\log\lam)^2)$ in 
\refeq(tB3) for $\tB_3(\lam)^{-1}$ 
produces GPR by the reason and  
\bqn \lbeq(9-last)
\Og(\tc_r)(\lam) \equiv 
\lam^{-2}M_v S_4 T_4^{-1}S_4 
(\tg_2(\lam) G_4^{(v)} + G_{4,l}^{(v)})S_2 D_2 S_2 M_v\,.
\eqn 
We express the right of \refeq(9-last) via the basis 
$\{\z_1, \dots, \z_m\}$ of $S_2\HL$ such that 
$\{\z_{r'} \dots, \z_m\}$ is the one of $S_4\HL$. We obtain 
\bqn 
\Og(\tc_r)(\lam)\equiv 
\sum_{j=r'}^m \sum_{k=1}^m \lam^{-2} \c_{jk}(\lam)(v\z_j)\otimes (v\z_k), 
\quad \c_{jk}(\lam)= c_{jk}\log\lam + d_{jk}.
\eqn 
Thus, the operator \refeq(sta0-low) produced by $\Og(\tc_r)(\lam)$ 
becomes 
\bqn \lbeq(ocr)
\sum_{j=1}^l \sum_{k=1}^n 
\int_0^\infty (R^{+}_0 (\lam^4)M_v\z_j)(x) \la v\z_k, \Pi(\lam)u \ra
\lam \c_{jk}(\lam) \chi_{\leq{a}}(\lam)d\lam 
\eqn 
Since $\int_{\R^4} {v}(z)\z_k(z) dz =0$, we may replace 
$\Pi(\lam)u$ by \refeq(reason) in \refeq(ocr). Then each summand 
becomes $i \sum_{m=1}^4 \int_0^1 d\th\int_{\R^4} z_m (v\z_k)(z) dz$ of 
\bqn \lbeq(ocr-jkm)
\int_0^\infty 
(R^{+}_0 (\lam^4)M_v\z_j)(x)\Pi(\lam)(\tau_{-\th{z}} R_m u)(0) 
\lam ^2 \c_{jk}(\lam) \chi_{\leq{a}}(\lam)d\lam \,.
\eqn 
If we replace $\c_{jk}(\lam)$ by the constant $d_{jk}$,  
the operator \refeq(ocr-jkm) becomes the trivial modification of 
\refeq(B-op) and it is bounded in $L^p(\R^4)$ for $1<p<4$. 
Thus, the next lemma with Minkowski's inequality 
completes the proof of \reflm(tcr) 
\edpf 

\bglm \lblm(last-lm) 
Let $\z\in S_4\HL$. Then, the operator 
\bqn \lbeq(Zadd)
Z_{add}u(x)= \int_0^\infty R^{+}_0 (\lam^4)(v\z)(x)\Pi(\lam)u(0) 
\lam^2 (\log \lam) \chi_{\leq{a}}(\lam)d\lam,
\eqn 
is bounded in $L^p(\R^4)$ for $1<p<4$,
\edlm 
\bgpf 
The proof is the modification of the one of \reflms(Bad,Bl-p). \\[3pt]
(1) We first show that 
$\chi_{\geq 4a}(|D|)Z_{add}\in \Bb(L^p)$ for $1<p<4$.
In view of the proof of \reflm(Bad), \refeq(chge2a) in particular, 
it suffices to show that 
the following linear functional is a bounded  on $L^p(\R^4)$ for $1<p<4$:
\[
\tilde{l}(u)
=\int_0^\infty \Pi(\lam)u(0)\lam^2 (\log \lam)\chi_{\leq{a}}(\lam) d\lam 
= \frac1{(2\pi)^2}\int_{\R^4}u(x)f(x)dx, 
\]
where 
$f(x)=\Fg\left(\chi_{\leq{a}}(|\xi|)(|\xi|^{-1}\log|\xi|\right)(x)$. 
However, this is obvious by H\"older's inequality since 
$\Fg(|\xi|^{-1}(\log|\xi|))(x) =|x|^{-3}(\a \log |x|+\b)$
for constants $\a$ and $\b$ (cf. \cite{Grafakos-1}, Theorem 2.4.6 )
and, for $4/3<p\leq \infty$, 
\bqn \lbeq(log-Fourier)
|f(x)|\leq C \ax^{-3}\la \log(1+|x|) \ra 
\in L^p(\R^4) \,.
\eqn 

(2) We next show that 
$\chi_{\leq 4a}(|D|)Z_{add}u(x)$ is bounded in $L^p(\R^4)$ for $1<p<4$ 
which is equal to 
\[
\int_0^\infty \chi_{\leq 4a}(|D|)R^{+}_0 (\lam^4)(v\z)(x)\Pi(\lam)u(0) 
\lam^2 (\log \lam) \chi_{\leq{a}}(\lam)d\lam.
\]
We follow the argument of \reflm(Bl-p).
Let $\ph(x)  = v(x) \z(x)$. Since $\z\in S_4\HL$.   
$\int_{\R^4}x^\a \ph(x)dx=0$ for $|\a|\leq 2$, we have   
\[
\hat{\ph}(\xi)= \frac{(-iz\xi)^3}{2(2\pi)^2}
\int_0^1 (1-\th)^2\left(\int_{\R^4} e^{-i{\th}z\xi}\ph(z) dz \right) d\th
\]
and $\chi_{\leq 4a}(|D|)R_0(\lam^4+i\ep)\ph(x)$ is equal to 
$\sum_{|\a|=3}C_\a \int_0^1 (1-\th)^2 d{\th} $ of 
\bqn \lbeq(64-a)
\int_{\R^4} z^\a \ph(z) 
R^\a \tau_{\th{z}} \left(\int_{\R^4}\frac{e^{ix\xi}|\xi|^3\chi_{\leq{4a}}(|\xi|)}
{|\xi|^4-\lam^4-i\ep} \frac{d\xi}{(2\pi)^4}\right)dz.   
\eqn 
We substitute    
\bqn \lbeq(mult-2)
\frac{|\xi|^3}{|\xi|^4-\lam^4-i\ep}
= \frac{\lam^3}{|\xi|^4-\lam^4-i\ep}+ 
\frac{|\xi|^3-\lam^3}{|\xi|^4-\lam^4-i\ep} 
\eqn 
and take the limit as $\ep\to  0$ . Then, the inner integral 
in \refeq(64-a)
is uniformly bounded by $C\ax^{-\frac32}$ and converges to 
\bqn \lbeq(two-terms)
\lam^3\chi_{\leq 4a}(|D|)\Rg_\lam(x) 
+ \frac1{2(2\pi)^4}\int_{\R^4}
\Big(\frac{1}{|\xi|+ \lam}+
\frac{|\xi|+\lam}{|\xi|^2+\lam^2}\Big)e^{ix\xi}
\chi_{\leq{4a}(|\xi|)}d\xi.
\eqn 
Then the first term of \refeq(two-terms) 
contributes to $\chi_{\leq 4a}(|D|)Z_{add}u(x)$ by 
the superposition by 
$\sum_{|\a|=3}C_\a \int_0^1 (1-\th)^2 d{\th}$ of  
\begin{align}
& \int_{\R^4} z^\a \ph(z) 
R^\a \tau_{\th{z}}\chi_{\leq{4a}}(|D|)
\left(\int_0^\infty \Rg_{\lam}(x) \Pi(\lam)\m_a(|D|)u(0)
\lam^3 d\lam\right)dz  \notag \\
& = \int_{\R^4} z^\a \ph(z) 
R^\a \tau_{\th{z}}\chi_{\leq{4a}}(|D|) K \m_a(D) u(x)dz, \lbeq(64-b)
\end{align}
where $\m_a(\lam)= \chi_{\leq{a}}(\lam) (\log\lam)\lam^2 $ is GMU. 
Thus, \refeq(64-b) is GOP by virtue of \reflm(allp).  

If we ignore harmless constants, 
the last integral term of \refeq(two-terms) 
contributes to $\chi_{\leq 4a}(|D|)Z_{add}u(x)$ by  the superposition by 
$\sum_{|\a|=3}C_\a \int_0^1 (1-\th)^2 d{\th}\int_{\R^4}
z^\a \ph(z)R^\a \tau_{\th{z}}dz $ of  
\begin{align*}
& \int_0^\infty \left( 
\int_{\R^4}
\Big(\frac{1}{|\xi|+ \lam}+
\frac{|\xi|+\lam}{|\xi|^2+\lam^2}\right)e^{ix\xi}\chi_{\leq{4a}}d\xi\Big)
\Pi(\lam)u(0)\lam^2\log\lam \chi_{\leq{a}}(\lam) d\lam 
\\ 
& \hspace{1cm} 
= \int_{\R^4}(L_(x,y)+ L_2(x,y))u(y) = \colon L_1 u(x) + L_2 u(x).
\end{align*}
Here, considering $\lam \w=\eta$ as polar coordinates of $\eta\in \R^4$, 
we comupte  
\begin{align}
L_1(x,y)& =\int_{\R^8} \frac{
e^{ix\xi-iy\eta}\chi_{\leq{4a}}(|\xi|)\chi_{\leq{a}}(|\eta|)
}
{(2\pi)^2 (|\xi|+|\eta|)}
\frac{\log|\eta|}{|\eta|} d\xi d\eta, \lbeq(L1-def)\\
L_2(x,y)& =\int_{\R^8} \frac{
e^{ix\xi-iy\eta}(|\xi|+|\eta|)
\chi_{\leq{4a}}(|\xi|)\chi_{\leq{a}}(|\eta|)}
{(2\pi)^2 (|\xi|^2+|\eta|^2)}
\frac{\log|\eta|}{|\eta|} d\xi d\eta\,. \lbeq(L2-def)
\end{align}
The proof will be finished if we prove that 
$L_1$ and $L_2$ are bounded in $L^p(\R^4)$ for $1<p<4$. Let $q=p/(p-1)$. 
Obviously $4/3<q<\infty$. 

(i) The obvious modification of the proof of \reflm(2-step) by using 
\refeq(log-Fourier) in stead of 
$|\Fg(|\eta|^{-1}\hat{\chi}_{\leq a}(\eta))(y)|\leq C\ay^{-3}$ implies 
\[
|L_1(x,y)|\leq \frac{C \log (|y|+2)}{\ax^3\ay (1+|x|+|y|)^2}\,.
\] 
It is obvious that 
$\|L_1(x,\cdot)\|_{q}\leq C_q\ax^{-3}$ for $q>4/3$ and we need more 
decay as $|x|\to \infty$. Assume first 
$|x|\geq 10^{10}+ \exp(2q/(3q-4))$. Then, 
\bqn \lbeq(L1xleqy)
\left(\int_{|x|\leq |y|} |L_1(x,y)|^q dy \right)^{\frac1{q}}
\leq \frac{C}{\ax^3}
\left(\int_{|x|}^\infty \frac{(\log r)^q}{r^{3q-3}}dr\right)^{\frac1{q}} 
\leq \frac{C\log |x|}{\ax^{2+\frac{4}{p}}}. 
\eqn 
Here we used that the second term on the right of  
\[
\int_{|x|}^\infty \frac{(\log r)^q}{r^{3q-3}}dr= 
\frac1{3q-4}\frac{(\log |x|)^q}{|x|^{3q-4}} 
+\frac{q}{3q-4}\int_{|x|}^\infty \frac{(\log r)^{q-1}}{r^{3q-3}}dr
\]
is smaller than one-half of the left hand side for 
$x$ under consideration. For the integral on $|y|\leq |x|$ we have 
\begin{align} \lbeq(L1yleqx)
& \left(\int_{|y|\leq |x|} |L_1(x,y)|^q dy \right)^{\frac1{q}}
\leq \frac{C\log|x|}{\ax^5}
\left(\int_0^{|x|} \frac{r^3 dr}{\la r\ra^{q}}dr\right)^{\frac1{q}} \\
& \hspace{2cm} \leq C \log |x| \left\{\begin{array}{ll} 
\ax^{-5}, \ & q>4, \\
\ax^{-2-\frac{4}{p}} , \  & q<4.
\end{array}
\right.  \notag 
\end{align}
Estimates \refeq(L1xleqy) and \refeq(L1yleqx) imply 
\bqn 
\|L_1 u\|_p \leq  
\left(\int_{\R^4}\|L_1(x,y)\|_{L^q(\R^4}^p dx\right)^\frac1{p} \|u\|_p 
\leq C \|u\|_p
\eqn 
(ii) Let $\n_{8a}(\xi)= |\xi|\chi_{\leq{8a}}(|\xi|)$,  
$\n_{2a}(\eta)= |\eta|\chi_{\leq{2a}}(|\eta|)$ and 
\[
{L}_{3}(x,y) =\int_{\R^8} \frac{
e^{ix\xi-iy\eta}\chi_{\leq{4a}}(|\xi|)\chi_{\leq{a}}(|\eta|)}
{|\xi|^2+|\eta|^2}
\frac{\log|\eta|}{|\eta|} d\xi d\eta 
\]
We have, replacing $|\xi|+|\eta|$ by $\n_{8a}(\xi)+ \n_{2a}(\eta)$ 
in \refeq(L2-def), 
\[
L_{2}=\n_{8a}(D){L}_{3}+ {L}_{3}\n_{2a}(D);
\]
$\hat{\n}_b \in L^1(\R^4)$ for any $b>0$ and $\n(|D|)\in \Bb(L^p)$, 
$1\leq p \leq \infty$; by virtue of \refeq(log-Fourier), 
\[
|{L}_{3}(x,y)|
\leq 
\int_{\R^4} \frac{\log(2+|y-z|)dz}{(1+|x|+|z|)^6 \la y-z\ra^3}\,.
\]
Then, H\"older's inequality implies 
\[
\sup_{y \in \R^4} \int_{\R^4} |{L}_{3}(x,y)| dx 
\leq \sup_{y \in \R^4}
\int_{\R^4}\frac{\log(2+|y-z|)dz}{(1+|z|)^2 \la y-z\ra^3} \\
<\infty 
\]
and ${L}_{3}$ is bounded in $L^1(\R^4)$. 
For $2<p<4$, $4/3<q<2$ and 
\[
|{L}_{3}u(x)|  
\leq \|u\|_p \left\|\frac{\log (1+|y|)}{\ay^3}\right\|_{q}
\int_{\R^4}\frac{dz}{(1+|x|+|z|)^6}
\leq C \frac{\|u\|_p}{\ax^{2}} \in L^p(\R^p) 
\]
Thus, $L_{3}$ is bounded in $L^p(\R^4)$ also for $2<p<4$ 
and, hence, for all $1\leq p<4$ by interpolation. Thus, so is $L_2$ 
and the lemma is proved. 
\edpf

\section{Appendix, Proof of \reflm(intL)}

We substitute \refeq(spect-proj) for $\Pi(\lam)u(0)$, 
use polar coordinates $\eta= \lam\w$  and change 
the order of integration. This implies that 
$L$ is the integral operator with kernel 
\bqn \lbeq(Lkernel)
L(x,y) = \iint_{\R^8}
\frac{e^{ix\xi+iy\eta} \chi_{\leq 4a}(|\xi|)\chi_{\leq{a}}(|\eta|)}
{(|\xi|^2+|\eta|^2)(|\xi|+|\eta|)|\eta|}d\xi{d\eta}\,.
\eqn 

We admit the following lemma for the moment and complete the proof 
of \reflm(intL). 
\bglm \lblm(last-lamma-a)
There exists constant $C>0$ such that 
\bqn \lbeq(int-L)
|L(x,y)|\leq \frac{C}{\ax(1+|x|+|y|)^3}\,.
\eqn 
\edlm

The right side of \refeq(int-L) is ``almost'' 
homogeneous in $(x,y)$ and we apply the well-known technique 
for the homogenous kernel: Change variables $y= (1+|x|){z}$ 
and integrate over the spherical variables first
\[
|Lu(x)| 
\leq C \int_{\R^4} \frac{|f((1+|x|)z)|dz}{(1+|z|)^3} \\ 
= 
C\c_3 \int_{0^\infty} \frac{M_{|f|}((1+|x|)r) r^3 dr }{(1+r)^3}
\]
where $\c_3={\rm meas}({\mathbb S}^3)$. 
Then H\"older's inequality implies 
\begin{align*}
& \int_{\R^4} M_{|f|}((1+|x|)r)^p dx =  
\c_3 \int_{0}^\infty M_{|f|}((1+\r){r})^p \r^3 d\r \\
& \leq \c_3 \int_{1}^\infty (M_{|f|}(\r{r}))^p \r^3 d\r 
\leq \c_3 r^{-4}\int_{\R^4} M_{|f|}({\r})^p \r^3 d\r 
\leq C r^{-4}\|f\|_p^p  .
\end{align*}
Hence, Minkowski's inequality implies that for $1<p<4$
\bqn 
\|L f\|_p \leq C \c_3
\int_0^{\infty} \frac{r^{3-\frac{4}{p}} \|f\|_p dr }{(1+|r|)^3}
\leq C' \|f\|_p .
\eqn 
This completes the proof of \reflm(Bl-p). 

We show \reflm(last-lamma-a). Note that $L(x,y)$ is 
the convolution of Fourier transforms of 
\[
\frac{\chi_{\leq a}(|\xi|)\chi_{\leq{a}}(|\eta|)}{|\xi|+|\eta|}, \quad 
\frac{\chi_{\leq{a}}(|\eta|)}{|\eta|}, \quad 
\frac{\chi_{\leq a}(|\xi|)\chi_{\leq{a}}(|\eta|)}{|\xi|^2+|\eta|^2}.  
\]

\bglm \lblm(9-1) 
For $a>0$, there exists a constant $C>0$ such that 
\bqn \lbeq(9-1) 
\iint_{\R^8}e^{ix\xi-iy\eta}\frac{\chi_{\leq a}(|\xi|)\chi_{\leq{a}}(|\eta|)}
{|\xi|+|\eta|}d\xi{d\eta} \absleq 
\frac{C}{\ax^{2}\ay^{2}(\ax+\ay)^3}.
\eqn
\edlm 
\bgpf Following the argument in \cite{Stein-old}, pp. 61-62, we have 
\begin{align*} 
& \iint_{\R^8}\frac{e^{ix\xi-iy\eta}}{|\xi|+|\eta|}d\xi{d\eta}
= \int_0^\infty 
\left(\int_{\R^4}e^{ix\xi-t|\xi|}d\xi\right)
\left(\int_{\R^4}e^{-iy\eta-t|\eta|}d\eta\right)dt \\
& = \frac{c_4}{(|x|^2+|y|^2)^{\frac72}} 
\int_0^\infty\frac{s^2ds}{(s^4+ s^2  + F^2)^\frac52},  
\quad 0\leq F= \frac{|x||y|}{|x|^2+ |y|^2}\leq \frac12.
\end{align*} 
The last integral is bounded by  
$\int_0^\infty s^2(s^2  + F^2)^{-\frac52}ds = C F^{-2}$ 
and  
\bqn \lbeq(F-transform-1)
\iint_{\R^8}\frac{e^{ix\xi-iy\eta}}{|\xi|+|\eta|}d\xi{d\eta}
\absleq \frac{C}{|x|^{2}|y|^{2}(|x|+|y|)^3}.
\eqn 
Then, for large $N$, 
\bqn \lbeq(fi)
\iint_{\R^8}\frac{\la x-z\ra^{-N}\la y-w \ra^{-N}}
{|z|^{2}|w|^{2}(|z|+|w|)^3}dw dz \leq
\frac{C}{\ax^{2}\ay^{2}(\ax+\ay)^3}
\eqn 
and \refeq(9-1) follows. 
\edpf

\bglm \lblm(2-step)
For $a>0$, there exists a constant $C>0$ such that 
\[ 
\iint_{\R^8}e^{ix\xi-iy\eta}\frac{\chi_{\leq a}(|\xi|)\chi_{\leq{a}}(|\eta|)}
{(|\xi|+|\eta|)|\eta|}d\xi{d\eta} \absleq 
\frac{C}{\ax^{3}\ay(\ax+\ay)^2}.
\] 
\edlm 
\bgpf It is well-known that 
$|\Fg(|\eta|^{-1}\hat{\chi}_{\leq a}(\eta))(y)|\leq C\ay^{-3}$ 
and, in virtue of \reflm(9-1), it suffices to prove  
\bqn \lbeq(9-1-a)
\int_{\R^4}\frac{dz}{\az^{2}(1+|x|+|z|)^3\la y-z \ra^3}
\leq \frac{C}{\ax \ay(\ax+\ay)^2}.
\eqn 
Let  $\D_1=\{|y-z|\leq |y|/2\}$, $\D_2 =\{|z|\geq 2|y|\}$ 
and $\D_3=\{|y|/2<|y-z|, \ |z|\leq 2|y|\}$such that 
$\R^4= \D_1 \cup \D_2 \cup \D_3$.
Denote the integrand of \refeq(9-1-a) by $F(x,y,z)$. 
Since $|y|/2\leq |z| \leq 3|y|/2$ on $\D_1$, by using polar coordinates 
we have 
\[
\int_{\D_1} F(x,y,z) dz \leq 
\int_0^{|y|/2}\frac{r^3 dr}{\ay^{2}(1+|x|+|y|)^3 \la r\ra^3}
\leq \frac{C}{\ay(1+|x|+|y|)^3}.
\]
Since $|y-z|\geq |z|/2$ on $\D_2$, 
\[
\int_{\D_2} F(x,y,z) dz \leq 
\int_{|z|>2|y|}\frac{dz}{\az^{5}(1+|x|+2|y|)^3}
\leq \frac{C}{\ay (1+|x|+|y|)^3}.
\]
Since $|y|/2<|z-y|\leq 5|y|/2$ on $\D_3$, 
\[
\int_{\D_3}F(x,y,z) dz 
\leq \frac{C}{\ay^3}\int_{0}^{2|y|} \frac{rdr}{(1+|x|+r)^3} 
\leq \frac{C}{\ay \ax(1+|x|+|y|)^2}\,.
\]
Summing up, we obtain \refeq(9-1-a). 
\edpf 

\paragraph{\bf Completion of proof of \reflmb(last-lamma-a)}

It is well known that 
\[
\Fg\left(
\frac{\chi_{\leq a}(|\xi|)\chi_{\leq{a}}(|\eta|)}{|\xi|^2+|\eta|^2}\right),  
\absleq \frac{C}{(1+|x|+|y|)^6}\,.
\]
Hence, by virtue of \reflm(2-step), it suffices to show 
\[
\int_{\R^8}
\frac{dw dz}{(\la x-w\ra+\la y-z\ra)^6 \la w \ra^{3}\az(\aw+\az)^2} 
\leq \frac{C}{\ax(1+|x|+|y|)^3}.
\]
Denote the integrand by $F= F(x, y, w,z)$ and split 
$\R^4_w= \D_1 \cup \D_2 \cup \D_3$ and 
$\R^4_z= \D_1' \cup \D_2' \cup \D_3'$ where 
$w\in \D_1, \D_2, \D_3$ and 
$z\in \D_1', \D_2', \D_3'$ respectively satisfies 
\begin{gather*}
|w-x|\leq |x|/2, \ |x|/2 <|w-x|\leq 2|x|, \ |w-x| \geq 2|x| ; \\
|z-y|\leq |y|/2, \ |y|/2 <|z-y|\leq 2|y|, \  |z-y| \geq 2|y| 
\end{gather*} 
We shall separately prove for $1\leq j,k \leq 3$ that 
\bqn 
L_{jk}(x,y)= \int_{\D_j \times \D_k'}F(x, y, w,z)dwdz 
\leq \frac{C}{\ax(1+|x|+|y|)^3}. \lbeq(Ljk) 
\eqn 
(11) Since $|x|/2\leq |w|<(3/2)|x|$ and $|y|/2\leq |z|<(3/2)|y|$ on 
$\D_1\times \D_1'$, 
\[
L_{11}(x,y) 
\absleq \frac{1}{\la x \ra^{3}\ay(1+|x|+|y|)^2}
\int_0^{|x|/2}\int_0^{|y|/2}\frac{r^3 \r^3 dr d\r}{(1+r+\r)^6} 
\]
and by changing variables $r=\s^\frac14$, $\r=\t^\frac14$ we estimate the 
integral by $\int_0^{|x|^4}\int_0^{|y|^4}(1+\s+\t)^{-\frac32}d\s d\t$ 
which is equal to 
\[
4C((1+|x|^4)^{\frac12}+ (1+|y|^4)^{\frac12} - (1+|x|^4+|y|^4)^{\frac12}-1)
\] 
and bounded by $C \ax^2 \ay^2 (\ax^2+ \ay^2)^{-1}$.   
\refeq(Ljk) for $L_{11}(x,y)$ follows. \\[3pt]
(12) Let $I_1(x,y)$ and $I_2(x,y)$ be defined by 
\[
I_1(x,y)=\int_{\D_1}\frac{dw}{(1+|x-w|+|y|)^6}, 
\ \  
I_2(x,y)= \int_{\D_2'} \frac{dz}{\az(1+|x|+|z|)^2}.
\]
Changing the variables $r=\s^\frac14$ and $\r=\t^\frac13$, 
we estimate 
\begin{align*}
& I_1(x,y)
= \c_3 \int_{0}^{|x|/2} \frac{r^3 dr}{(1+r+|y|)^6}
\leq  C \frac{|x|^4} {(1+|y|^4)^\frac12 (1+|x|^4+|y|^4)}. \\
& I_2(x,y) 
\leq C \int_{0}^{3|y|} 
\frac{\r^2 d\r}{(1+|x|+\r)^2} \leq \frac{C |y|^3}{(1+|x|^2+|y|^2 )}\,.
\end{align*}
From that  $|x|/2\leq |w|<(3/2)|x|$, 
$|y|/2\leq |y-z|\leq 2|y|$ and $|z|<3|y|$ on $\D_1\times \D_2'$ 
we then obtain 
\begin{align*}
& L_{12}(x,y)\leq 
\frac{C}{\ax^3}
\int_{\D_1 \times \D_2'}\frac{C dw dz}{(1+|x-w|+|y|)^6 \az(1+|x|+|z|)^2} \\
& \hspace{1cm} \leq \frac{C}{\ax^3}I_1(x,y) I_2(x,y) 
\leq \frac{C }{\ax(1+|x|+|y|)^3}.
\end{align*}

\noindent 
(13) Since $|x|/2\leq |w| \leq 3|x|/2$ on $\D_1$, we have  
\[
L_{13}(x,y) \leq \frac{C}{\ax^3} \int_{\D_3'} \left(
\int_{\D_1}\frac{dw}{(1+|x-w|+|y-z|)^6}
\right)
\frac{dz}{\az(1+|x|+|z|)^2}
\]
The integral by $dw$ produces $I_1(x,y-z)$ of the part (12) and, 
since $|z|/2\leq |y-z|\leq 2|z|$ for $z\in \D_3'$   
\[
I_1(x,y-z)\leq \frac{C |x|^4} {(1+|z|^4)^\frac12 (1+|x|^4+|z|^4)}.
\]
Since $|z|>|y|$ for $z\in \D_3'$, it follows that 
\begin{align*}
L_{13}(x,y) \leq C|x| 
\int_{|z|>|y|}\frac{dz}{\az^ 3 (1+|x|+|z|)^6} \leq \frac{C}{\ax(1+|x|+|y|)^3}.
\end{align*}
 
(21) Since $|x|/2\leq |x-w|\leq 2|x|$, we have 
\begin{align*}
& L_{21}(x,y) \leq  \int_{\D_2\times \D_1'}
\frac{C dw dz}{(1+|x|+|y-z|)^6 \la w \ra^{3}\ay(1+|w|+|y|)^2} \\
& \leq \frac{C}{\ay}
\left(\int_{D_2}\frac{dw}{\la w \ra^{3}(1+|w|+|y|)^2}\right) 
\left(\int_{\D_1'}\frac{dz}{(1+|x|+|y-z|)^6}\right).
\end{align*}
Since $|w|\leq 3|x|$ on $D_2$, the first integral by $dw$ on the right 
is bounded by a constant times  
\[
\int_0^{3|x|}\frac{dr}{(1+r+|y|)^2}
\leq \frac{C|x|}{(1+|y|)(1+|x|+|y|)}
\]
and the second is equal to $I_1(y,x)$. Hence  
\[
L_{21}(x,y) \leq \frac{C|y|^2}{\ax (1+|x|+|y|)^5} 
\leq \frac{C}{\ax (1+|x|+|y|)^3}
\]
as desired. 

\noindent 
(22)  
On $\D_2\times \D_2'$, $|x|/2<|w-x|<2|x|$ and $|y|/2<|z-y|<2|y|$ 
and $|w|\leq 3|x|$ and $|z|\leq 3|y|$, we have 
\[
L_{22}(x,y)\leq  \frac{C}{(1+|x|+|y|)^6}
\int_{\D_2\times \D_2'}
\frac{dw dz}{\la w \ra^{3}\az(1+|w|+|z|)^2}.
\]
Using polar coordinates $w=r\w$ and $z=\r\w'$, we estimate the integral by   
\[
C \int_{0}^{3|y|}\int_0^{3|x|}\frac{\r^3 drd\r}{(1+r+\r)^2(1+\r)} 
\leq C \int_{0}^{3|y|}\int_0^{3|x|} dr d\r =9C|x||y|.
\]
It follows that 
\[
L_{22}(x,y)\leq  
\frac{C|x||y|}{(1+|x|+|y|)^6}
\leq \frac{C}{\ax (1+|x|+|y|)^3}.
\] 

\noindent 
(23) We have $|x|/2<|w-x|<2|x|$ and $|w|<3|x|$ on $\D_2$ and 
$|z|/2 \leq |z-y|<2|z| $ and $|z|>|y|$ on $\D_3'$. Hence 
\begin{align*}
& L_{23}(x,y) \leq 
\int_{|w|<3|x|, |z|>|y|}
\frac{C dw dz}{(1+|x|+|z|)^6 \la w \ra^{3}\az(1+|w|+|z|)^2} \\
& \leq C \int_{|y|}^\infty 
\left(\int_0^{3|x|} \frac{dr}{(1+r+\r)^2 }\right)
\frac{\r^3 d\r}{(1+|x|+\r)^6(1+\r)} \\
& \leq C |x| \int_{|y|}^\infty 
\frac{\r^3 d\r}{(1+|x|+\r)^7(1+\r)^2 } 
\leq C |x| \int_{|y|}^\infty \frac{{\r}d\r}{(1+|x|+\r)^7 } 
\end{align*}
We estimate the last member as previously by 
\[
C|x| \int_{|y|^2}^\infty \frac{d\r}{(1+|x|^2+\r)^{\frac72}} 
= \frac{C|x|}{(1+|x|^2+|y|^2)^{\frac52}} 
\leq \frac{C}{\ax (1+|x|+|y|)^3}
\]

(31) We have $|w|>|x|$ and $|w|/2 \leq |w-x|<2|w| $ on $\D_3$  
and $|z-y|<|y|/2$ and $|y|/2<|z|<3|y|/2$ on $\D_1'$. Hence 
\[
L_{31}(x,y) \leq C \int_{\D_3}\left(
\int_{\D_1}
\frac{dz}{(1+|w|+|y-z|)^6} \right)
\frac{dw}{\la w \ra^{3}\ay(1+|w|+|y|)^2}.
\]
The integral by $dz$ is equal to $I_1(y,w)$ in part (12). 
It follows that 
\begin{align*}
L_{31}(x,y) & \leq  \frac{C|y|^4}{\ay} 
\int_{|w|>|x|}\frac{dw}{\la w\ra^5 (1+|w|+|y|)^6} \\
& \leq  \frac{C}{(1+|x|+|y|)^3}\int_{|w|>|x|}\frac{dw}{\la w\ra^5 }
\leq  \frac{C}{\ax (1+|x|+|y|)^3}.
\end{align*}

\noindent 
(32) On $\D_3$ we have $|w|>|x|$ and $|w|/2 \leq |w-x|<2|w| $; 
on $\D_2'$ we have $|y|/2<|z-y|<2|y|$ and $|z|<3|y|$ on $\D_2'$ 
\[
L(x,y) \leq  
\int_{|w|>|x|} \left(\int_{|z|<3|y|}
\frac{dz}{\az(1+|w|+|z|)^2}\right) \frac{dw}{(1+|w|+|y|)^6 \la w \ra^{3}}.
\]
Using the estimate for $I_2(x,y)$ in the part (12), we have 
\begin{align*}
L(x,y) 
& \leq C |y|^3 \int_{|w|>|x|} \frac{dw}{(1+|w|+|y|)^8 \la w \ra^{3}} \\
& \leq C \int_{|x|}^\infty  \frac{dr}{(1+ r+|y|)^5} 
\leq \frac{C}{\ax (1+ |x|+|y|)^3} . 
\end{align*}

(33) Since $|w|>|x|$ and $|w|/2 \leq |w-x|<2|w| $ on $\D_3$  
and $|z|>|y|$ and $|z|/2 \leq |z-y|<2|z| $ on $\D_3'$, we have 
It follows that 
\begin{align*}
L(x,y) & \leq  \int_{|w|>|x|, |z|>|y|}
\frac{C dw dz}{(1+|w|+|z|)^8 \la w \ra^{3} \az} \\
& = \int_{|z|>|y|} 
\left(\int_{|w|>|x|}\frac{C dw}{(1+|w|+|z|)^8 \la w \ra^{3}}
\right)\frac{C dz}{(1+|z|)} \\
& \leq \int_{|z|>|y|}
\frac{C dz}{(1+|z|)(1+|x|+|z|)^7} \\
& \leq C \int_{|y|}^\infty \frac{dr}{(1+|x|+r)^5}
\leq \frac{C}{\ax(1+|x|+|y|)^3}.
\end{align*}
This completes the proof. \qed

\end{document}